\documentclass[11pt]{article}
\usepackage[centertags]{amsmath}
\usepackage{amsfonts} \usepackage{amssymb} \usepackage{amsthm}

\def\one{{\rm 1\kern -.9mm l}}

\def\beq{\begin{equation}}
\def\eeq{\end{equation}}
\def\beqa{\begin{eqnarray}}
\def\eeqa{\end{eqnarray}}

\newcommand{\bea}{\begin{eqnarray}}
\newcommand{\ena}{\end{eqnarray}}

\usepackage{graphicx}
\usepackage{dcolumn}
\def\one{{\hbox{ 1\kern-.8mm l}}}
\newcommand{\drawsquare}[2]{\hbox{%
\rule{#2pt}{#1pt}\hskip-#2pt
\rule{#1pt}{#2pt}\hskip-#1pt
\rule[#1pt]{#1pt}{#2pt}}\rule[#1pt]{#2pt}{#2pt}\hskip-#2pt
\rule{#2pt}{#1pt}}
\def\drawbox#1#2{\hrule height#2pt
        \hbox{\vrule width#2pt height#1pt \kern#1pt
              \vrule width#2pt}
              \hrule height#2pt}
\def\Asym#1#2{\vcenter{\vbox{\drawbox{#1}{#2}
              \kern-#2pt 
               \drawbox{#1}{#2}}}}
\newcommand{\Yfund}{\raisebox{-.5pt}{\drawsquare{6.5}{0.4}}}
\newcommand{\Ysymm}{\Yfund\hskip-0.4pt%
                    \Yfund}
\def\symm{\Ysymm}
\def\asymm{\Asym{6.5}{0.4}}
\def\Fund#1#2{\vcenter{\vbox{\drawbox{#1}{#2}}}}
\def\funda{\Fund{6.5}{0.4}}
\def\bfunda{\overline{\funda}}

\begin{document}
\begin{titlepage}
\rightline{NORDITA-2005/30 HE} \rightline{DSF-7/2005} \vskip 3.0cm
\centerline{\Large \bf { On the Gauge/Gravity Correspondence and}}
\vskip .5cm \centerline{\Large \bf {the Open/Closed String
Duality}} \vskip .5cm \vskip 1.4cm \centerline{\bf Paolo Di
Vecchia $^a$, Antonella Liccardo $^b$, Raffaele Marotta $^b$ and
Franco Pezzella $^b$} \vskip .8cm \centerline{\sl $^a$ NORDITA,
Blegdamsvej 17, DK-2100 Copenhagen \O,
Denmark}\centerline{e--mail: {\tt divecchi@alf.nbi.dk}} \vskip
.4cm \centerline{\sl
 $^b$ Dipartimento di
Scienze Fisiche, Universit\`{a} di Napoli and INFN, Sezione di
 Napoli} \centerline{\sl Complesso Universitario Monte
S. Angelo, ed. G  - via Cintia -  I-80126 Napoli, Italy}
\centerline{e--mail: {\tt
name.surname@na.infn.it}}

 \vskip 2cm
\begin{abstract}
In this article we review  the conditions for the  validity of the
{\em gauge/gravity correspondence} in both supersymmetric and
non-supersymmetric string models. We start by  reminding what
happens in  type IIB theory on the orbifolds
$\mathbb{C}^2/\mathbb{Z}_2$ and $\mathbb{C}^3/(\mathbb{Z}_2{\times}
\mathbb{Z}_2)$, where this correspondence beautifully works. In
these cases, by performing a complete stringy calculation of the
interaction among D3-branes, it has been shown that the fact that
this correspondence works is a consequence of the
 open/closed  duality  and of the absence of threshold
corrections. Then we review the construction of type 0 theories with
their orbifolds and orientifolds having spectra free from both open
and closed string tachyons and for such models we study the validity
of the gauge/gravity correspondence, concluding that this is not a
peculiarity of supersymmetric theories, but it may work also for
non-supersymmetric models.
Also in these cases, when it works, it is  again a consequence of
the open/closed string duality and of vanishing threshold
corrections.
\end{abstract}
\end{titlepage}

\newpage
\tableofcontents
\vskip 1cm

\section{Introduction}
\label{sez0} A D-brane is characterized by the fundamental
properties of being a solution of the low-energy string effective
action that is given by supergravity (SUGRA) and of having open
strings with their endpoints attached to its world-volume. In
particular, the lightest open string excitations correspond to a
gauge field and its supersymmetric partners if the theory is
supersymmetric. These two complementary descriptions of a D-brane
provide a powerful tool for deriving the {\em quantum} properties of
the gauge theory living on the D-brane world-volume from the {\em
classical} brane dynamics and viceversa. In particular, the fact
that one can determine the gauge-theory quantities in terms of the
supergravity solution goes under the name of {\em gauge/gravity
correspondence}. This has allowed to derive properties of
${\cal{N}}=4$ super Yang-Mills as one can see for example in
Ref.~\cite{CJ} and, by the addition of a decoupling limit, also to
formulate the Maldacena conjecture of the equivalence between
${\cal{N}}=4$ super Yang-Mills and type IIB string theory
compactified on $AdS_5 \otimes S^5$.~\cite{MALDA}

Although it has not been possible to extend the Maldacena conjecture
to non-conformal and less supersymmetric gauge theories,
nevertheless a lot of apriori unexpected informations on these
theories have been obtained from the gauge/gravity
correspondence~\footnote{For general reviews on various approaches
see for instance Ref.s~\cite{REV,KLEBA,MA,MILA,DL0307}.}. These more
realistic gauge theories can be identified with the ones living on
the world-volume of D5 branes wrapped either on a non-trivial
2-cycle of a non-compact Calabi-Yau space or on the ``shrinking''
2-cycle located at the fixed point of an orbifold background. This
second kind of wrapped D5 branes are called {\em fractional}
D3 branes. They are branes stuck at the fixed point of an orbifold
and are the ones that we will concentrate on in this review article
because they admit an explicit stringy description. In this case,
the role of the orbifold background is to reduce supersymmetry,
while the one of fractional branes is to break conformal invariance.
We must, however,  stress that the conclusions we will draw from the
fractional branes of an orbifold background seem also to be valid in
the case of the wrapped branes described for instance by the
Maldacena-N{\'{u}}{\~{n}}ez~\cite{MN,CV} and
Klebanov-Strassler~\cite{KLESTRA} classical solutions, although in
these cases this cannot be checked because of the lack of an
explicit stringy description.

In particular, it has been shown that the classical SUGRA solutions
corresponding to those D-branes encode perturbative and
non-perturbative properties of non-conformal and less supersymmetric
gauge theories living on their world-volume as the chiral and scale
anomalies and the superpotential.~\cite{KOW,ANOMA,IL0310,MS0207,PARK} It
was of course expected that the perturbative properties could be
derived from studying in string theory the gauge theory which lives
on those D-branes by taking the field theory limit of one-loop open
string annulus diagram using methods as those described for instance
in Ref.~\cite{FILIM}. But it came as a surprise that these
properties were also encoded in the SUGRA solution, especially after
the formulation of the Maldacena conjecture which relates the SUGRA
approximation to the strong coupling regime of the gauge theory.

The explanation of this fact was given in Ref.~\cite{LMP} where it
was shown, in the case of the two orbifolds
$\mathbb{C}^2/\mathbb{Z}_2$ and $\mathbb{C}^3/(\mathbb{Z}_2{\times}
\mathbb{Z}_2)$ of type IIB, where a complete stringy calculation of the
interaction among D-branes is possible, that the contribution of the
massless open string states to the coefficient of the gauge kinetic
term obtained from the annulus diagram, providing the inverse of the
squared gauge coupling constant, is exactly equal, under open/closed
string duality, to the contribution of the massless closed string
states. Actually, in the cases that we have considered, it can also
be shown that the contribution of the
massive states is identically zero giving no threshold corrections.
Hence the absence of threshold corrections makes open/closed string
duality to work at the level of massless string states. This is the
reason  why one can use the SUGRA solutions to derive the
perturbative behaviour of the dual gauge theory. In conclusion, it
turns out that the validity of the gauge/gravity correspondence in
these non-conformal and less supersymmetric theories that we have
considered, is a direct consequence of the vanishing of threshold
corrections. Actually, reading Ref.~\cite{SS} where considerations
similar to ours were made and applied to non-commutative gauge
theories, we have become aware  that the absence of
threshold corrections in the case of the orbifold
$\mathbb{C}^2/\mathbb{Z}_2$ was already noticed in Ref.~\cite{DL} and
was explained, following Ref.~\cite{HM}, as due to the fact that the
open string massive states exchanged in the loop belong to supersymmetric
long multiplets of ${\cal{N}}=2$ that are actually equal to short multiplets of
${\cal{N}}=4$. But, since short multiplets of ${\cal{N}}=4$ never
contribute to the gauge coupling constant, one can conclude that in
the orbifold $\mathbb{C}^2/\mathbb{Z}_2$ there cannot be any
contribution from massive open string states circulating in the loop.
The new point, as far as we know, is that these considerations can
be directly extended to the case of the
orbifold $\mathbb{C}^3/(\mathbb{Z}_2{\times} \mathbb{Z}_2)$ that has
fractional branes having ${\cal{N}}=1$ super Yang-Mills living on
their world-volume, because the three twisted sectors of this orbifold
are just three copies  of the twisted sector of the orbifold
$\mathbb{C}^2/\mathbb{Z}_2$. Therefore also in this case there
are no threshold corrections.

In this review article we start by reminding what happens in the
theories where the gauge/gravity correspondence beautifully works
and then we try to see what happens in a certain number of
non-supersymmetric theories. In these cases, however, we do not
construct the ``supergravity dual'' solution corresponding to a
system of D3 branes, in order to reproduce from it the perturbative
properties of the gauge theory, but we write down instead the
one-loop vacuum amplitude of an open string stretching between a
D3 brane with a background gauge field turned-on on its world-volume
and a stack of $N$  D3 branes. This procedure is in fact closely
related to the construction of the ``supergravity dual'' because,
under  open/closed string duality, one can also regard the previous
amplitude as the interaction of the ``dressed'' brane and the stack
of $N$ D3 branes via the exchange of a closed string propagator and
this in turn, in the low-energy limit, encodes the information about
the large distance behaviour of the classical supergravity
solution.~\cite{DL0307,tesia}

The first simple non-supersymmetric string theory we consider is the
bosonic string theory in the orbifold ${\mathbb C}^{{\delta}/2}
/{\mathbb Z}_{2}$, with $\delta \leq 22$.
Explicit calculations  show that, differently from the type IIB
case, the threshold corrections to the gauge kinetic term do not
vanish. Moreover the contribution of massless states in the closed
channel turns out to be zero in the field theory limit, showing that
supergravity does not give any information about the gauge theory
parameters.  Therefore the gauge/gravity correspondence does not
hold in this case. On the other hand, this theory is not consistent
because of the presence of both open and closed string tachyons and
therefore we do not discuss it any further.

Other natural candidates for non-supersymmetric theories are the
type $0$ ones that have been studied by constructing their
supergravity duals in Ref.s~\cite{KT9811,KT9812,KT9901}.
Such theories exhibit, however, also the problem of having a tachyon
in the closed string NS-NS sector.
Moreover one finds that the zero-force condition among identical
branes is not satisfied.
This problem can be solved by considering a dyonic configuration of
branes, made of $N$ electric  and $N$ magnetic D3 branes. The gauge
theory living on the world-volume of such a brane configuration is a
$U(N) {\times} U(N)$ gauge theory with one gauge vector,
six adjoint scalars for each
gauge factor and four Weyl fermions in the bifundamental
representation of the gauge group $(N, \bar{N})$ and $(\bar{N}, N)$.
It exhibits a Bose-Fermi degeneracy at each mass level of  the open
string spectrum, which therefore guarantees the absence of
interaction among the branes, at  least at the lowest order in
$g_s$. The interaction between a stack of $N$ dyonic D3 branes and
one extra dyonic D3 brane dressed with an $SU(N)$ gauge field turns
out to be, in this theory, twice the corresponding one
in type IIB. In particular, the coefficient of the gauge kinetic
term is identically zero, yielding the right vanishing beta function
both in the open and in the closed channel. Hence, for dyonic branes
the gauge/gravity correspondence holds. When this theory is put in
the orbifold $\mathbb{C}^{2}/\mathbb{Z}_{2}$ or
$\mathbb{C}^{3}/(\mathbb{Z}_2 {\times} \mathbb{Z}_2)$, then the
interaction between a stack of $N$ dyonic fractional branes and an electric or
magnetic fractional brane dressed with an $SU(N)$ gauge field turns out to be
the same as in type IIB in that orbifold. Therefore, all the
features discussed for that theory, as the validity of the
gauge/gravity correspondence, are also shared by these
non-supersymmetric models. The gauge theories living on the
world-volume
of such a brane configuration provide an example of the
so-called {\it orbifold field theories}\cite{KS,BKV9803,KS183,BJ249}
which are non-supersymmetric gauge theories that in the planar limit
are perturbatively equivalent to some supersymmetric one.

Tachyon free orientifolds of type $0$ theories, called $0'$
theories, were introduced in Ref.~\cite{SAGNOTTI} and their
properties were extensively studied from different points of view in
Ref.s~\cite{BFL9904,AA9912,DM0010}. This is the
theory we consider next in the orbifold
$\mathbb{C}^{2}/\mathbb{Z}_{2}$. When one computes in type $0'$ the
one-loop vacuum amplitude of an open string stretching between a
stack of $N$ fractional D3 branes and one brane of the same kind
dressed with an external gauge field $SU(N)$, the result obtained in
type IIB in the same orbifold is recovered. In particular, as in
type IIB, the threshold corrections to the gauge theory parameters
vanish: we find also in this case that this condition is crucial for
the validity of the gauge/gravity correspondence.

Recently non-supersymmetric and non-conformal theories have been
studied that in the large number of colours are equivalent
to supersymmetric theories~\footnote{See the recent review by
Armoni, Shifman and Veneziano Ref.~\cite{ASV} and Ref.s
 therein. In Ref.~\cite{SS0309} $1/N$ corrections are analysed.}.
They are based on orientifolds of the 0B theory and go under the
name of {\it orientifold field theories}. In Ref.s \cite{AA9912} and
\cite{BFL9906}, non-supersymmetric gauge theories that are conformal
in the planar limit have been discussed. One of them lives on the
world-volume of $N$ D3 branes of the orientifold $\Omega' I_6
(-1)^{F_L}$ of the 0B theory, where $\Omega'$ is the world-sheet
parity~\footnote{We denote the world-sheet parity by $\Omega '$
because its
action on the string states is not quite the same as the world-sheet
parity $\Omega$ that is usually used for constructing type I
ten-dimensional theory.}, $I_6$ the inversion of the coordinates
orthogonal to the world-volume of the D3 branes and $F_L$ is the
space-time fermion number operator in the left sector. This gauge
theory  is an example of orientifold field theory being, in the
large $N$ limit, equivalent to ${\cal{N}}=4$ super Yang-Mills. It
contains one gluon, six scalars in the adjoint representation and
four Dirac fermions transforming according to the two-index
(anti)symmetric representation of the gauge group
$U(N)$~\footnote{The gravity dual of this theory has been
constructed in Ref.~\cite{AA0003}.}.

More recently some attention has been paid to the orientifold field
theories that contain a gluon and a fermion transforming according
to the two-index symmetric or antisymmetric representation of the
gauge group $SU(N)$ ~\cite{ASV2} and that in the large $N$ limit are
equivalent to ${\cal{N}}=1$ SYM.

In Ref. \cite{DLMP0407} the complete stringy description of the
orientifold field theory whose spectrum has, in the large $N$ limit,
the same number of degrees of freedom as ${\cal{N}}=2,1$ super
Yang-Mills, is provided by considering the orbifold projections
$\mathbb{C}^2/ \mathbb{Z}_2$ and $\mathbb{C}^3/(\mathbb{Z}_2{\times}
\mathbb{Z}_2)$ of the orientifold  0B$/\Omega' I_6 (-1)^{F_L}$. In
Ref. \cite{ASV} the latter theory has been shown to be planar
equivalent, both at perturbative and non-perturbative level, to
${\cal N}=1$ SYM .

In Ref. \cite{DLMP0407} the running coupling constant has been
computed in the open string framework and it has been shown that in
the large $N$ limit, where the Bose-Fermi degeneracy of the gauge
theory is recovered and the threshold corrections
vanish, one can obtain the perturbative behaviour
of the orientifold field theories also from the closed string
channel. However the next-to-leading term in the large $N$ expansion
of the $\beta$-function cannot be obtained from the closed string
channel. This means that, as far as the running coupling constant is
concerned, the gauge/gravity correspondence holds only in the planar
limit. When considering the $\theta$-angle instead, one can see that
both the leading and the next-to-leading terms can be equivalently
determined from the open and the closed string channel. This follows
from the fact that in the string framework  the $\theta$-angle does
not admit threshold corrections.

{From} the analysis of the above models we can conclude that the
gauge/gravity correspondence is not a property concerning only
supersymmetric theories, as type IIB. It may work as well in
non-supersymmetric models and when the threshold corrections vanish
(i.e. the contributions of massive states to the gauge theory
parameters are zero)  it admits a stringy description in terms of
open/closed string duality. When this condition is satisfied,
indeed, the contribution of the massless states in the open channel
is mapped, under the modular transformation representing the open/closed
string duality, into the corresponding
one in the closed channel, allowing the gauge/gravity correspondence
to hold.

The paper is organized as follows. In Sect. \ref{philo} we review
the philosophy of the gauge/gravity correspondence, deriving general
expressions for the holographic identifications valid both for
fractional D3 branes and wrapped D5 branes. We also illustrate
our procedure to get a stringy interpretation of this correspondence
at the perturbative level in terms of open/closed string duality and
its connection with the background field method. Sect. \ref{sez1} is
devoted to the analysis of the gauge/gravity correspondence in the
supersymmetric cases of type IIB in the orbifolds
$\mathbb{C}^2/\mathbb{Z}_2$ and $\mathbb{C}^3/(\mathbb{Z}_2{\times}
\mathbb{Z}_2)$. We first use the holographic relations to derive the
gauge theory parameters from the SUGRA solution and then we show
that this correspondence is a direct consequence of open/closed
string duality by evaluating the one-loop vacuum amplitude of an
open string stretching between a stack of $N$ D3 fractional branes
and a further brane dressed with an $SU(N)$ background field. {From}
Sect. \ref{Bosonic} to Sect. \ref{orienti} we apply the same
procedure to non-supersymmetric string models. In Sect.
\ref{Bosonic} we discuss the case of the bosonic string in the orbifold
$\mathbb{C}^{{\delta}/2}/\mathbb{Z}_2$ showing that the presence of
threshold corrections and  of the closed string tachyon do
not allow the massless states in the closed string channel to
reproduce the behaviour of the gauge-theory parameters. Sect.
\ref{0B} is devoted to the case of type 0B string: we first review
the structure of its open and closed string spectrum and that of the
boundary state and then we explore the gauge/gravity correspondence
in the case of dyonic branes configurations. In Sect. \ref{0'} we
analyse the case of type $0'$ theories, discussing in some detail
the structure of its open and closed string spectrum and the
boundary state description of the branes. Then we show that, also in
this case, the gauge/gravity correspondence holds and it follows
from open/closed string duality. In Sect. \ref{orienti} we discuss
another orientifold of type $0$B which is type
$0$B$/\Omega'I_6(-1)^{F_L}$ and its orbifolds
$\mathbb{C}^2/\mathbb{Z}_2$ and $\mathbb{C}^3/(\mathbb{Z}_2{\times}
\mathbb{Z}_2)$. We analyse the open and closed string spectrum of
these theories, their interpretation as orientifold field theories
and then explore the gauge/gravity correspondence with the usual
strategy. In Sect. \ref{conclu} we summarize the main results and
illustrate the conclusions of our work. Finally there are three
appendices devoted respectively to the $\Theta$-functions and their
properties under modular transformations, to the
explicit derivation of some results mentioned in various parts of the
paper and to the Euler-Heisenberg
actions that can be obtained for the various theories discussed in
this article by performing the field theory limit.

\section{The Philosophy of Gauge/Gravity Correspondence
and its Stringy Interpretration} \label{philo}

The gauge/gravity correspondence follows from the twofold nature of
D$p$ branes which admit two alternative descriptions: a closed
string description in which they appear as sources of closed strings
that are emitted in the entire ten-dimensional space and an open string
description in which they appear as hyperplanes where open strings
are attached with their end-points satisfying appropriate boundary
conditions. In the first perspective the massless closed string
states emitted in the bulk generate non trivial SUGRA profiles,
while in the second one the open string massless fluctuations give
rise to the existence of a $(p+1)$-dimensional gauge theory living
on the brane world-volume. This twofold nature allows one to derive
the gauge-theory quantum properties  from the knowledge of the D$p$ brane
classical geometry
leading to the existence of some {\em holographic identifications}
which relate the gauge theory parameters to the supergravity fields:
\begin{equation}
\frac{1}{g^2_{YM}}=f({\rm SUGRA\,\,\,fields}) \,\,\,\,\,\,\,\,\, ,
\,\,\,\,\,\,\,\,\,\,\, \theta_{YM}=g({\rm SUGRA\,\,\,fields})~~,
\label{hi}
\end{equation}
where $f$ and $g$ are some particular functions. In particular, this
correspondence relates the weak coupling regime of the gauge theory
to the long distance behaviour of the SUGRA solution and the strong
 coupling regime to its near horizon limit. Indeed the typical
structure of the SUGRA solution involves harmonic functions
depending on the ratio $g_sN/r^{7-p}$ and then a large $r$
expansion is formally equivalent to an expansion for small values
of $g_sN$ (weak 't Hooft coupling) and viceversa. Therefore the
amount of information that SUGRA can give about the gauge theory,
by means of the holographic relations, depends on the specific
case one is dealing with. Generally speaking, whenever the SUGRA
solution is well-defined everywhere, holographic identifications
should give (in principle) both perturbative and non-perturbative
information about the gauge theory. This is what happens in the
case of the Maldacena-N{\'{u}}{\~{n}}ez solution~\cite{MN} for
which the SUGRA solution - which is not affected by any
singularity - has been shown to encode the presence of a gaugino
condensate \cite{ABCPZ0112} and has been used to derive the
complete perturbative NSVZ $\beta$-function of the pure
${\cal{N}}=1$ SYM theory with gauge group $SU(N)$ \cite{NSVZ}
with, in addition, non-perturbative corrections due to fractional
instantons. \cite{DLM,Mueck,BM} These properties of ${\cal{N}}=1$
super Yang-Mills have also been  derived from the regular
Klebanov-Strassler solution~\cite{KLEBA,EMILIANO,PAOLO,GOMEZ} that
is also free of singularities. Instead in the cases of fractional
branes in orbifolds the SUGRA solutions are affected by naked
singularities and thus they cannot be trusted in the near horizon
limit. Therefore it is not possible to use them in order to get
non-perturbative information about the dual gauge theory: the
holographic identification may be used only at the perturbative
level (unless one considers specific deformations of the singular
spaces as in Ref.s \cite{geotrans,IL}).

Let us briefly illustrate how to derive these gauge/gravity
relations for the gauge theory living on fractional D3 and wrapped
D5 branes using supergravity calculations. Since also the fractional
D3 branes are D5 branes wrapped on a vanishing 2-cycle located at
the orbifold fixed point,  we can start from the {\it world-volume
action} of a D5 brane, that is given by:
\begin{equation}
S = S_{BI} + S_{WZW}~~,
\end{equation}
where the {\it Born-Infeld action} $S_{BI}$ reads as:
\begin{equation}
S_{BI}  = -\tau_5 \int d^{6} \xi {e}^{-\phi} \sqrt{- \det ( G_{IJ} +
B_{IJ} + 2 \pi \alpha' F_{IJ})}~~,~~ \tau_5 = \frac{1}{g_s
\sqrt{\alpha'} (2 \pi \sqrt{\alpha'})^5}~~,\label{boinfe45}
\end{equation}
while the {\it Wess-Zumino-Witten action} $S_{WZW}$ is given by:
\begin{equation}
S_{WZW} = \tau_{5} \int_{V_{6}} \left[ \sum_{n} C_{n}\wedge e^{2 \pi
\alpha' F + B_{2}} \right].
\end{equation}
We divide the six-dimensional world-volume into four flat directions
in which the gauge theory lives and two directions on which the
brane is wrapped. Let us denote them with the indices $ I,J = (
\alpha, \beta; A,B)$ where $\alpha$ and $\beta$ denote the flat
four-dimensional ones and $A$ and $B$ the wrapped ones. As usual, we
assume the supergravity fields to be independent from the
coordinates $\alpha,\beta$. We also assume that the determinant in
Eq. (\ref{boinfe45}) factorizes into a product of two determinants,
one corresponding to the four-dimensional flat directions where the
gauge theory lives and the other one corresponding to the wrapped
ones where we have only the metric and the NS-NS two-form field. By
expanding the first determinant and keeping only the quadratic term
in the gauge field we obtain:
\begin{equation}
(S_{BI})_2 = - \tau_5 \frac{(2 \pi \alpha')^2}{8} \int d^{6} \xi
  {e}^{-\phi}
  \sqrt{-\det G_{\alpha \beta}} G^{\alpha \gamma} G^{\beta \delta}
  F_{\alpha \beta}^{ a} F_{\gamma \delta}^{ a} \sqrt{\det{( G_{AB} + B_{AB})}},
\label{binf74}
\end{equation}
where we have included a factor $1/2$ coming from the normalization
of the gauge group generators ${\rm Tr} [T^a
T^b]=\frac{\delta^{ab}}{2}$.

We assume that along the flat four-dimensional directions the metric
is the Minkowski one apart from the warp factor, while along the
wrapped ones, in addition to the warp factor, there is also a
non-trivial metric. This means that the longitudinal part of the
metric can be written as
\begin{equation}
ds^2 = H^{-1/2}\left(  d x_{3,1}^{2} + ds_{2}^{2} \right).
\label{warp943}
\end{equation}
By inserting this metric in Eq. (\ref{binf74}) we see that the warp
factor cancels out in the Yang-Mills action and from it  we
can then extract the inverse of the squared gauge coupling
constant as the coefficient of
the gauge kinetic term $-\frac{1}{4} \int d^{4}x F^{a\,\alpha\beta}
F_{\alpha\beta}^{a}$:
\begin{equation}
\frac{4 \pi}{g_{YM}^{2}} =  \frac{1}{g_s ( 2 \pi
\sqrt{\alpha'})^{2}} \int_{{\cal{C}}_2} d^{2} \xi {e}^{- \phi}
\sqrt{\det{( G_{AB} + B_{AB})}}~~ .
\label{volu87}
\end{equation}
This formula is valid for both wrapped and fractional branes of the
orbifolds having only one vanishing 2-cycle as the orbifold
$\mathbb{C}^2 / \mathbb{Z}_2$. ${{\cal{C}}_2}$ is the cycle around
which the branes are wrapped.  The $\theta$ angle, in the case of
both fractional D3 branes and wrapped D5 branes, can be obtained by
extracting the coefficient of the term $\frac{1}{32\pi^2} \int d^{4}x
F_{\alpha\beta}^
a\tilde F^{a\,\alpha\beta}$ from $S_{WZW}$ getting:
\begin{equation}
\theta_{YM} = \tau_5 (2 \pi \alpha')^2   (2 \pi)^2
\int_{{\cal{C}}_2} ( C_2 + C_0  B_2) = \frac{1}{2 \pi \alpha' g_s}
\int_{{\cal{C}}_2} ( C_2 + C_0  B_2) ~~. \label{theta56}
\end{equation}
Eq.s (\ref{volu87}) and (\ref{theta56}) provide an explicit realization
of the holographic identifications (\ref{hi}), establishing a
relation between quantities peculiar of the gauge theory living on
the world-volume of the D3 branes and the supergravity

fields. We want to stress that these relations are not based on the
probe analysis; they have a more general validity as stressed in
Ref.~\cite{ANOMA}
and are therefore also valid in the case of supersymmetric
${\cal{N}}=1$ theories where the probe analysis cannot be done.
Before proceeding further, it is interesting to notice that
in the case of fractional branes Eq.s (\ref{volu87}) and
(\ref{theta56}) can be written in a single expression:
\begin{eqnarray}
\tau_{YM} \equiv \frac{\theta_{YM}}{2 \pi} + i \frac{4
  \pi}{g^{2}_{YM}} = \frac{1}{(2 \pi \sqrt{\alpha'})^2 g_s}
  \int_{{\cal{C}}_2} ( C_2 + \tau B_2 )~~,~~ \tau = C_0 + i {e}^{- \phi}.
\label{tauym39}
\end{eqnarray}
After defining the quantity $G_3 \equiv d ( C_2 + \tau B_2)$, Eq.
(\ref{tauym39}) can be rewritten in the following form:
\begin{eqnarray}
\tau_{YM} = \frac{1}{(2 \pi \sqrt{\alpha'})^2 g_s} \int_{\cal B} G_3~~,
\label{g3}
\end{eqnarray}
where $\cal B$ is the 3-cycle given by the direct product of the
original 2-cycle ${\cal C}_2$ with a suitable non-compact 1-cycle
living in the plane orthogonal to both the branes and  the orbifold.
Eq.s (\ref{tauym39}) and (\ref{g3}) can also be extended to the case
of fractional branes in the orbifold
$\mathbb{C}^3/(\mathbb{Z}_2{\times} \mathbb{Z}_2)$ where  the gauge
theory living on the world-volume of the D3 branes preserves four
supersymmetry charges. In this case we have \cite{NAPOLI,FERRO}
\begin{eqnarray}
\tau_{YM}  = \frac{1}{2(2 \pi \sqrt{\alpha'})^2 g_s}
\sum_{i=1}^3\int_{{\cal C}_2^{i}} (C_2+\tau B_2)~~, \label{g3b1}
\end{eqnarray}
being ${\cal C}_2^{i}$ the exceptional shrinking 2-cycle of the
orbifold geometry. By defining the non-compact 3-cycle $\cal B$
\begin{eqnarray}
{\cal B}\equiv \bigcup_{\ell=1}^3{\cal B}_{\ell} \qquad{\rm
with}\qquad{\cal B}_{\ell}\equiv \bigcup_{i=1}^3{{\cal
C}_2^{i}}{\times}{\beta}_{\ell}~~, \label{g3b2}
\end{eqnarray}
where ${\beta}_{\ell}$ is a suitable non-compact 1-cycle living in
the plane $z^{\ell}=x^{2 \ell+2}+ix^{2 \ell+3}$ ($\ell=1,2,3$)
orthogonal to the brane, we can write
\begin{eqnarray}
\tau_{YM} = \frac{1}{2(2 \pi \sqrt{\alpha'})^2 g_s} \int_{{\cal B}}
G_3 ~~.\label{g3b3}
\end{eqnarray}
This provides a generalization to the case of a non constant axion
and dilaton of the formulas used in computing the
parameters of the gauge theory after the geometric
transition.~\cite{geotrans}

The aim of this review is to discuss the stringy interpretation of
the gauge/gravity correspondence. In particular in Ref.~ \cite{LMP}
we have elaborated a strategy which has allowed us to understand why
the SUGRA solution is able to reproduce the gauge theory parameters,
at the first order of their perturbative expansion. Let us review
the main features of this procedure.

We first remind that in field theory the one-loop running coupling
constant may be determined through the background field method, by
calculating the one-loop correction to the two-point function involving
two external background field strengths $F$ (see Fig. 1) and reading
its contribution to the gauge kinetic term. In the context of string
theory, the open string amplitude reducing to the previous one in
the field theory limit
is given by the one-loop vacuum amplitude of an open string
stretching between a stack of $N$ D3 branes and a further D3 brane
with a background $SU(N)$ gauge field turned-on on its world-volume.
{From} it we can  extract the second order term in the background
field, selecting the amplitude shown in Fig. 2, which gives the
full open string one-loop correction to the two-point function with
two external background field strengths $F$. This means that, as in
field theory, one can read the gauge theory
parameters with all the string corrections as follows:
\begin{equation}
\frac{1}{g^2_{YM}}\,\,\,\,\,\, {\rm as ~ the ~coefficient ~of ~the~
gauge~ kinetic~ term~}-\frac{1}{4}\int d^4 x F^{a\,\alpha\beta}
F_{\alpha\beta}^{a},
\label{gym}
\end{equation}
\begin{equation}
\theta_{YM}~~ {\rm as~ the ~coefficient ~of ~the~ topological~
charge~ term~}\frac{1}{32\pi^2}\int d^4 x F^{a\,\alpha\beta}{\tilde
F}_{\alpha\beta}^{ a}.
\label{theym}
\end{equation}
Obviously, by performing the field theory limit in the open string
channel, only  massless open
string states (i.e. the gauge degrees of freedom) propagate in the
loop, and therefore this procedure to evaluate the gauge theory
parameters coincides with the one of the background field method in
field theory.
\begin{figure}
\null
\begin{center}
\includegraphics[width=10cm]{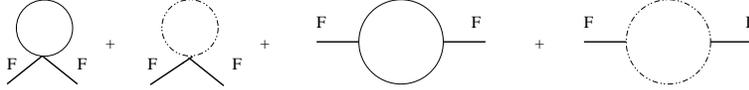}
\label{fig1}
\end{center}
\caption{\baselineskip=13pt {One-loop correction to the two-point
function with two external background field strengths. These are the
diagrams one has to consider when evaluating the one loop running
coupling constant of the gauge theory via the background field
method. Dashed lines denote the ghost fields. }}
\end{figure}
\begin{figure}
\null
\begin{center}
\includegraphics[width=10cm]{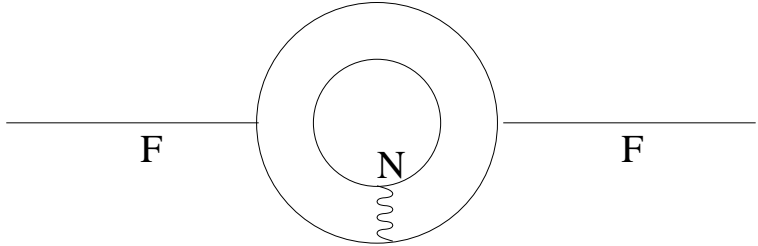}
\label{fig2}
\end{center}
\caption{\baselineskip=13pt {Second order expansion in the
background field of the one-loop vacuum amplitude of an open string
stretching between a stack of $N$ D3 branes and a further brane
dressed with an $SU(N)$ background field. The number $N$ of
D3 branes encodes the information about the non-abelian nature of
the quantum gauge fields propagating in the loop. In the field
theory limit this diagram reduces to those in Fig. 1}}
\end{figure}

One can also rewrite the one-loop vacuum amplitude in the closed
string channel and identify again the running coupling constant
and the $\theta$-angle, at the full closed string level,
respectively as the coefficient of the gauge kinetic term and of
the topological charge term in the $F$ expansion, obtaining for
those quantities the same result as in the open string channel.
This is a direct consequence of the open/closed string duality.
What is not obvious and actually in general not true, however, is
that the contribution of the massless open string states
circulating in the loop is equal to that of the massless closed
string states exchanged between the branes and selected by
performing the large distance limit between the branes
.\cite{DL0307,tesia} We show that this is exactly what happens in those
cases in which
the supergravity solution is able to reproduce the
gauge theory parameters. Hence, this means that if the contribution of the
open massless string states is mapped, under open/closed string
duality, into the contribution of closed string massless states,
then the supergravity solution contains the full information
about the gauge theory parameters. Notice that the above technique
gives a quantitative
explanation of why the SUGRA fields reproduce the $\beta$-function
and $\theta$-angle at one loop.

In this section we have computed the gauge parameters in two apriori
different ways obtaining respectively Eq.s (\ref{volu87}),
(\ref{theta56}) and Eq.s (\ref{gym}), (\ref{theym}). In the first
case we have used the Born-Infeld action with the inclusion of the WZW
term and we have determined the gauge parameters in terms of the
supergravity fields given by a classical solution of the SUGRA equations of
motion describing $N$ fractional D3 or wrapped D5 branes, while in the
second case we have performed a complete stringy calculation, that can
be done only in the case of fractional branes, computing the
interaction between a fractional D3 brane having a non-abelian gauge
field on its world-volume and a system of $N$ fractional D3 branes. In
both cases no use has been done of the probe technique and in general
we expect to obtain two different results because the stringy
calculation includes also the contribution of the massive string
states and in general the contribution of the massless open string
states circulating in the loop is not mapped under open/closed string
duality into that of the massless closed string states that are the
only ones appearing in the approach based on supergravity. We will
see, however, that in many interesting cases the massless states
appearing in the two channels are precisely mapped into each other and
in this case the supergravity approach provides the correct
perturbative behaviour of the gauge theory living on the world-volume
of $N$ D3 branes.

One could in principle generalize it
to the full perturbative level by considering multiloop open string
amplitudes and converting them into the corresponding
multiboundaries tree level amplitudes in the closed channel, but
calculations would be of course much more involved.

\section{Gauge/Gravity Correspondence in Supersymmetric String Theory}
\label{sez1} In this section we first use fractional branes of the
orbifold $\mathbb{C}^2 /\mathbb{Z}_2$ to show that the perturbative
behaviour of the gauge theory living on their world-volume, namely
${\cal{N}}=2$ super Yang-Mills, can be reproduced from their
corresponding classical solution through the gauge/gravity
relations. Then  we show that, working in the pure string framework
and using only open/closed string duality,  the perturbative
properties of the gauge theory living on the fractional D-branes can
be   derived  not only in the open string channel, as expected, but
also  in the closed string channel. We finally show that this at
first sight surprising result turns out to be, in these models, a
direct consequence of the absence of threshold corrections to the
running gauge coupling constant.

\subsection{Gauge/gravity correspondence in the orbifold
$\mathbb{C}^2 /\mathbb{Z}_2$}

In this subsection we consider fractional D3 and D7 branes of the
non-compact orbifold $\mathbb{C}^2 /\mathbb{Z}_2$  in order to study the
properties of ${\cal{N}}=2$ super QCD. We group the coordinates of
the directions ($x^4, \dots, x^9 $) transverse to the world-volume of
the D3 brane where the gauge theory lives, into three complex
quantities:
\begin{equation}
z_1 = x^4 + i x^5~~,~~z_2 = x^6+i x^7~~,~~ z_3 = x^8 + i x^9.
\label{z1z2z3}
\end{equation}
The non trivial generator $h$ of $Z_2$ acts as
\begin{eqnarray}
z_2  \rightarrow - z_2~~~,~~~z_3 \rightarrow  - z_3 \label{z2z3}
\end{eqnarray}
leaving $z_1$ invariant, showing the presence of one
fixed point at the origin corresponding to a vanishing 2-cycle
located at $z_2=z_3 =0$.  Fractional D3 branes are D5 branes wrapped
on the vanishing 2-cycle and therefore are, unlike bulk branes,
stuck at the orbifold fixed point. By considering $N$ fractional D3
and $M$ fractional D7 branes of the orbifold $\mathbb{C}^2
/\mathbb{Z}_2$ we are able to study  ${\cal{N}}=2$  super QCD with
$M$ hypermultiplets.
 In order to do
that, we need to determine the classical solution corresponding to
the previous brane configuration. For the case of the orbifold
$\mathbb{C}^2 /Z_2$ the complete classical solution was found in
Ref.~\cite{d3d7}~\footnote{See also
Ref.s~\cite{KLENE,d3,polch,grana2,marco} and Ref.~\cite{REV} for a
review on fractional branes.}.  In the following we write it
explicitly for  a system of $N$  fractional D3 branes with their
world-volume along the directions $x^0, x^1 , x^2 , $ and $ x^3$ and
$M$ fractional D7 branes containing the D3 branes in their
world-volume and having the remaining four world-volume directions
along the orbifolded ones. The metric, the $5$-form field strength,
the axion  and the dilaton are given by~\footnote{We denote by
$\alpha$ and $\beta$ the four directions corresponding to the
world-volume of the fractional D3 brane, by $\ell$ and $m$ those
along the four orbifolded directions $x^6 , x^7, x^8 $ and $x^9$ and
by $i$ and $j$ the directions $x^4$ and $x^5$ that are transverse
to both the D3 and the D7 branes.}:
\begin{eqnarray}
ds^2 &=& H^{-1/2}\, \eta_{\alpha\beta}\,d x^\alpha dx^\beta +
H^{1/2} \,\left(\delta_{\ell m}\,dx^\ell dx^m + {\rm e}^{-\phi}
\delta_{ij}
 dx^i dx^j\right)  ~~,
\label{met48} \\
{\widetilde{F}}_{(5)} &=&  d \left(H^{-1} \, dx^0 \wedge \dots
\wedge dx^3 \right)+ {}^* d \left(H^{-1} \, dx^0 \wedge \dots \wedge
dx^3 \right) ~~, \label{f5ans}
\end{eqnarray}
\begin{equation}
\tau \equiv C_0 + i {\rm e}^{- \phi} = {\rm i}\left(1 -\,
\frac{Mg_s}{2\pi}\, \log \frac{z}{\epsilon}\right)~~,~~ z \equiv x^4
+ i x^5 = y {\rm e}^{i\theta}~~,\label{tausol}
\end{equation}
where the self-dual field strength ${\tilde{F}}_{(5)}$ is given in terms
of the NS-NS and R-R 2-forms $B_2$ and $C_2$ and of
the 4-form potential $C_4$ by   $\tilde{F}_{(5)}=dC_4+C_2\wedge dB_2$.
The warp factor $H$ is a function of the coordinates ($x^4,\ldots, x^9$) and
$\epsilon$ is an infrared cutoff.
The twisted fields
are instead given by  $B_2 = \omega_2 b$, $C_2 = \omega_2 c$ where
$\omega_2$ is the volume form of the vanishing 2-cycle
and
\begin{equation}
b {\rm e}^{-\phi} = \frac{(2 \pi \sqrt{\alpha'})^2}{2} \left[ 1 +
\frac{2N-M}{\pi} g_s \log \frac{y}{\epsilon} \right]~~,~~ c + C_0 b
= - 2 \pi \alpha' \theta g_s (2N-M) ~~.\label{bc63}
\end{equation}
It can be seen that the previous solution has  a naked singularity
of the repulson type at short distances. But, on the other hand, if
we use a brane probe approaching from infinity the stack of branes,
described by the previous classical solution,
it can also be seen that the tension of the probe vanishes at a
distance that is larger than that of the naked singularity.
The point where the probe brane becomes tensionless is called in the
literature {\em enhan{\c{c}}on}~\cite{enhancon} and at this point
the classical solution does not describe anymore the stack of
fractional branes.

Let us now exploit the gauge/gravity relations derived in the previous
section, to determine the
coupling constants of the world-volume theory from the supergravity
solution. In the case of fractional D3 branes of the orbifold
$\mathbb{C}^2 /\mathbb{Z}_2$, that is characterized by one single
vanishing 2-cycle ${\cal{C}}_2$, the gauge coupling constant
 given in Eq. (\ref{volu87}) reduces to
\begin{equation}
\frac{1}{g_{YM}^{2}} =  \frac{\tau_5 (2 \pi \alpha')^2}{2}
\int_{{\cal{C}}_2} {e}^{- \phi} B_2 = \frac{1}{4 \pi g_s (2 \pi
\sqrt{\alpha'})^2} \int_{{\cal{C}}_2} {e}^{- \phi} B_2 ~~.
\label{coupli98}
\end{equation}
By inserting in Eq.s (\ref{coupli98}) and (\ref{theta56}) the
classical solution we get the following expressions for the gauge
coupling constant and the $\theta_{YM}$ angle:~\cite{d3d7}
\begin{eqnarray}
\frac{1}{g_{YM}^{2}} = \frac{1}{8 \pi g_s} + \frac{2N-M}{16 \pi^2}
\log \frac{y^2}{\epsilon^2}~~,~~ \theta_{YM} =- \theta (2N-M)~~.
\label{thetaym76}
\end{eqnarray}
Notice that the gauge coupling constant appearing in the previous
equation is the \underline{bare} gauge coupling constant computed at
the scale $m
\sim {y}/{\alpha'}$, while the square of the bare gauge coupling
constant computed at the cut-off $\Lambda \sim
{\epsilon}/{\alpha'}$ is equal to $8 \pi g_s$.

In the case of an ${\cal{N}}=2$ supersymmetric gauge theory the
gauge multiplet contains  a complex scalar field $\Psi$  whose action term
can be found  when deriving the Yang-Mills action from the
Born-Infeld one. In fact the derivation of the kinetic term for the
scalar field is obtained from the term in the Born-Infeld action
depending on the brane coordinates $x^4$ and $x^5$ that are
transverse to both the branes and  the orbifold. This implies that
the complex scalar field of the gauge supermultiplet is related to
the coordinate $z$ of supergravity through the following
gauge/gravity relation $\Psi \sim \frac{z}{2 \pi \alpha'}$. This is
another example of holographic identification between a quantity,
$\Psi$, peculiar of the gauge theory living on the fractional D3
branes and another one, the coordinate $z$, peculiar of
supergravity. It allows one to obtain the gauge
theory anomalies from the supergravity background. In fact, since we
know how the scale and $U(1)$ transformations act on $\Psi$, from
the previous gauge/gravity relation we can deduce how they act on
$z$, namely
\begin{equation}
\Psi \rightarrow s {\rm e}^{2i \alpha} \Psi \Longleftrightarrow z
\rightarrow s  {\rm e}^{2 i \alpha} z   \Longrightarrow y
\rightarrow s y~~,~~  \theta \rightarrow \theta + 2 \alpha ~~.
\label{gau78}
\end{equation}
Those transformations  do not leave invariant the supergravity
background in Eq. (\ref{bc63}) and when we use them in Eq.s
(\ref{coupli98}) and (\ref{theta56}), they generate the anomalies of
the gauge theory living on the fractional D3 branes. In fact, by acting
with those transformations in Eq.s (\ref{thetaym76}), we get:
\begin{equation}
\frac{1}{g_{YM}^{2}} \rightarrow \frac{1}{g_{YM}^{2}} +\frac{2N-M}{8
\pi^2} \log s~~,~~ \theta_{YM} \rightarrow \theta_{YM} - 2 \alpha
(2N-M)~~. \label{tra38}
\end{equation}
The first equation generates the $\beta$-function of ${\cal{N}}=2$
super QCD with $M$ hypermultiplets given by:
\begin{equation}
\beta (g_{YM} ) = - \frac{2N-M}{16 \pi^2} g_{YM}^{3}~~, \label{beta35}
\end{equation}
while the second one reproduces the chiral $U(1)$
anomaly.~\cite{KOW,ANOMA} In particular, if we choose $\alpha =
\frac{2 \pi}{2(2N-M)}$, then $\theta_{YM}$ is shifted by a factor
$2 \pi$. But since $\theta_{YM}$ is periodic of $2 \pi$, this
means that the subgroup $Z_{2(2N-M)}$ is not anomalous in perfect
agreement with the gauge theory results.

{From} Eq.s (\ref{thetaym76}) it is easy to compute the combination:
\begin{equation}
\tau_{YM} \equiv \frac{\theta_{YM}}{2 \pi} + i \frac{4
\pi}{g_{YM}^{2}} = i \frac{2N-M}{2 \pi} \log \frac{z}{ y_{e}}~~,~~
y_e = \epsilon {\rm e}^{-\pi/[(2N-M)g_s]}~~,\label{tauym67}
\end{equation}
where $y_e$ is the enhan{\c{c}}on radius
and corresponds, in the gauge theory, to the dimensional scale
generated by dimensional transmutation that we call $\Lambda_{QCD}$ in
order not to confuse it with the cut-off $\Lambda$.
Eq. (\ref{tauym67}) reproduces the perturbative moduli space of
${\cal{N}}=2$ super QCD, but not the instanton corrections. This is
consistent with the fact that the classical solution is reliable for
large distances in supergravity corresponding to short distances in
the gauge theory, while it cannot be used below the enhan{\c{c}}on
radius where non-perturbative physics is expected to show up. Notice
that the quantity $G_{3}$, defined in the previous section, results to be:
\begin{eqnarray}
G_3 \equiv d ( C_2 + \tau B_2 ) =  i (2 \pi \sqrt{\alpha'})^2 g_s
{\cdot} \frac{2N-M}{2 \pi} \omega_2 \wedge \frac{dz}{z} \label{g356}
\end{eqnarray}
and we get the following identities
\begin{eqnarray}
\frac{1}{(2 \pi \sqrt{\alpha'})^2 g_s} \int_{\cal B} G_3 =
\tau_{YM}~~~,~~~ \frac{1}{(2 \pi \sqrt{\alpha'})^2 g_s} \int_{\cal
A} G_3 = N - \frac{M}{2}~~. \label{g384}
\end{eqnarray}
Hence Eq.s (\ref{tauym67}) and (\ref{g356}) provide an explicit realization of the
holographic identity given in Eq. (\ref{g3}).
$\cal B$ is the non-compact 3-cycle consisting of
${\cal{C}}_2$ times a 1-cycle along which we have to integrate
between the IR cutoff $y_e$
and an UV one  $ z \equiv y {e}^{i \theta}$. $\cal A$ is a compact
3-cycle consisting of ${\cal{C}}_2$ and a 1-cycle along which we
have to integrate between $-y_e$ and $y_e$. Notice, however, that what we call
UV cutoff according to the notation followed in the literature, should
be more properly called the scale  $m \sim {y}/{\alpha'}$ at which
we compute the \underline{bare} gauge coupling constant, while the UV
cutoff of the gauge theory is actually $\Lambda \sim
{\epsilon}/{\alpha'}$.

Eq.s (\ref{g384}) are precisely the ones used in the approach
followed in Ref.~\cite{geotrans} for $M=0$.

The previous results can also be extended to the case of fractional
D$3$/D$7$ branes of the orbifold ${\mathbb C^3}/(\mathbb{Z}_2
{\times}\mathbb{Z}_2)$. In this case only the large distance
behaviour of the classical solution is known.\cite{NAPOLI} In
particular the solution for the twisted fields is:
\begin{eqnarray}
G_3 =  \frac{i}{\pi} (2 \pi \sqrt{\alpha'})^2 g_s {\cdot}
\sum_{i=1}^3  \left[ \omega_2^i\wedge\left( N
\frac{dz_i}{z_i}-\frac{M}{6}\frac{dz_1}{z_1} -
\frac{M}{2}\delta_{i1} \frac{dz_1}{z_1}  \right) \right],
\label{g3561}
\end{eqnarray}
where $N$ is the number of fractional D$3$ branes of type $1$ and
$M$ is the number of D$7$ branes of type $3$ and $4$ in the notation
of Ref. \cite{NAPOLI} and the following expression for the
background value of the $B_2$ field has been used:
\begin{eqnarray}
\int_{{\cal C}_2^i} B_2=\frac{4\pi^2\alpha'}{6}~~. \label{g3562}
\end{eqnarray}
By integrating the three form in Eq.(\ref{g3561}) on the non-compact
${\cal B}$ cycle defined in Eq. (\ref{g3b2}) we get:
\begin{eqnarray}
\tau_{YM}=\frac{1}{2(2\pi{\sqrt \alpha'})^2g_s}\int_{{\cal B}}
G_3=\frac{i}{2\pi} (3N-M) \log(z/y_e)~~. \label{g3563}
\end{eqnarray}
that is the correct one-loop expression of the gauge coupling constant
of ${\cal{N}}=1$ super QCD.
Moreover one has:
\begin{eqnarray}
\frac{1}{2(2\pi{\sqrt \alpha'})^2g_s}\int_{{\cal A}_{\ell}}
G_3=N-M\delta_{\ell 1}~~, \label{g3564}
\end{eqnarray}
where  $A_{\ell}$ is the compact 3-cycle consisting of $\bigcup_{i=1}^3
C_2^i$ and of a 1-cycle along the direction $\ell$ which we have to
integrate between $-y_e$ and $y_e$, being $y_e=\epsilon
e^{-\pi/[2(3N-M)g_s]}$. The last two equations give a generalization
of the relations obtained in Ref. \cite{geotrans} to a case with
running dilaton and axion.

In conclusion, we have derived the perturbative behaviour of
${\cal{N}}=2$ super QCD with $M$ flavours by using the holographic
identifications, given in Eq.s (\ref{coupli98}) and (\ref{theta56}).
This apriori unexpected result will be clarified in
the next subsection by computing the annulus diagram in the full
string theory and showing that the threshold corrections to the
gauge coupling constant vanish. This means that, under open/closed
string duality, the contribution of the massless open string states
is precisely mapped into that of the massless closed string states
and that therefore the supergravity solution is sufficient to derive
the perturbative behaviour of the gauge theory.

\subsection{Gauge/gravity correspondence from open/closed duality:
${\cal N}$=~2 case} \label{subsez1}

In this section we compute the one-loop vacuum amplitude of an open
string stretching between a fractional D$3$ brane of the orbifold
$\mathbb{C}^2/\mathbb{Z}_2$ {\em dressed} with a background $SU(N)$ gauge field
on its world-volume and
a stack of  $N$ ordinary fractional D$3$ branes. This can be
equivalently done
by computing the tree closed string diagram containing two boundary
states and a closed string propagator.

We are interested in the case of parallel fractional D3 branes with
their world-volume along the directions $x^0 , x^1 , x^2, x^3$, that
are completely external to the space on which the orbifold acts. The
background gauge field lives on the four-dimensional world-volume of
the fractional D3 brane and, without loss of generality, it can be
taken to have the following form:
\begin{eqnarray}
\hat{F}_{\alpha \beta } \equiv 2 \pi \alpha' F_{\alpha \beta}  =
\left(
\begin{array}{cccc}
0  & f & 0 & 0\\
-f & 0 & 0 & 0\\
0  & 0 & 0 & g\\
0  & 0 &- g& 0
\end{array}
\right) ~~.\label{effe}
\end{eqnarray}
The free energy of an open string stretched  between a dressed D3 brane and
a stack of $N$ D3 branes located at a distance $y$ in the plane ($x^4
, x^5$) that is orthogonal to both the world-volume of the
 D3 branes and
the four-dimensional space on which the orbifold acts, is given by:
\begin{eqnarray}
Z  = N\int_{0}^{\infty} \frac{d \tau}{\tau} Tr_{NS-R} \left[
\left(\frac{e+h}{2} \right) (-1)^{F_s}(-1)^{G_{bc}}P_{GSO}
{ e}^{- 2 \pi \tau L_0} \right]
 \equiv & Z_{e}^o + Z_{h}^o
\label{Z1}
\end{eqnarray}
where $F_s$ is the space-time fermion number, $G_{bc}$ is the
ghost number and the GSO projector is given by:
\begin{eqnarray}
P_{GSO}= \frac{(-1)^{G_{\beta \gamma}} + (-1)^F }{2} ~~,\label{GSO}
\end{eqnarray}
with $G_{\beta\gamma}$ being the superghost number:
\begin{eqnarray}
\! G_{\beta \gamma}  =   -\!\!\! \sum_{m=1/2}^{\infty} \left( \gamma_{-m}
\beta_{m}+ \beta_{-m} \gamma_{m} \right)~,~ G_{\beta \gamma}  =
-\gamma_{0} \beta_{0} -\!\! \sum_{m=1}^{\infty} \left( \gamma_{-m}
\beta_{m} + \beta_{-m} \gamma_{m} \right)\label{sghonu}
\end{eqnarray}
respectively in the NS and in the R sector.
$F$ is the world-sheet fermion number defined by
\begin{eqnarray}
F= \sum_{t=1/2}^\infty \psi_{-t} {\cdot} \psi_t -1 \label{fns78}
\end{eqnarray}
in the NS sector and by
\begin{eqnarray}
(-1)^{F} = \Gamma^{11}(-1)^{{F}_R}~~,~~ \Gamma^{11} \equiv
\Gamma^{0} \Gamma^{1} \dots \Gamma^{9}~~,~~ F_{R} =
\sum_{n=1}^{\infty} \psi_{-n} {\cdot} \psi_{n} \label{fr79}
\end{eqnarray}
in the R sector.

The superscript $o$ stands for open because we are computing the
annulus diagram in the open string channel. The fact that we are
considering a string theory in the orbifold $\mathbb{C}^2
/\mathbb{Z}_2$ is encoded in the presence of the orbifold projector
$P=(e+h)/2$ in the trace. The explicit computation can be found in
Appendix B of Ref.~\cite{LMP}. Here we give only the final results:
\begin{eqnarray}
Z_e^o & = & - \frac{N}{(8 \pi^2 \alpha')^{2}} \int d^4x \sqrt{
-\mbox{det}(\eta+\hat{F})}\nonumber\\
&\times& \int_{0}^\infty
\frac{d\tau}{\tau}e^{-\frac{y^2 \tau}{2\pi\alpha'}} \frac{\sin \pi
\nu_f  \sin \pi \nu_g}{f_1^4(e^{-\pi \tau})
\Theta_{1}(i\nu_f\tau|i\tau)
  \Theta_{1}(i\nu_g\tau|i\tau)
} \nonumber \\
&  {\times}& \left[f_3^4(e^{-\pi \tau}) \Theta_{3}(i\nu_f \tau|i
\tau) \Theta_{3}(i\nu_g\tau|i\tau) - f_4^4(e^{-\pi \tau})
\Theta_{4}(i\nu_f\tau|i\tau)\Theta_{4}(i\nu_g\tau|i\tau)  \right.
\nonumber
\\
&-&\left.  f_2^4(e^{-\pi \tau}) \Theta_{2}(i\nu_f
\tau|i\tau)\Theta_{2}(i\nu_g\tau|i\tau) \right] \label{zetae}
\end{eqnarray}
and\footnote{Notice that the sign of the topological term is
opposite to the one appearing in the Ref.s \cite{DLMP0407,LMP}
because of the different definition of the operator $N_0$ given in
Eq. (\ref{ni56bis}) with respect to the corresponding operator given in
Eq. (113) of Ref. \cite{LMP}.}
\begin{eqnarray}
Z_h^o & = & -\frac{N}{(8 \pi^2 \alpha')^2} \int d^4x \sqrt{
-\mbox{det}(\eta+\hat{F})}\nonumber\\
&\times &\int_{0}^\infty
\frac{d\tau}{\tau}e^{-\frac{y^2 \tau}{2\pi\alpha'}} \left[ \frac{
4\, \sin \pi \nu_f  \sin \pi \nu_g} {\Theta_{2}^2(0|i\tau)
\Theta_{1}(i\nu_f\tau|i\tau)\Theta_{1} (i\nu_g\tau|i\tau)} \right]
\nonumber \\
{} &\times& \left[ \Theta_{4}^2(0|i\tau) \Theta_{3}(i\nu_f\tau|i\tau)
\Theta_{3}(i\nu_g\tau|i\tau)-
\Theta_{3}^2(0|i\tau)\Theta_{4}(i\nu_f\tau|i\tau)
\Theta_{4}(i\nu_g\tau|i\tau)
 \right]  \nonumber \\
&-&   \frac{iN}{32\pi^2}\int d^4x\, F_{\alpha\beta}^a {\tilde
F}^{a\,\alpha\beta} \int_{0}^\infty \frac{d\tau}{\tau}e^{-\frac{y^2
\tau}{2\pi\alpha'}}~~,\label{zetatet}
\end{eqnarray}
where $\tilde{F}_{\alpha \beta }=\frac{1}{2}\epsilon _{\alpha \beta
\delta \gamma }F^{\delta \gamma }$. The $\Theta$-functions are
listed in \ref{app0}. In the previous equations we have
defined:
\begin{eqnarray}
\nu_{f} \equiv \frac{1}{2 \pi i}\log \frac{1+ 2 \pi \alpha'
{\hat{f}}}{1- 2 \pi \alpha' {\hat{f}}} \,\,\,\,\,\,\,\,\,\,\,{\rm
and}\,\,\,\,\,\,\,\,\,\,\, \nu_{g} \equiv \frac{1}{2 \pi i}\log
\frac{1-i 2 \pi \alpha' {\hat{g}}}{1+i 2 \pi \alpha'{\hat{g}}} ~~.
\label{nufnug2}
\end{eqnarray}
The calculation of $Z_{e}$ for the untwisted sector was originally
done in Ref.~\cite{BAPO} for the case of a D9 brane. The three terms
in Eq. (\ref{zetae})  come respectively  from the $NS$, $NS(-1)^F$
and $R$ sectors, while the contribution from the $R(-1)^F$ sector
vanishes. In Eq. (\ref{zetatet}) the three terms come respectively
from the $NS, NS(-1)^F$ and $R(-1)^F$ sectors, while the $R$
contribution vanishes because the projector $h$ annihilates the
Ramond vacuum.

The above computation can also be  performed in the {\em closed
string channel} where $Z_e^c$ and $Z_h^c$ are now given by the tree
level closed string amplitude between two untwisted and two twisted
boundary states respectively:
\begin{eqnarray}
Z_e^c = \frac{ \alpha' \pi N }{2} \int_{0}^{\infty} dt \, \, \,
{}^U \langle D3; F | e^{-\pi t ( L_0 + \bar{L}_{0}) } |D3
\rangle^{U} \label{ze32}
\end{eqnarray}
and
\begin{eqnarray}
Z_{h}^c = \frac{ \alpha'\pi N}{2} \int_{0}^{\infty} dt \, \, \,
{}^T \langle D3; F | e^{- \pi t (L_0 + \bar{L}_{0}) } |D3
\rangle^{T}~~,\label{zg32}
\end{eqnarray}
where $ |D3; F > $ is the boundary state dressed with the gauge
field $F$. The details of this calculation are presented in Appendix
C of Ref.~\cite{LMP}. Here we give only the final results that are
\begin{eqnarray}
Z_e^c & = & \frac{N\, }{(8\pi^2\alpha')^2} \int d^4x \sqrt{
-\mbox{det}(\eta+\hat{F})} \int_{0}^\infty
\frac{dt}{t^3}e^{-\frac{y^{2} }{2\pi\alpha't}} \frac{\sin\pi\nu_f
\sin\pi\nu_g}{ \Theta_{1}(\nu_f|it)\Theta_{1}(\nu_g|it)
f_1^4(e^{-\pi t})}
\nonumber\\
&& \times \left\{f_3^4(e^{-\pi t}) \Theta_{3}(\nu_f|it)\Theta_{3}(\nu_g|it)
-f_2^4(e^{-\pi t})\Theta_{2}(\nu_f|it)\Theta_{2}(\nu_g|it)\right.\nonumber\\
&&\left. -f_4^4(e^{-\pi t})\Theta_{4}(\nu_f|it) \Theta_{4}(\nu_g|it)
\right\} \label{op2}
\end{eqnarray}
and
\begin{eqnarray}
Z_h^c & = & \frac{N}{(8\pi^2\alpha')^2} \int d^4x \sqrt{
-\mbox{det}(\eta+\hat{F})} \int_{0}^\infty
\frac{dt}{t}e^{-\frac{y^{2} }{2\pi\alpha't}} \frac{4\,\sin\pi\nu_f
\sin\pi\nu_g}{ \Theta_{4}^2(0|it)
\Theta_{1}(\nu_f|it)\Theta_{1}(\nu_g|it)}
\nonumber\\
&&\times \left\{\Theta_{2}^2(0|it) \Theta_{3}(\nu_f|it)\Theta_{3}(\nu_g|it)
-\Theta_{3}^2(0|it) \Theta_{2}(\nu_f|it)\Theta_{2}(\nu_g|it)\right\}\nonumber\\
&&-\frac{iN}{32\pi^2} \int d^4x F^a_{\alpha\beta}\tilde
F^{a\alpha\beta}\int \frac{dt}{t} e^{-\frac{y^{2} }{2\pi\alpha't}} ~~.
\label{closed}
\end{eqnarray}
The three terms in Eq. (\ref{op2}) respectively come from the
$NS$-$NS$, $R$-$R$ and $NS$-$NS(-1)^ {F }$ sectors, while those in
Eq.~(\ref{closed}) from the $NS$-$NS$, $R$-$R$ and $R$-$R (-1)^ {F }$
sectors. In particular, the twisted odd $R$-$R (-1)^F$ spin structure
gets a nonvanishing contribution only from the zero modes, as
explicitly shown in Ref. \cite{LMP}.

It goes without saying that the two expressions for $Z$ separately
obtained in the open and  the closed string channels are, as
expected, equal to each other. This equality goes under the name of
open/closed string duality and can be easily shown  by using how
the $\Theta$ functions transform (see Eq. (\ref{modtras})) under the
modular transformation that relates the modular parameters in the
open and closed string channels, namely $\tau=\frac{1}{t}$. It can
be easily  seen that, in going from the open (closed) to the closed
(open) string channel, we have the following correspondence between
the various non vanishing spin structures:~\cite{DL99121}
\begin{eqnarray}
&& NS \leftrightarrow  NS-NS~, ~~~~~~~~NS (-1)^F
\leftrightarrow R-R~~\nonumber\\
&& R \leftrightarrow  NS-NS (-1)^F~,~~R (-1)^F \leftrightarrow R-R
(-1)^F ~~.\label{spinstruco}
\end{eqnarray}
The distance $y$ between the dressed
D3 brane and the stack of the $N$ D3 branes  makes the integral in
Eq.~(\ref{closed}) convergent for small values of $t$, while in the
limit $t \rightarrow \infty$, the integral is logarithmically
divergent. This divergence is  due to a twisted tadpole
corresponding to the exchange of massless closed string states
between the two boundary states in Eq. (\ref{zg32}). We would like
here to stress  that the presence of the gauge field makes the
divergence to appear already {\em at the string level} before any
field theory limit ($\alpha ' \rightarrow 0$) is performed. When $F$
vanishes, the divergence is eliminated by the integrand being
identically zero as a consequence of the fact that the fractional
branes are BPS states.

As observed in Ref.s~\cite{PC,LR,KAKURO,BM0002,NAPOLI} tadpole
divergences  correspond in general to the presence of gauge
anomalies that make the gauge theory inconsistent and have to be
eliminated by drastically modifying the theory or by fixing
particular values of the parameters. For instance in type I
superstring they are eliminated by fixing the gauge group to be
$SO(32)$. As stressed in Ref.s~\cite{LR,KAKURO,BM0002,NAPOLI}
logarithmic tadpole divergences do not instead correspond to gauge
anomalies.  In the bosonic string they have been cured in different
ways.~\cite{DIL,SHA,LOVE,FS,CLNY,MT} It turns out, in our case, that
the logarithmic divergent tadpoles correspond to the fact that the
gauge theory living on the brane is not conformal invariant and in
fact they provide the correct one-loop running coupling constant.
Following the suggestion of Ref.s~\cite{LR,KAKURO}, we cure these
divergences just by introducing in Eq. (\ref{closed}) an infrared
cutoff that regularizes the contribution of the massless closed
string states. Since in the open/closed string duality an infrared
divergence in the closed string channel corresponds to an
ultraviolet divergence in the open string channel it is easy to see
that the expression in Eq. (\ref{zetatet}) is divergent for small
values of $\tau$ and needs an ultraviolet cutoff. It will turn out
that this divergence is exactly the one-loop divergence that one
gets in ${\cal{N}}=2$ super Yang-Mills that is the gauge theory
living on the world-volume of the fractional D3 brane. Our results
are also consistent with the approach of Ref.~\cite{BIANCHI}.

In the following we want to use the previous string calculation for
deriving the coefficient of the quadratic term involving the gauge
field and to show that only massless states give a non-vanishing
contribution to it. The contribution of the massive states is
identically zero and this implies the absence of threshold
corrections.

To this aim it is useful to write Eq. (\ref{zetatet}) in a more
convenient way.
Using the notation for the $\Theta$-functions  given in Eq.
(\ref{theab98}) and the identity in Eq. (\ref{kir}) with
\begin{equation}
\left\{
\begin{array} {ll}
h_i=g_1=g_2=0
\nonumber\\
g_3=-g_4=1
\nonumber\\
\nu_1=i\nu_f\tau\,\,;\,\,
\nu_2=i\nu_g\tau\,\,;\,\,\nu_3=\nu_4=0\nonumber\\
\nu'_1=-\nu'_2=\frac{i}{2}\left(\nu_g-\nu_f\right)\tau\,\,;\,\,
\nu'_3=\nu'_4=\frac{i}{2}\left(\nu_g+\nu_f\right)\tau
\end{array}
\right.
\end{equation}
we can rewrite Eq. (\ref{zetatet}) as follows:
\begin{eqnarray}
Z_h^o &= & \frac{N}{(8 \pi^2 \alpha')^2} \int d^4x \sqrt{
-\mbox{det}(\eta+\hat{F})} \int_{0}^\infty
\frac{d\tau}{\tau}e^{-\frac{y^2 \tau}{2\pi\alpha'}}\nonumber\\
&\times& \left[ \frac{
4\, \sin \pi \nu_f  \sin \pi \nu_g} {\Theta_{2}^2(0|i\tau)
\Theta_{1}(i\nu_f\tau|i\tau)\Theta_{1} (i\nu_g\tau|i\tau)} \right]
 \left[ \Theta_{2}^2(0|i\tau)\Theta_{1}(i\nu_f\tau|i\tau)
\Theta_{1}(i\nu_g\tau|i\tau)\right.\nonumber\\
&-&\left.
2\Theta_{1}\left(i\frac{\nu_g-\nu_f}{2}\tau|i\tau\right)
\Theta_{1}\left(i\frac{\nu_f-\nu_g}{2}\tau|i\tau\right)
\Theta_{2}^2\left(i\frac{\nu_f+\nu_g}{2}\tau|i\tau\right) \right]  \nonumber \\
 {} & -& \frac{iN}{32\pi^2}\int d^4x\, F_{\alpha\beta}^a {\tilde
F}^{a\,\alpha\beta} \int_{0}^\infty \frac{d\tau}{\tau}e^{-\frac{y^2
\tau}{2\pi\alpha'}}~~,\label{zetatet1}
\end{eqnarray}
which turns out to be equal to
\begin{eqnarray}
&&Z_h^o  =  -\frac{2N}{(8 \pi^2 \alpha')^2} \int d^4x \sqrt{
-\mbox{det}(\eta+\hat{F})} \int_{0}^\infty
\frac{d\tau}{\tau}e^{-\frac{y^2 \tau}{2\pi\alpha'}} \left[ \frac{
4\, \sin \pi \nu_f  \sin \pi \nu_g} {\Theta_{2}^2(0|i\tau)
\Theta_{1}(i\nu_f\tau|i\tau)\Theta_{1} (i\nu_g\tau|i\tau)} \right]
\nonumber \\
&& \hspace{.5cm} \times \Theta_{1}\left(i\frac{\nu_g-\nu_f}{2}\tau|i\tau\right)
\Theta_{1}\left(i\frac{\nu_f-\nu_g}{2}\tau|i\tau\right)
\Theta_{2}^2\left(i\frac{\nu_f+\nu_g}{2}\tau|i\tau\right)
\label{zetatet2} .
\end{eqnarray}
because the first and the third terms in Eq. (\ref{zetatet1}) cancel
each other.
By expanding the previous equation up to the second order in $F$ and
using Eq. (\ref{usere34})
together with $\nu_f\simeq-i\frac{f}{\pi}$ and
$\nu_g\simeq-\frac{g}{\pi}$, we get:
\begin{eqnarray}
Z_h^o & = & \frac{N}{ 32 \pi^2} \int d^4x (F_{\alpha\beta}^a
{F}^{a\,\alpha\beta}-iF_{\alpha\beta}^a {\tilde F}^{a\,\alpha\beta})
\int_{0}^\infty \frac{d\tau}{\tau}e^{-\frac{y^2 \tau}{2\pi\alpha'}},
\label{zetatet3}
\end{eqnarray}
which reduces to
\begin{eqnarray}
Z_h^o(F)\!\!&\rightarrow&\!\!\left[- \frac{1}{4} \int d^4 x
F_{\alpha \beta}^{a} F^{a\,\alpha \beta } \right] \left\{
\frac{1}{g_{YM}^{2} (\Lambda )}
 - \frac{N}{8 \pi^2}
\int_{ { \frac{1}{\alpha' \Lambda^2}} }^{\infty} \frac{d
\tau}{\tau} {e}^{-\frac{y^{2} \tau}{2 \pi \alpha' } }
\right\} \nonumber\\ && - iN \left[\frac{1}{32\pi^2} \int d^4x
F^a_{\alpha\beta}\tilde F^{a\,\alpha\beta} \right] \int_{  {
\frac{1}{\alpha' \Lambda^2}} }^{\infty}
\frac{d\tau}{\tau}e^{-\frac{y^{2} \tau}{2\pi\alpha'}} ~~.
\label{gau57}
\end{eqnarray}
In the closed string channel we get instead:
\begin{eqnarray}
Z_h^c(F)\!\!&\rightarrow&\!\!\left[- \frac{1}{4} \int d^4 x
F_{\alpha \beta}^{a} F^{a\,\alpha \beta }\right] \left\{
\frac{1}{g_{YM}^{2}(\Lambda ) } - \frac{N}{8 \pi^2}
\int_{0}^{\alpha' \Lambda^2} \frac{dt}{t}  {e}^{- \frac{y^{2}}{2
\pi\alpha' t} }
\right\}
\nonumber\\
&& - iN \left[ \frac{1}{32\pi^2} \int d^4x F^a_{\alpha\beta}\tilde
F^{a\,\alpha\beta} \right] \int_{0}^{ {\alpha' \Lambda^2 }}
\frac{dt}{t} e^{-\frac{y^{2} }{2\pi\alpha't}} ~~. \label{F285}
\end{eqnarray}
Notice that in the two previous equations we have also added the
contribution, coming from the tree diagrams, that contains the bare
coupling constant. In an ultraviolet finite theory as string theory
we should not deal with a bare and a renormalized coupling. On the
other hand, we have already discussed the fact that the introduction
of a gauge field produces a string amplitude that is already
divergent at the string level and that therefore must be regularized
with the introduction of a cutoff.

We have already mentioned that Eq.s (\ref{gau57}) and (\ref{F285})
are equal to each other as one can see by performing the modular
transformation $\tau = \frac{1}{t}$. Actually one can see that,
under such a transformation, the contribution of the massless open
string states gets transformed into that of the massless closed
string states and viceversa. This follows from the fact that the
threshold corrections vanish in the two channels. This also means
that the open/closed string duality exactly maps the ultraviolet
divergent contribution coming from the massless open string states
circulating in the loop and that reproduces the divergences of
${\cal{N}}=2$ super Yang-Mills, living on the world-volume of the
fractional D3 branes, into the infrared divergent contribution due
to the massless closed string states propagating between the two
boundary states.

In the open string channel the integrals are naturally regularized
in the infrared ($\tau \rightarrow \infty$) by the fact that the two
stacks of branes are at a finite distance $y$.
In the closed string channel the presence of a distance $y$ between
the branes  makes the integral convergent in the ultraviolet ($t
\rightarrow 0$), but instead  an infrared cutoff $\Lambda$ is
needed. If we identify the two cutoffs $\Lambda$'s, we see that the
expressions  in the two field theory limits are actually equal! This
observation clarifies now why the supergravity solution gives the
correct answer for the perturbative behaviour of the non-conformal
world-volume theory as found in
Ref.s~\cite{d3,polch,marco,grana2,d3d7} and reviewed in
Ref.~\cite{REV}. In fact, we can extract the coefficient of the term
$F^2$ from either of the two Eq.s~(\ref{gau57}) and ~(\ref{F285})
getting the following expression:
\begin{eqnarray}
\frac{1}{g_{YM}^{2} (\epsilon) } + \frac{N}{8\pi^2} \log
\frac{y^{2}}{\epsilon^2}  \equiv \frac{1}{g_{YM}^{2} (y ) }~~,~~
\epsilon^2 \equiv 2 \pi (\alpha' \Lambda )^2 ~~,\label{run2}
\end{eqnarray}
where the integral appearing in Eq.~(\ref{gau57}) has been
explicitly computed:
\begin{eqnarray}
I (\Lambda,y) \equiv \int_{1/\alpha'\Lambda^2}^{\infty}
\frac{d\tau}{\tau} e^{-\frac{y^2 \tau}{2\pi (\alpha')}} \simeq \log
\frac{2 \pi
  (\alpha' \Lambda)^2}{y^2} ~~.
\label{comple67}
\end{eqnarray}
Eq. (\ref{run2}) is equal to the first equation in (\ref{thetaym76})
for $M=0$
where we identify the square of the bare coupling constant
$g_{YM}^{2} ( \Lambda)$ computed at the cutoff $\Lambda \sim
{\epsilon}/{\alpha'}$ with $8 \pi g_s$.
The previous derivation makes it clear why the running coupling
constant of ${\cal{N}}=2$ super Yang-Mills can be obtained from the
supergravity solution corresponding to $N$ fractional D3 branes of
the orbifold $\mathbb{C}^2 /\mathbb{Z}_2$.

Eq.~(\ref{run2}) gives the one-loop correction to the bare gauge
coupling constant $g_{YM} (\Lambda )$  in the gauge theory
regularized with the cutoff $\Lambda$. The renormalization procedure
can then be performed by introducing the renormalized coupling
constant $g_{YM}^{ren} (\mu)$ given in terms of the bare one by the
relation:
\begin{eqnarray}
 \frac{1}{g_{YM}^{2} (\Lambda)} =
\left( \frac{1}{g_{YM}^{2} (\mu) } \right)^{ren} + \frac{N}{8\pi^2} \log
\frac{\Lambda^{2}}{\mu^2}~~,\label{run23}
\end{eqnarray}
with $\mu$ being the renormalization scale. Using the previous
equation in Eq. (\ref{run2}) one can rewrite the coefficient of the
$F^2$ term in terms of the renormalized gauge coupling constant
\begin{eqnarray}
\left( \frac{1}{g_{YM}^{2} (\mu)} \right)^{ren}  + \frac{N}{8\pi^2} \log
\frac{m^{2}}{\mu^2} = \left( \frac{1}{g_{YM}^{2} (m)}
\right)^{ren}~~~;~~~ m^2 \equiv
\frac{y^2}{2 \pi \alpha^{'2}} ~~.\label{run24}
\end{eqnarray}
{From} it, or equivalently from Eq.
(\ref{run23}), we can now determine the one-loop $\beta$-function:
\begin{eqnarray}
\beta \equiv \mu \frac{\partial}{\partial \mu} g_{YM}^{ren}(\mu) =
-\frac{g^{ren\,3}_{YM} N}{8 \pi^2}~~,\label{betafun}
\end{eqnarray}
that is the correct one for ${\cal{N}}=2$ super Yang-Mills.

Let us turn now to the vacuum angle $\theta_{YM}$ that is provided
by the terms in Eq.s~(\ref{gau57}) and ~(\ref{F285}) with the
topological charge. If we extract it from either of the two
Eq.s~(\ref{gau57}) and ~(\ref{F285}) we find that it is imaginary
and moreover must be renormalized as the coupling constant. A way of
eliminating these problems is to introduce a complex cutoff, to
allow the gauge field to be in either one of the two stacks and by
taking the symmetric combination:
\begin{eqnarray}
\frac{1}{2} \left[\langle D3; F | D  |D3 \rangle + \langle D3 | D  |
D3; F \rangle \right] = \frac{1}{2} \left[\langle D3; F | D  |D3
\rangle + \overline{\langle D3; F | D  | D3 \rangle} \right]~~.
\label{sym78}
\end{eqnarray}
If we introduce a complex cutoff $ \Lambda \rightarrow \Lambda
{e}^{-i
  \theta} $ the divergent
integral in Eq.~(\ref{gau57}) becomes:
\begin{eqnarray}
I (z) \equiv \int_{1/( \alpha'\Lambda^2 e^{-2i\theta}) }^{\infty}
\frac{d\tau}{\tau} e^{-\frac{y^2 \tau}{2\pi \alpha'}} \sim \log
\frac{2 \pi
  (\alpha' \Lambda)^2}{y^2 {e}^{ 2i \theta}}=\log
\frac{2 \pi
  (\alpha' \Lambda)^2}{z^2}~~.
\label{comple67bis}
\end{eqnarray}
This procedure leaves unchanged all the previous considerations
concerning the gauge coupling constant because in this case one
gets as before:
\begin{eqnarray}
\frac{1}{2} \left[ I (z) + I ({\bar{z}}) \right] = \log \frac{2 \pi
  (\alpha')^2 \Lambda^2}{y^{2}}~~.
\label{gau73}
\end{eqnarray}
For the $\theta_{YM}$ angle one gets instead:
\begin{eqnarray}
\theta_{YM} = - i\frac{N}{2} \left [ I (z) - I ({\bar{z}}) \right] =
-2 N\theta~~, \label{theta45}
\end{eqnarray}
that exactly reproduces the result given in the second equation in
(\ref{thetaym76}) for $M=0$. Remember, however, that in Eq. (\ref{thetaym76})
$\theta$ is the phase of the complex quantity $z = y {e}^{i
\theta}$. But, as it can be seen in Eq. (\ref{comple67bis}), giving
a phase to the cutoff corresponds to give the opposite phase to the
distance $y$ between the branes. We prefer to complexify the cutoff
rather than $y$ in order to keep the open string Virasoro operator
$L_0$ real, but the effect is in fact the same.

In the second part of this subsection we consider a bound state of
$N$ D3 branes and $M$ D7 branes in order to add matter
hypermultiplets to the pure gauge theory considered until now. In
particular we consider the same brane configuration as the one
analyzed in the previous subsection.

Let us start by giving the spectrum of the massless open strings stretched
between the two stacks of D3 and D7 branes. Physical states are taken in
the  picture $-1$ for the NS sector and in the picture $-1/2$ for the
R one. The massless ones come from the lowest level and are given by:
\begin{eqnarray}
\lambda_{\rm NS}|0;\,k \rangle_{-1} \otimes
|s_0=1/2,\,s_3,\,s_4\rangle \qquad \lambda_{\rm R} |s_{0}=1/2,
\, s_1,\,s_2 \rangle_{-1/2} \otimes |0;\,k \rangle ~~.\label{sd3d7}
\end{eqnarray}
The structure of the states in Eq. (\ref{sd3d7}) can be
easily understood if one considers that
the orbifold projection changes the modding of the oscillators
along the four directions spanned by the orbifold. Moreover the
orbifold breaks the Lorentz group $SO(1,9)$ to $SO(1,5) \otimes
SO(4)$ and in Eq. (\ref{sd3d7}) we have correspondingly  splitted
the string ground states.
In Eq.s  (\ref{sd3d7}) $\lambda$'s denote the Chan-Paton factors and
$s_{i}$ ($i=1, \dots, 4$) are the eigenvalues of the ``number
operators''$N_{i}$ so defined\footnote{In this paper the
  definition of number operators
differs by a factor 2 from the corresponding expressions given in
Ref.s \cite{DLMP0407,LMP} and, as previously noticed, $N_{0}$ has an opposite sign.} :
\begin{eqnarray}
\Gamma^0 \Gamma^1 = -2 N_0~~~,~~~\Gamma^{2i}\Gamma^{2i+1}=2\,i\,
N_{i} \qquad\mbox{ with}\,\,\,i=1,\dots,4~~. \label{ni56bis}
\end{eqnarray}
We remind that $s_{0}=1/2$ and that the GSO projectors are defined
as:
\begin{eqnarray}
P_{GSO}^{\rm NS}=\frac{(-1)^{G_{\beta\gamma}}
-\Gamma^6\dots\Gamma^9(-1)^F}{2} =\frac{(-1)^{G_{\beta\gamma}}
+2^2N_3N_4(-1)^F}{2} \label{GSO1}
\end{eqnarray}
and
\begin{eqnarray}
P_{GSO}^{\rm R}=\frac{(-1)^{G_{\beta\gamma}} -
\Gamma^0\dots\Gamma^5(-1)^{F_R}}{2}=\frac{(-1)^{G_{\beta\gamma}}
-2^3N_0N_1N_2(-1)^{F_R}}{2} \label{GSO1b}
\end{eqnarray}
where the fermion numbers $F$ and $F_R$
are obtained from the corresponding ones defined in Eq.s
(\ref{fns78}) and (\ref{fr79}) by changing modding of the
oscillators along the mixed $6\dots9$ directions, i.e. in the NS
sector the modding of the fermionic [bosonic] oscillators is integer
[half-integer]  while in the R one is half-integer [half-integer].

Eq.s (\ref{GSO1}) and (\ref{GSO1b})  impose that $s_{1}=-s_{2}$ and
$s_{3}=-s_4$. In the NS sector we have two real scalars while in the
R sector we have two Weyl fermions. Altogether they form an
hypermultiplet.

The physical states are the ones left invariant by the orbifold. In
the R sector the non trivial generator $h$ of the  orbifold group
acts as:
\begin{eqnarray}
 &\lambda^{\rm R}_{ij} |s_0=1/2,\,s_1,\,s_2 \rangle_{-1/2}
\otimes |0,\,k \rangle \rightarrow &
\nonumber \\
& \left(\gamma^{\rm D3}_h \right)_{ih}\lambda^{\rm
R}_{hk}(\gamma^{\rm D7}_h)_{kj}^{-1}|s_0=1/{2},\,s_1,\,s_2 \rangle_{-1/2}
\otimes |0;\,k \rangle & ~~,\label{orbac}
\end{eqnarray}
where $\gamma_{h}$ is the orbifold action on the Chan-Paton factors
and  can be taken as $ \pm \mathbb{I}$ depending on the kind of
fractional brane considered. The previous expression implies that
the surviving open strings are those stretched between fractional
branes of the same kind for which one has $\gamma^{\rm
D3}_h=\gamma^{\rm D7}_h$, while strings stretched between different
kinds of branes are projected out. Analogous considerations hold in
the NS sector where the non trivial element $h$ of the orbifold
group acts on the oscillator vacuum as $h=-4N_{3}N_{4}$.

The annulus diagram for open strings stretched between a dressed D3
and a bunch of $M$ D7 branes is given, as in the previous case, by
two contributions, $Z_{37}=Z^{o}_{e;37} + Z^{o}_{h;37}$, with:
\begin{eqnarray}
\!\!Z^{o}_{e;37}\! &=&\!\!-\frac{M}{(8 \pi^2 \alpha' )^2}\!\! \int\! d^4 x  \sqrt{-
\det ( \eta + {\hat{F}})}\!\!
\int_{0}^{\infty}\!\! \frac{d
  \tau}{\tau} {e}^{- \frac{y^2 \tau}{2\pi
  \alpha'}}
\frac{2 \sin \pi \nu_f \,2 \sin \pi \nu_g }{  \Theta_1( i \nu_f \tau
| i \tau ) \Theta_1 ( i \nu_g \tau | i \tau ) \Theta_4^2 (0 | i \tau
)}
\nonumber\\
&{\times}&
 \frac{1}{4}
 \left[ \Theta_2^2(0 | i\tau)  \Theta_3 ( i \tau \nu_g | i \tau )
\Theta_3 ( i \tau \nu_f | i \tau )  - \Theta_3^2(0|i\tau) \Theta_2 (
i \tau \nu_g | i
  \tau)
\Theta_2 ( i \tau \nu_f | i \tau ) \right]
\nonumber\\
&+&i\frac{M}{4}\left[\frac{1}{32 \pi^2} \int d^4x F_{\alpha\beta}^a
\tilde{F}^{a\,\alpha\beta}\right]\int^\infty_{1/(\alpha'\Lambda^2)}
\frac{d\tau}{\tau} e^{-\frac{y^2}{2\pi\alpha'}}
 \label{d3d71}
\end{eqnarray}
and
\begin{eqnarray}
\!\!Z^{o}_{h;37}\! & = &\! \mp\frac{M}{(8 \pi^2 \alpha')^2}\!\! \int\! d^4x \sqrt{
-\mbox{det}(\eta+\hat{F})}\!\! \int_{0}^\infty\!\!
\frac{d\tau}{\tau}e^{-\frac{y^2 \tau}{2\pi\alpha'}} \frac{
2\, \sin \pi \nu_f 2\, \sin \pi \nu_g} {\Theta_{3}^2(0|i\tau)
\Theta_{1}(i\nu_f\tau|i\tau)\Theta_{1} (i\nu_g\tau|i\tau)}
\nonumber \\
{} &{\times}& \frac{1}{4}\left[
\Theta_{2}^2(0|i\tau)\Theta_{4}(i\nu_f\tau|i\tau)
\Theta_{4}(i\nu_g\tau|i\tau) - \Theta_{4}^2(0|i\tau)
\Theta_{2}(i\nu_f\tau|i\tau)
\Theta_{2}(i\nu_g\tau|i\tau) \right]\nonumber\\
&\pm& i\frac{M}{4}\left[\frac{1}{32 \pi^2} \int d^4x
F_{\alpha\beta}^a
\tilde{F}^{a\,\alpha\beta}\right]\int^\infty_{1/(\alpha'\Lambda^2)}
\frac{d\tau}{\tau} e^{-\frac{y^2}{2\pi\alpha'}}~~.\label{gauge451}
\end{eqnarray}
In the latter equation the upper [lower]  sign refers to the case of
open strings stretched between fractional branes of the same
[different] kind.

In Eq. (\ref{d3d71}) the first  and the second term come
respectively from the NS  and R sector, while the last one comes
from the R$(-1)^F$ sector. The $NS(-1)^{F}$ sector does not
contribute due to the vanishing of the trace over the zero modes.
The first two terms of Eq. (\ref{d3d71})  can be obtained
respectively from the first and the third term in Eq. (\ref{zetae})
by changing, as explained after Eq. (\ref{GSO1b}), the modding of
the string oscillators along the directions $6\dots 9$. Such a
modification transforms $\Theta_{3}(0|i\tau)$ into
$\Theta_2(0|i\tau)$, leaving unchanged the $\Theta$'s having in their
argument $\nu_f$ or $\nu_g$, and maps the NS and R terms of Eq.
(\ref{zetae})  into the first two terms of Eq. (\ref{d3d71}).
Analogous considerations can be extended to Eq. (\ref{gauge451})
where the first, the second and the third terms correspond
respectively to the NS$(-1)^F$, R and R$(-1)^F$ sector. The first term
of Eq. (\ref{gauge451}) can be obtained from the second term in Eq.
(\ref{zetatet}) with the substitution $\Theta_3 \rightarrow
\Theta_2$ as before. The second term of Eq. (\ref{gauge451}) is
obtained from the first one in Eq. (\ref{zetatet}) by changing first
$\Theta_3 \rightarrow \Theta_2$ and $\Theta_4 \rightarrow \Theta_1$
which map the NS in the R sector and then by performing the
transformation $\Theta_1 (0|i\tau) \rightarrow \Theta_4(0|i\tau)$
that corresponds to change modding along the mixed directions.

Eq. (\ref{d3d71}) can be rewritten in a more compact form by using
Eq. (\ref{kir}) with:
\begin{eqnarray}
g_i=h_3=h_4=0,\,\,\,\,\,h_1=-h_2=1, \,\,\,\,\, \nu_1=\nu_2=0,
\,\,\,\,\,\nu_3=i\tau \nu_f, \,\,\,\,\,\nu_4=i\tau\nu_g~.
\label{scelta}
\end{eqnarray}
In this way one gets the following identity:
\begin{eqnarray}
&&\Theta_2^2(0 | i\tau)   \Theta_3 ( i \tau \nu_g | i \tau )
\Theta_3 ( i \tau \nu_f | i \tau )  - \Theta_3^2(0|i\tau) \Theta_2 (
i \tau \nu_g | i
  \tau)
\Theta_2 ( i \tau \nu_f | i \tau )=\nonumber\\
&& \Theta_4^2( 0 |i\tau) \Theta_1(i\tau \nu_f |i\tau)
\Theta_1(i\tau \nu_g
|i\tau)+2\Theta_4^2(i\frac{\tau}{2}(\nu_f+\nu_g) |i\tau)
 \Theta_1^2(i\frac{\tau}{2}(\nu_f-\nu_g) |i\tau)~,\nonumber\\
&&\label{id}
\end{eqnarray}
which allows us to rewrite Eq. (\ref{d3d71}) as:
\begin{eqnarray}
Z^{o}_{e;37} &=& -\frac{M}{(8 \pi^2 \alpha' )^2} \int d^4 x \sqrt{
{\rm det}\left(\eta+ \hat{F}\right)} \sin \pi \nu_f\sin \pi \nu_g
\int_{0}^{\infty}  \frac{d
  \tau}{\tau} {e}^{- \frac{y^2 \tau}{2\pi\alpha'}} \nonumber \\
&&{\times}\left\{ 1+ 2 \frac{\Theta_4^2(i\frac{\tau}{2}(\nu_f+\nu_g)
|i\tau)
 \Theta_1^2(i\frac{\tau}{2}(\nu_f-\nu_g) |i\tau)}{
\Theta_4^2(0|i\tau) \Theta_1(i\tau\nu_f|i\tau)
\Theta_1(i\tau\nu_g|i\tau)}\right\}\nonumber\\
&&+i\frac{M}{4}\left[\frac{1}{32 \pi^2} \int d^4x F_{\alpha\beta}^a
\tilde{F}^{a\,\alpha\beta}\right]\int^\infty_{1/(\alpha'\Lambda^2)}
\frac{d\tau}{\tau} e^{-\frac{y^2}{2\pi\alpha'}} ~~.
\label{ze37}
\end{eqnarray}
Analogously, by using again Eq. (\ref{kir}), but this time with:
\begin{eqnarray}
&& h_3=-h_4=g_3=-g_4=1 \qquad h_1=h_2=g_1=g_2=0  \nonumber\\
&&\nu_1=i \nu_f \tau  \qquad \nu_2=i \nu_g \tau \qquad \nu_3=\nu_4=0~~,
\end{eqnarray}
one gets the following identity:
\begin{eqnarray}
& \Theta_2^2\left(0
|i\tau\right)\Theta_4\left(i\nu_f \tau |i\tau\right)
\Theta_4\left(i\nu_g \tau |i\tau\right)- \Theta_4^2\left(0
|i\tau\right)\Theta_2\left(i\nu_f \tau |i\tau\right)
\Theta_2\left(i\nu_g \tau |i\tau\right)&\nonumber\\
&  = \Theta_3^2\left(0
|i\tau\right)\Theta_1\left(i\nu_f \tau |i\tau\right)
\Theta_1\left(i\nu_g \tau |i\tau\right)&\nonumber\\
&  +2\Theta_1^2\left(i\frac{\tau}{2}(\nu_f-\nu_g) |i\tau\right)
\Theta_3^2\left(i\frac{\tau}{2}(\nu_f+\nu_g)\tau |i\tau\right)~~, &
\end{eqnarray}
which allows one to write Eq. (\ref{gauge451}) as follows:
\begin{eqnarray}
 Z_{h;37}^o & = & \mp\frac{M}{(8 \pi^2 \alpha')^2} \int d^4x
\sqrt{ -\mbox{det}(\eta+\hat{F})} \sin \pi \nu_f\sin \pi \nu_g
\int_{0}^\infty \frac{d\tau}{\tau}e^{-\frac{y^2 \tau}{2\pi\alpha'}}
\nonumber \\
&&{\times}\left\{1 +2 \frac{
\Theta_1^2\left(i\frac{\tau}{2}(\nu_f-\nu_g) |i\tau\right)
\Theta_3^2\left(i\frac{\tau}{2}(\nu_f+\nu_g)
|i\tau\right)}{\Theta_{3}^2(0|i\tau)
\Theta_{1}(i\nu_f\tau|i\tau)\Theta_{1}
(i\nu_g\tau|i\tau)} \right\} \nonumber\\
&&\pm i\frac{M}{4}\left[\frac{1}{32 \pi^2} \int d^4x
F_{\alpha\beta}^a
\tilde{F}^{a\,\alpha\beta}\right]\int^\infty_{1/(\alpha'\Lambda^2)}
\frac{d\tau}{\tau} e^{-\frac{y^2}{2\pi\alpha'}}~~.
\label{zh37}
\end{eqnarray}
Expanding Eq.s (\ref{ze37}) and (\ref{zh37}) up to the
quadratic terms in the gauge field, according to the procedure above
explained, yields:
\begin{eqnarray}
\!\!\!\!\!\!Z^{o}_{h;37}= \pm Z^{o}_{e;37}  &=&  \mp
\frac{M}{2(8 \pi^2 \alpha' )^2} \int d^4 x  [ 2\,if\,g + (if-g)^2]
\int^{\infty}_{1/(\alpha'\Lambda^2)}  \frac{d
  \tau}{\tau} {e}^{- y^2 \tau/(2\pi\alpha')}\nonumber\\
&\pm& i\frac{M}{4}\left[\frac{1}{32 \pi^2} \int d^4x
F_{\alpha\beta}^a
\tilde{F}^{a\,\alpha\beta}\right]\int^\infty_{1/(\alpha'\Lambda^2)}
\frac{d\tau}{\tau} e^{-\frac{y^2}{2\pi\alpha'}}\nonumber\\
&=& \pm \frac{M}{2(4 \pi )^2}\!\!\int\! d^4
x\!\left[-\frac{1}{4}\!\left( F_{\mu\nu}^a F^{a\,\mu\nu} -i
F_{\alpha\beta}^a \tilde{F}^{a\,\alpha\beta}\right)\!  \right]\! \nonumber \\
&& \times
\int^{\infty}_{1/(\alpha'\Lambda^2)}  \frac{d
  \tau}{\tau} {e}^{- \frac{y^2 \tau}{2\pi\alpha'}} .
\end{eqnarray}
{From} it we can extract the gauge coupling constant:
\begin{eqnarray}
&&\frac{1}{g_{YM}^2} = \frac{M}{(4 \pi)^2} \left( \frac{1}{2} \pm
\frac{1}{2} \right) \int_{1/(\alpha' \Lambda^2)}^{\infty} \frac{d
\tau}{\tau} {e}^{- y^2 \tau/(2\pi
  \alpha')} \nonumber \\
&&=   \frac{M}{(4 \pi)^2}  \left( \frac{1}{2} \pm \frac{1}{2}
\right) \log \frac{2 \pi (\alpha'
  \Lambda)^2}{y^2}
\label{d3d74}
\end{eqnarray}
and by taking,  as explained in Eq. (\ref{sym78}), the symmetric
combination, we obtain for the $\theta_{\rm YM }$ angle:
\begin{eqnarray}
\theta_{\rm YM}=  \left( \frac{1}{2} \pm \frac{1}{2} \right) M
\theta~~,\label{theta37}
\end{eqnarray}
where the positive [negative] sign corresponds to open strings
stretched between fractional branes of the same [different] kind.

These are just the expected running coupling constant and chiral
anomaly and, when added respectively to
 Eq. (\ref{run2}) and (\ref{theta45}),
coincide with Eq. (\ref{thetaym76}) obtained via supergravity.
 It is important to observe that also in this case
threshold corrections vanish and this provides the reason why the
contribution of the massless string states gets transformed into
that of the massless closed string states and viceversa, making the
gauge/gravity correspondence to hold.

\subsection{Gauge/gravity correspondence from open/closed duality:
${\cal N} =1$ case}

In the following we extend the analysis performed in the previous
section to the case of the orbifold $\mathbb{ C}^3/(\mathbb{
Z}_2{\times} \mathbb{Z}_2)$ that preserves four supersymmetry
charges. The orbifold group $\mathbb{ Z}_2{\times}\mathbb{Z}_2$
contains four elements whose action on the three complex
coordinates:
\begin{eqnarray}
&&z_1=x_4+ix_5 \qquad z_2=x_6+ix_7 \qquad z_3=x_8+ i x_9
\label{z123}
\end{eqnarray}
is chosen to be:
\begin{eqnarray}
R_e= \left[
\begin{array}{ccc}
~~1 & ~0& ~~0
\\ ~~0& ~1 & ~~0 \\ ~~0 & ~0  & ~~1
\end{array}
\right],  && \quad  R_{h_1}=\left[
\begin{array}{ccc}
1 & 0& 0
\\ 0& -1 & 0 \\ 0 & 0  & -1
\end{array}
\right], \nonumber\\
 \quad R_{h_2}= \left[
\begin{array}{ccc}
-1 & 0& 0
\\ 0& 1 & 0 \\ 0 & 0  & -1
\end{array}
\right], &&\quad   R_{h_3}= \left[
\begin{array}{ccc}
-1 & 0& 0
\\ 0& -1 & 0 \\ 0 & 0  & 1
\end{array}
\right] \ .\label{erre}
\end{eqnarray}
The orbifold group acts also on the Chan-Paton factors. Fractional
branes are defined as branes for which these latter transform
according to irreducible representations of the orbifold group.
$\mathbb{ Z}_2{\times} \mathbb{ Z}_2$ has four irreducible
one-dimensional representations that correspond to four different
kinds of  fractional branes. The orbifold $\mathbb{ C}^3/(\mathbb{
Z}_2{\times} \mathbb{ Z}_2)$, as already explained in
Ref.s~\cite{NAPOLI,FERRO}, can be seen as obtained by three copies
of the orbifold $\mathbb{ C}^2/\mathbb{ Z}_2$ where the $i$-th
$\mathbb{Z}_2$ contains the elements $(e, h_i)$ $(i=1,2,3)$. This
means that the boundary states associated to each fractional brane
are:
\begin{eqnarray}
&&| Dp>_1 =| Dp>_u+ | Dp>_{t_1}+ | Dp>_{t_2} +| Dp>_{t_3} \ , \nonumber\\
&&| Dp>_2 =| Dp>_u+ | Dp>_{t_1}- | Dp>_{t_2} -| Dp>_{t_3} \ ,\nonumber \\
&&| Dp>_3 =| Dp>_u- | Dp>_{t_1}+ | Dp>_{t_2} -| Dp>_{t_3} \ ,\nonumber \\
&&| Dp>_4 =| Dp>_u- | Dp>_{t_1}- | Dp>_{t_2} +| Dp>_{t_3} \ ,\label{bs}\\
\nonumber
\end{eqnarray}
where $| Dp>_u$ is the untwisted boundary state that, apart from an
overall factor $\frac{1}{2}$ due to the orbifold projection, is the
same as the one in flat space and $ | Dp>_{t_i}$ $(i=1,2,3)$ are
exactly the same as the twisted boundary states on the orbifold
$\mathbb{ C}^2/\mathbb{ Z}_2$, apart from a factor $\frac{1}{\sqrt
2}$. The signs in front of each twisted term in Eq. (\ref{bs})
depend on the irreducible representation chosen for the orbifold
group action on the Chan-Paton factors.

In order to keep the forthcoming discussion as general as possible,
we study the one-loop vacuum amplitude of an open string stretching
between a stack of $N_I$ ($I=1, \dots, 4$) branes of type $I$ and a
D3 fractional brane, for example of type $I=1$, with a background
$SU(N)$ gauge field turned-on on its world-volume. Due to the
structure of the orbifold  $\mathbb{ C}^3/(\mathbb{ Z}_2{\times}
\mathbb{ Z}_2)$, this amplitude is the sum of four terms:
\begin{eqnarray}
Z= Z_{e} + \sum_{i=1}^{3} Z_{h_i},    \label{ZN=1}
\end{eqnarray}
where $Z_e$ and $Z_{h_i}$ are obtained in the open [closed]  channel
by multiplying Eq.s~(\ref{zetae}) and  (\ref{zetatet}) [Eq.s~(\ref{op2}) and
(\ref{closed})] by an extra 1/2
factor due to the orbifold projection. Since, even in this case, we
are interested in analyzing the meaning of the divergences in the
non-trivial twisted contributions $Z_{h_i}$, we focus on these
latter. In particular, in the open string channel, $Z^{o}_{h_i}$ is:
\begin{eqnarray}
\!Z_{h_i}^o &\!\! = &\! \frac{f_i(N)}{2\,(8\pi^2\alpha')^2}\! \int\!
d^4x \sqrt{
-\mbox{det}(\eta+\hat{F})}\! \int_{0}^\infty\! \frac{d
\tau}{\tau}e^{-\frac{y^{2}_i \tau }{2\pi\alpha'}}
\frac{2\,\sin\pi\nu_f 2\, \sin\pi\nu_g}{ \Theta_{2}^2(0|i \tau)
\Theta_{1}(i \nu_f \tau| i \tau) \Theta_{1}(i \nu_g \tau|i \tau)}
\nonumber\\
&\times&\left\{\Theta_{3}^2(0|i \tau) \Theta_{4}(i \nu_f \tau|i
\tau)\Theta_{4}(i \nu_g \tau |i \tau) -\Theta_{4}^2(0|i \tau)
\Theta_{3}(i \nu_f \tau|i \tau)
\Theta_{3}(i \nu_g \tau |i \tau)\right\}\nonumber\\
&-&\frac{i\, f_i(N)}{64\pi^2} \int d^4x F^a_{\alpha\beta}\tilde F^{a
\, \alpha\beta}\int \frac{d\tau}{\tau} e^{-\frac{y_i^{2} \tau
}{2\pi\alpha'}}~~. \label{ztet1}
\end{eqnarray}
In the previous expression we should put to zero the distance $y_i$
between the stack of the $N_I$ branes and the dressed one, since the
fractional branes are constrained to live at the orbifold fixed
point $z_1=z_2=z_3=0$. However $y_i$ provides a natural infrared
cut-off in Eq.~(\ref{ztet1}). Therefore we keep this quantity small
but finite and we are going to put it to zero just at the end of the
calculation. The functions $f_i(N)$ introduced in Eq.~(\ref{ztet1})
depend on the number of the different kinds of fractional branes
$N_I$ and their explicit expressions are: \cite{NAPOLI,FERRO}
\begin{eqnarray}
&& f_1(N_I)= N_1 + N_2 - N_3 - N_4~~,\nonumber\\  && f_2(N_I)= N_1 -
N_2 + N_3 - N_4~~, \nonumber\\  && f_3(N_I)= N_1 - N_2 - N_3 +
N_4~~. \label{coupling}
\end{eqnarray}
Let us now extract in both channels the quadratic terms in the gauge
field $F$. In the open sector, we get:
\begin{eqnarray}
Z_h^o(F) &\rightarrow& \left[- \frac{1}{4} \int d^4 x
F_{\alpha \beta}^{a} F^{a\,\alpha \beta } \right] \left\{
\frac{1}{g_{YM}^{2} (\Lambda )}
 - \sum_{i=1}^{3} \frac{f_i(N) }{16 \pi^2} \left[
\int_{1/(\alpha' \Lambda^2)}^{\infty} \frac{d \tau}{\tau}
{e}^{-\frac{y^{2}_i \tau}{2 \pi \alpha' } }
\right] \right\}
\nonumber\\
&&-i \left[\frac{1}{32\pi^2} \int d^4x F^a_{\alpha\beta}\tilde
F^{a\,\alpha\beta} \right] \sum_{i=1}^3 \frac{f_i(N)}{2}
\int_{\frac{1}{\alpha'\Lambda^2}}^{\infty}
\frac{d\tau}{\tau}e^{-\frac{y^{2}_i \tau}{2\pi\alpha'}}~~,
\label{gau571}
\end{eqnarray}
while in the closed string channel we obtain:
\begin{eqnarray}
Z_h^c(F)&\rightarrow&        \left[- \frac{1}{4}
\int d^4 x F_{\alpha \beta}^{a} F^{a\,\alpha \beta }\right] \left\{
\frac{1}{g_{YM}^{2}(\Lambda ) } - \sum_{i=1}^{3} \frac{f_i(N)}{16
\pi^2} \left[ \int_{0}^{\alpha' \Lambda^2} \frac{dt}{t}  {e}^{-
\frac{y^{2}_i}{2 \pi\alpha' t} }
\right] \right\}
\nonumber\\
&& - i \left[ \frac{1}{32\pi^2} \int d^4x F^a_{\alpha\beta}\tilde
F^{a\,\alpha\beta} \right] \sum_{i=1}^{3} \frac{f_i(N)}{2}
\int_{0}^{\alpha' \Lambda^2} \frac{dt}{t} e^{-\frac{y_i^{2}
}{2\pi\alpha't}} ~~. \label{F2851}
\end{eqnarray}
Analogously to the case of the previous orbifold
the divergent contribution is due to the massless states in both
channels. We have also introduced the one coming from the tree
diagrams. The main properties exhibited by the interactions in the
$\mathbb{C}^2 /\mathbb{Z}_2$ orbifold, given in Eq.s~(\ref{gau57})
and (\ref{F285}), are also shared by the interactions in the
$\mathbb{C}^3/(\mathbb{Z}_2 {\times} \mathbb{Z}_2)$ orbifold, given
in Eq.s~(\ref{gau571}) and (\ref{F2851}). In particular, also in
this case, one can see that the contribution  of the massive states
vanishes and the open string massless contribution is transformed into
the closed string massless one. This confirms the main result
obtained in the previous subsection, i.e. the open/closed string
duality exactly maps the ultraviolet divergent contribution coming
from the massless open string states, which reproduces the
divergences of ${\cal N}=1$ super Yang-Mills, into the infrared
divergent contribution due to the massless closed string states. By
extracting the coefficient of the term $F^2$ in Eq. (\ref{gau571})
or Eq. (\ref{F2851}), we get:
\begin{eqnarray}
\frac{1}{g^2_{\rm YM} (\epsilon)}+ \frac{1}{16 \pi^2} \sum_{i=1}^3
f_i(N_I) \log\frac{y_{i}^{2}}{\epsilon^{2}} \equiv \frac{1}{g^2_{\rm
YM}}(y)\qquad \epsilon^2= 2\pi (\alpha' \Lambda)^2~~. \label{run21}
\end{eqnarray}
Eq.~(\ref{run21}) reproduces Eq.~(3.14) of Ref.~\cite{ANOMA}
clarifying again why the supergravity solution, dual to ${\cal N}=1$
super Yang-Mills theory, gives the correct answer for the
perturbative behaviour of the non-conformal world-volume theory, as
found in Ref.s~\cite{ANOMA,NAPOLI,FERRO}.

In performing the renormalization procedure, we introduce the
renormalized coupling constant $g_{\rm YM}^{ren} (\mu)$ given in terms of
the bare one by the relation:

\begin{eqnarray}
 \frac{1}{g_{YM}^{2} (\Lambda)}& = &
\left( \frac{1}{g_{YM}^{2} (\mu) } \right)^{ren} +
\sum_{i=1}^{3}\frac{f_i(N)}{16\pi^2}
\log \frac{\Lambda^{2}}{\mu^2} \nonumber \\
& = & \left( \frac{1}{g_{YM}^{2} (\mu) } \right)^{ren} +
\frac{3N_1 - N_2 -N_3 -N_4}{16\pi^2} \log \frac{\Lambda^{2}}{\mu^2}~~.
\label{run231}
\end{eqnarray}
{From} this equation we can obtain the $\beta$-function:
\begin{eqnarray}
\beta (g_{\rm YM}^{ren} ) \equiv \mu \frac{\partial}{\partial \mu}
g^{ren}_{YM} (\mu) =
-\frac{g^{ren\,3}_{YM}}{16 \pi^2} \, \, (3N_1 - N_2 - N_3 -N_4)~~,
\label{betafun1}
\end{eqnarray}
that is  the correct one for the world-volume theory living on the
dressed brane.

Finally, in the same spirit as in Sect. 3, we  consider the
symmetric combination given in Eq.~(\ref{sym78}) and by introducing
a complex cut-off $\Lambda e^{-i\theta}$, or equivalently the
complex coordinates $z_i= y_i e^{i\theta}$, we get the following
expression for $\theta_{\rm YM}$:
\begin{eqnarray}
\theta_{\rm YM}=- \sum_{i=1}^3 f_i(N_I)\theta ~~,\label{theta451}
\end{eqnarray}
that is again in agreement with the result given in Eq.~(3.14) of
Ref.~\cite{ANOMA}.

\section{Gauge/Gravity Correspondence in Bosonic String Theory}
\label{Bosonic}

In this section we study the validity of the gauge/gravity
correspondence in the 26-dimensional bosonic string and in order to
compare it with the supersymmetric case discussed in the previous
section, we consider it in the orbifold $C^{\delta/2}/{\mathbb Z}_2$
with $\delta < 22$. As in the previous section we consider the
one-loop vacuum amplitude of an open string stretching  between a D3
brane dressed with a  background gauge field and a system of $N$
undressed D3 branes. It is given by:
\begin{eqnarray}
Z  = N\int_{0}^{\infty} \frac{d \tau}{\tau} Tr \left[
\left(\frac{e+h}{2} \right) (-1)^{G_{bc}} { e}^{- 2 \pi \tau L_0 }
\right] \equiv & Z_{e}^o + Z_{h}^o~~,\label{frene28}
\end{eqnarray}
where $L_0$ includes the ghost and the matter contribution, the
first one having the same structure as in flat space, while the
matter part being derived in Appendix B of Ref.~\cite{LMP}. By
performing the explicit calculation of the one-loop vacuum amplitude
one gets:
\begin{eqnarray}
Z_e^o  = - \frac{N}{(8 \pi^2 \alpha')^{2}} \int d^4x \sqrt{
-\mbox{det}(\eta+\hat{F})} \int_{0}^\infty
\frac{d\tau}{\tau}e^{-\frac{y^2 \tau}{2\pi\alpha'}} \nonumber \\
 \times\,\, \frac{2 {e}^{\pi
\tau (\nu_{f}^{2} +\nu_{g}^{2})} \sin \pi \nu_f  \sin \pi
\nu_g}{f_1^{18} (e^{-\pi \tau}) \Theta_{1}(i\nu_f\tau|i\tau)
  \Theta_{1}(i\nu_g\tau|i\tau)}
\label{zetaebos}
\end{eqnarray}
and
\begin{eqnarray}
 Z_h^o & = &  - \frac{N}{ (8\pi^2\alpha^`)^2} \int d^4x \sqrt{
-\mbox{det}(\eta+ {\hat{F}})} \int_{0}^\infty \frac{d\tau}{\tau}
e^{-\frac{y^{2} \tau}{2\pi\alpha'}} \left[ \frac{ 2 {e}^{\pi \tau
(\nu_{f}^{2} +\nu_{g}^{2})} \sin\pi\nu_f   \sin\pi\nu_g}{
\Theta_{1}(i\tau\nu_f|i\tau) \Theta_{1}(i\tau\nu_g|i\tau)} \right]
\nonumber \\
 && \hspace*{3cm}{\times}\,\, 2^{\frac{\delta}{2}}[f_{1} (k) ]^{- (18 - \delta) }
[f_{2}
(k) ]^{- \delta  },  \label{op56}
\end{eqnarray}
where the power $18$ is obtained from $d -8 $ for the value of the
critical dimension $d=26$. The calculation of the untwisted sector
was originally performed in Ref.~\cite{BAPO} for the case of D9
branes.

The previous expressions can also be rewritten in the closed string
channel and one gets:
\begin{eqnarray}
\!\!Z_e^c =\!  \frac{N\, }{(8\pi^2\alpha')^2}\!\! \int\!\! d^4\!x \sqrt{
-\mbox{det}(\eta+\hat{F})}\! \int_{0}^\infty\!\!
\frac{dt}{t^{11}}e^{-\frac{y^{2} }{2\pi\alpha't}} \frac{2
\sin\pi\nu_f \sin\pi\nu_g}{f_1^{18}(e^{-\pi t})
\Theta_{1}(\nu_f|it)\Theta_{1}(\nu_g|it)} \label{op2bos}
\end{eqnarray}
for the untwisted sector and
\begin{eqnarray}
Z_h^c&  = &   \frac{N }{ (8\pi^2\alpha^`)^2} \int d^4x \sqrt{
-\mbox{det}(\eta+ {\hat{F}})} \int_{0}^\infty \frac{d t}{t^{11 -
\delta/2} } e^{-\frac{y^{2} }{2\pi\alpha' t}} \left[ \frac{ 2
\sin\pi\nu_f   \sin\pi\nu_g}{ \Theta_{1}( \nu_f|i t) \Theta_{1}(
\nu_g|it)} \right]
\nonumber \\
 && \hspace*{3cm}{\times} 2^{\delta/2} \,
[f_{1} (q) ]^{- (18 - \delta) }  [f_{4} (q) ]^{- \delta  }
\label{cl56}
\end{eqnarray}
for the twisted sector. They can be shown to be equal to Eq.s
(\ref{zetaebos}) and (\ref{op56}) respectively by using Eq.s
(\ref{modtras}) and (\ref{mtf1}) for $\Theta_1$, $f_1$ and $f_2$.

Eq. (\ref{th1ex}) and Eq.s (\ref{exp77}) allow one to
extract easily from Eq. (\ref{frene28}) the coefficient of the
kinetic term for the gauge field that turns out to be:
\begin{eqnarray}
\frac{1}{g^{2}_{YM}} =& - \frac{N }{2 (4\pi)^2} \int_{0}^{\infty}
\frac{d  \tau}{\tau} e^{-\frac{y^2 \tau} {(2 \pi \alpha')}}
\left[ \frac{1}{3 \tau^2}  + \frac{1}{\pi \tau} - 2 k \frac{d}{dk}
\log
  f_{1} (k) \right] f_{1}^{-24} (k) \nonumber\\
{}&\times \left[ 1+ 2^{\frac{\delta}{2}}
  \left(\frac{f_{1}(k)}{f_{2} (k)}\right)^{\delta} \right].
\label{op34}
\end{eqnarray}
The previous expression gives the running coupling constant
including all the threshold corrections coming from the massive
open string states. Eq. (\ref{op34}) can be written more
explicitly in the following form:
\[
\frac{1}{g^{2}_{YM}}= - \frac{N }{2 (4\pi)^2} \int_{0}^{\infty}
\frac{d  \tau}{\tau} {e}^{-y^2 \tau /(2 \pi \alpha')} \left[
k^{-2} \prod_{n=1}^{\infty} (1 - k^{2n})^{-24} \right] {\times}
\]
\begin{eqnarray}
{\times} \left[ \frac{1}{3 \tau^2}  + \frac{1}{\pi \tau} -
\frac{1}{6} + 4
  \sum_{n=1}^{\infty} \frac{k^{2n}}{(1 -
    k^{2n})^2}\right] \left[1+ \prod_{n=1}^{\infty} \left(\frac{1 -
    k^{2n}}{1+ k^{2n}} \right)^{\delta} \right],
\label{op35}
\end{eqnarray}
where we have used Eq. (\ref{rel82}). Notice that, differently from
the supersymmetric case, the threshold corrections to the gauge
kinetic term do not vanish. Indeed,  from Eq. (\ref{op35}) one can
see that both the massless and massive bosonic open string states
contribute to the gauge coupling constant. The $\beta$-function of
the gauge theory living on the stack of branes can be computed by
selecting only the contribution of the massless states. This can be
done by performing the field theory limit ($ \alpha' \tau \equiv
\sigma$ fixed with $\alpha' \rightarrow 0$ and $\tau \rightarrow
\infty$). In this way from Eq. (\ref{op35}) one gets:
\[
 \frac{1}{g^{2}_{YM}}  = - \frac{N }{2 (4\pi)^2}
\int_{1/(\alpha' \Lambda^2)}^{\infty} \frac{d  \tau}{\tau}
{e}^{-y^2 \tau /(2 \pi \alpha')} \left[ \frac{1+24 k^2}{k^2} ( -
\frac{1}{6} + 4 k^2 ) ( 2
  -2 \delta k^2)
+ \dots \right]
\]
\begin{eqnarray}
\simeq - \frac{N }{16 \pi^2} \frac{\delta}{6} \int_{1/(\alpha'
\Lambda^2)}^{\infty} \frac{d \tau}{\tau} {e}^{-\frac{y^2\tau}{ 2 \pi
\alpha'}} = - \frac{N}{(4 \pi)^2} \frac{\delta}{6} \log \frac{2 \pi
(\alpha' \Lambda)^2}{y^2}~~,\label{beta34}
\end{eqnarray}
where in the last step we have neglected the term  proportional to
$k^{-2}$ corresponding to the tachyon and used Eq. (\ref{comple67}).
Adding finally the contribution from the tree diagrams yields:
\begin{eqnarray}
\frac{1}{g^{2}_{YM}} = \frac{1}{g^{2}_{YM} (\Lambda)} + \frac{N }{16
\pi^2} \frac{\delta}{6} \log \frac{y^2 }{\epsilon^2}~~;~~ \epsilon^2
\equiv 2 \pi (\alpha' \Lambda)^2 ~~.\label{run64}
\end{eqnarray}
Notice that the contribution of massless states in Eq.
(\ref{beta34}) coming from the untwisted sector vanishes, in
agreement with the fact that it corresponds to the one-loop
$\beta$-function coefficient of a gauge theory with one gluon and 22
scalars~\footnote{See Ref.~\cite{MT} and Appendix B of
Ref.~\cite{FILIM}.}. One gets instead a non-zero contribution from
the twisted sector that reproduces the right
 one-loop $\beta$-function for a gauge theory with one vector field
and $N_s = 22 - \delta$, (i.e.  the number of directions orthogonal
both to the D$3$ brane and to the orbifold) scalars :
\begin{eqnarray}
\beta ( g_{YM} ) = \frac{g_{YM}^{3}}{(4 \pi)^2} \left[ -
\frac{11}{3} + \frac{N_s}{6} \right] = - \frac{\delta}{6}
\frac{g_{YM}^{3}}{(4 \pi)^2} ~~.\label{beta62}
\end{eqnarray}
Notice that in the bosonic case there are no terms proportional to
the topological charge coming from Eq.s (\ref{op2bos}) and
(\ref{cl56}), consistently with the fact that the world-volume
theory is a purely bosonic gauge theory and therefore not affected
by chiral anomaly.

One can also extract the coefficient of the gauge kinetic term from
the amplitude written in the closed string channel in Eq.s
(\ref{op2bos}) and (\ref{cl56}). Eq.s (\ref{exp77}) and
(\ref{exf2}) must be used, obtaining:
\begin{eqnarray}
\frac{1}{g^{2}_{YM}} = &- \frac{N }{2 (4\pi)^2} \int_{0}^{\infty}
\frac{dt}{t^{11}} e^{-\frac{y^2}{(2 \pi \alpha't)}} \left[
\frac{1}{3} + 2 q \frac{d}{dq} \log f_{1} (q)
\right] f_{1}^{-24} (q)\nonumber\\
&\times \left[ 1+ (2t)^{\delta /2}
  \left(\frac{f_{1}(q)}{f_{4} (q)}\right)^{\delta} \right].
\label{cl34}
\end{eqnarray}
Using the modular transformations of the various functions $f_i$ and
the relation:
\begin{eqnarray}
\tau^2 k \frac{d}{dk} =- q \frac{d}{dq} \label{rel98}
\end{eqnarray}
that implies
\begin{eqnarray}
-2 k \frac{d}{dk} \log f_{1} (k) + \frac{1}{\pi \tau} = 2 t^2
q\frac{d}{dq}\log f_{1} (q)~~, \label{rel43}
\end{eqnarray}
it is easy to see that Eq.s (\ref{op34}) and (\ref{cl34}) transform
into each other. It is convenient to write Eq. (\ref{cl34}) in the
following  more explicit way:
\[
\frac{1}{g^{2}_{YM}} = - \frac{N }{2 (4\pi)^2} \int_{0}^{\infty}
\frac{dt}{t^{11}} e^{-\frac{y^2}{(2 \pi \alpha't)}} q^{-2}
\prod_{n=1}^{\infty} (1 - q^{2n})^{-24}
\]
\begin{eqnarray}
{\times} \left[\frac{1}{2} - \sum_{n=1}^{\infty} \frac{4 n q^{2n}}{1
- q^{2n}}     \right] \left[ 1 + (2t)^{\delta/2} q^{\delta/8}
\prod_{n=1}^{\infty} \left( \frac{1 - q^{2n}}{1 -
      q^{2n-1} }
  \right)^{\delta} \right],
\label{cl34bis}
\end{eqnarray}
where we have used Eq. (\ref{rel82}). As in the open channel, we
find that there are threshold corrections to the running coupling
constant due to massive closed string states.

In order to understand the role of massless closed string states,
let us first examine the mass spectrum of closed strings, which is
given by:
\begin{eqnarray}
\frac{\alpha'}{2} M^2 = N + {\tilde{N}} - 2 + \frac{\delta}{8}~~,~~
N = {\tilde{N}}~~,\label{spe45}
\end{eqnarray}
where
\begin{equation}
N = \sum_{n=1}^{\infty} \left[ n a^{\dagger}_{n} {\cdot} a_{n} + ( n
  -\frac{1}{2} ) a^{\dagger}_{n - 1/2} {\cdot} a_{n- 1/2} \right],
\label{enne21}
\end{equation}
with the analogous expression for $\tilde N$ and  $\delta =0$
$(\delta \neq 0)$ in the untwisted (twisted) sector.
The intercept in Eq. (\ref{spe45}) comes from the zero-point energy:
\begin{eqnarray}
2 \frac{24  - \delta}{2} \sum_{n=1}^{\infty} n + 2 \frac{\delta}{2}
  \sum_{n=1}^{\infty} \frac{2n-1}{2} = - \frac{24 - \delta}{12} +
  \frac{\delta}{24} = -2 + \frac{\delta}{8}~~.
\label{zpe67}
\end{eqnarray}
The zero-point energy derived in the previous equation matches with
the power of $q$ appearing in Eq. (\ref{cl34bis}) both for the
twisted and the untwisted sector.

By performing the field theory limit ($t \rightarrow \infty, \alpha'
\rightarrow 0$ and $ \alpha' t$ fixed) and neglecting the divergent
contribution due to the closed string tachyon, one gets a vanishing
result in the untwisted sector. In the twisted sector one has again
a tachyon in the spectrum, if $N_s > 6$, that becomes massless if
$N_s =6$ and massive if $N_s < 6$.

Even in the case $N_s =6$ ($\delta =16$) in which the spectrum
admits massless states, by taking the field theory limit one gets
again a vanishing result, not reproducing Eq. (\ref{run64}) from the
closed channel.
 {From} the previous analysis it follows that the gauge/gravity
correspondence does not hold in the present case.

\section{Gauge/Gravity Correspondence in Type 0B String Theory}
\label{0B}

In this section we summarize the properties of type 0 string
theories~\cite{ASrep} and explore the gauge/gravity correspondence in this
framework. In particular we discuss the spectrum and D branes of
type 0B theory and compute the annulus diagram. Finally, we consider
the one-loop vacuum amplitude of an open string stretching  between
a stack of D branes and a brane having an external field on its
world-volume, and we study under which conditions the gauge/gravity
correspondence holds.

Type 0 string theories are non-supersymmetric closed string models
obtained by applying the following non-chiral diagonal projections
on the Neveu-Scharz-Ramond model:
\begin{eqnarray}
P_{\rm NS-NS}=\frac{1+(-1)^{F+\tilde{F} +G_{\beta \gamma}
    +{\tilde{G}}_{\beta \gamma} }}{2}\qquad
P_{\rm R-R}=\frac{1\pm(-1)^{F+\tilde{F} +G_{\beta \gamma}
    +{\tilde{G}}_{\beta \gamma} }}{2},
\label{gso}
\end{eqnarray}
where the upper [lower] sign in $P_{\rm R-R}$ corresponds to 0B [0A].
$G_{\beta \gamma}$ is defined in Eq.s (\ref{sghonu}), $F$ is
the world-sheet fermion number defined in Eq.s (\ref{fns78}) and
(\ref{fr79}) with analogous definitions for $\tilde{F}$ and
$(-1)^{\tilde{F}}$.
In addition it is imposed that the fermionic NS-R and R-NS sectors
are eliminated from the physical spectrum, obtaining a purely
bosonic string model.

\subsection{Closed string spectrum}

The closed string spectrum can be determined by keeping only the
string states that are left invariant by the action of the operators
given in Eq. (\ref{gso}). It results to be:
\begin{eqnarray}
&&\mbox{type 0A}\!\!\qquad ( {\rm NS}\,- \,,\,{\rm NS}\,-) \otimes (
{\rm NS}\,+\,,\,{\rm NS}\,+)\otimes ( {\rm R}\,-\,,\,{\rm R}\,+)
\otimes( {\rm R}\,+\,,\,{\rm R}\,-)\label{70as}\\
&& \mbox{type 0B}\!\!\qquad ( {\rm NS}\,-\,,\,{\rm NS}\,-) \otimes (
{\rm NS}\,+\,,\,{\rm NS}\,+)\otimes ( {\rm R}\,-\,,\,{\rm R}\,-)
\otimes( {\rm R}\,+\,,\,{\rm R}\,+)\label{70bs}
\end{eqnarray}
where the signs in the various sectors refer to the values
respectively taken by $(-1)^{F}$ and $(-1)^{\tilde{F}}$. In the (NS
$-$, NS $-$) sector the lowest state is a tachyon, while the
massless states live in the (NS +, NS +) sector. In the picture
$(-1,-1)$ they are described by:
\begin{equation}
\psi^{\mu}_{-\frac{1}{2}} {\tilde{\psi}}^{\nu}_{-\frac{1}{2}}
\hspace{.2cm} | 0,\, \tilde{0},\,k \rangle_{(-1,-1)} \label{nsmsl9}
\end{equation}
and  are the same as in type II theories, namely a graviton, a
dilaton and a Kalb-Ramond field. In the R-R sector, instead, we have
the following massless states in the picture $(-\frac{1}{2},
-\frac{1}{2})$:
\begin{equation}
u_A (k) {\tilde{u}}_{B} (k) |A  \rangle_{-\frac{1}{2}} |{\widetilde
{B }} \rangle_{-\frac{1}{2}} ~~.\label{rrsta7}
\end{equation}
Since the terms containing $G$ and ${\tilde{G}}$ in the second
equation in (\ref{gso}) act as the identity on the previous state,
the projector $P_{\rm R-R}$ imposes the existence of two kinds of
R-R $(p+1)$-potentials for any value of $p$ ($C_{p+1}$ and
${\bar{C}}_{p+1}$) characterized respectively by:
\begin{equation}
u_A \left(\frac{1+ \Gamma^{11}}{2} \right)^{A}_{\,\,B} =0
\hspace{.5cm}, \hspace{.5cm} {\tilde{u}}_A \left(\frac{1 \pm
\Gamma^{11}}{2} \right)^{A}_{\,\,B} =0 \label{pro74}
\end{equation}
and by
\begin{equation}
u_A \left(\frac{1- \Gamma^{11}}{2} \right)^{A}_{\,\,B} =0
\hspace{.5cm}, \hspace{.5cm} {\tilde{u}}_A \left(\frac{1 \mp
\Gamma^{11}}{2} \right)^{A}_{\,\,B} =0~~,\label{pro75}
\end{equation}
where the upper [lower] sign corresponds to 0B [0A]. The doubling of
the R-R potentials implies the existence of two kinds of  branes
that are charged with respect to both potentials. We follow the
convention of denoting by $p$ and $p'$ respectively branes having
equal or opposite charges with respect to the two $(p+1)$ R-R
potentials.
The $p$ and $p'$-branes
are called respectively electric and magnetic branes .\cite{KT9811}
In the case of a D3 brane the two potentials $C_4$ and ${\bar{C}}_4$
have field strengths $F_5 $ and ${\bar{F}}_5$ being the former
self-dual, $F_5 = {}^{*} F_5$, and the latter antiself-dual,
${\bar{F}}_5 = - {}^* {\bar{F}}_5$,  as follows from Eq.s
(\ref{pro74}) and (\ref{pro75}). This means that, if we
take the linear combinations:
\begin{equation}
\label{elmag} (C_4)^\pm=\frac{1}{{\sqrt 2}}(C_4\pm\bar C_4)~~,
\end{equation}
one can see that the Hodge duality transforms each field strength
into the other according to the relation
$\,^*F_{5}^{\pm}=F_{5}^{\mp}$. Therefore, while the D3 brane of type
IIB is naturally dyonic, in type 0B the dyonic D3 brane is
constructed as a superposition of an equal number of electric and
magnetic D3 branes.\cite{KT9901,M9811,TZ9902}

Type 0 string theory can also be thought as the orbifold type
IIB$/(-1)^{F_s}$, \cite{all} where  $F_s$ is the space-time fermion
number operator. {From} this point of view, the spectrum of the
physical closed string states, written in Eq.s (\ref{70as}) and
(\ref{70bs}), is made of the untwisted and twisted sectors of this
orbifold. Of course the untwisted spectrum coincides with the
bosonic states of type II theories. The twisted sector can be more
easily determined using the Green-Schwarz formalism, rather than the
NS-R one, due to the simple action of $(-1)^{F_{s}}$ on the
space-time fermionic coordinates $S^{Aa}$. Here $A=1,2$ and $a$ labels
the two spinor representations of the light-cone Lorentz group
$SO(8)$, namely it is either an $\bf 8_{s}$ or $\bf 8_{c}$ index. In
the twisted sector the boundary conditions on these coordinates are
antiperiodic rather than periodic. Hence, the Fourier expansion for
them  contains half-integer fermionic modes. The lowest level
corresponds to a tachyon while the first one, corresponding to the
massless states, is given by:
\begin{eqnarray}
&& S^{a}_{-\frac{1}{2}} \tilde{S}^{b}_{-\frac{1}{2}} |0 \rangle |
\otimes
 |\tilde{0} \rangle
\qquad \mbox{in type 0B}\\
&& S^{a}_{-\frac{1}{2}} \tilde{S}^{\dot{b}}_{-\frac{1}{2}} |0
\rangle \otimes
 | \tilde{0} \rangle
\qquad \mbox{in type 0A} .
\end{eqnarray}
These states provide the doubling of the R-R forms previously
discussed.

\subsection{Boundary state}

The presence of R-R potentials implies that type 0 theories contain
D-branes that, as in other string theories, admit a microscopic
description in the closed string channel in terms of boundary
states. Obviously, the boundary state describing a 0B (or 0A) brane
must be invariant under the GSO projectors defined in Eq.
(\ref{gso}).

In type II theories the boundary state is constructed in terms of
the state $|B,\,\eta\rangle$ with $\eta=\pm 1$ by imposing the
standard GSO projection  which selects the following invariant
combinations:
\begin{eqnarray}
\label{bs22ab} |B\rangle = {1\over 2} \Big( |B,+\rangle_{\rm NS-NS}
- |B,-\rangle_{\rm NS-NS} +|B,+\rangle_{\rm R-R} + |B,-\rangle_{\rm
R-R} \Big)~~.
\end{eqnarray}
In type 0 theories it is simple to verify that the boundary state
$|B,\,\eta\rangle$ that one uses, independently on the spin
structures, is already invariant under the GSO operators
(\ref{gso}).
This means that in type 0 string we have, for each $p$, four
different kinds of boundary states:
\begin{eqnarray}
|Bp,\,\eta,\eta'\rangle=|Bp,\,\eta\rangle_{\rm
NS-NS}+|Bp,\,\eta'\rangle_{R-R}~~. \label{binv1}
\end{eqnarray}
However the requirement of consistency of the various cylinder
amplitudes, giving in the closed channel the interaction between
electric and magnetic branes, with the corresponding amplitudes
computed in the open string channel, imposes the following
combinations:\cite{G0005,BCR9902}
\begin{eqnarray}
|Bp,\, \pm\rangle= \pm|Bp,\,\pm\rangle_{\rm
NS-NS}+|Bp,\,\pm\rangle_{R-R}~~,\label{btipoo}
\end{eqnarray}
which are indeed the boundary state descriptions of the $p$ and $p'$
branes (respectively for  $\eta=+,-$) already introduced, as one can
check by evaluating the coupling of the previous boundary states
with the two forms $C_{p+1}$ and $\bar{C}_{p+1}$.

Notice that, because of the ${\pm}$ sign in front of the NS-NS term
in Eq. (\ref{btipoo}), electric and magnetic branes have opposite
couplings with the tachyon, which implies that dyonic branes do not
couple to it.

The normalization coefficient of the boundary state (related to the
tension of the D$p$ brane~\footnote{See
  Ref.~\cite{DL99121} for details.}) in type 0 string is:\cite{KT9811}
\begin{eqnarray}
T_p=\frac{T_p^{II} }{\sqrt{2}}~~.\label{tens}
\end{eqnarray}
An easy way to see this is the following: in the open channel the
interaction between two D$p$ branes of the same type coincides with
the corresponding expression in the NS-sector of type II theories.
In the closed string channel, instead, in the expression for the
type~0 a factor 1/2 is missing  with respect to the one in type II,
due to the different GSO projection, and therefore the tension of
the branes must be $1/\sqrt{2}$ smaller.

\subsection{Open string spectrum}

The existence of two different kinds of branes in type 0 theories
(the $p$-brane and the $p'$-brane) implies the presence of four
distinct kinds of open strings: those stretching between two $p$ or
two $p'$-branes (denoted by $pp$ and $p'p'$) and those of mixed type
($pp'$ and $p'p$).  This means that the most general Chan-Paton
factor $\lambda$ in the expression of the open string states has the
following form:
\begin{eqnarray}
\lambda\equiv\left( \begin{array}{ll}
                     pp & pp'\\
                     p'p &p'p'
                     \end{array}
                     \right) .
\label{cpf}
\end{eqnarray}
Open/closed string duality makes the following spin structure
correspondence to hold:\cite{KT9811}
\begin{eqnarray}
\begin{array}{lll}
       {\rm Interactions} &
{\rm \,\,\,\,\,\,\,\,Closed\,\,states}  & {\,\,\,\,\,\,\,\,\rm Open\,\,states}\\
       pp\,\,\,\,p'p'& \,\,\,\,\,\,\,\,{\rm NS-NS}\,\,\,\,\,\,
&\,\,\,\,\,\,\,\,{\rm NS}\\
pp\,\,\,\,p'p'&\,\,\,\,\,\,\,\,{\rm R}\,-\,{\rm R} &\,\,\,\,\,\,\,\,{\rm NS}(-1)^F\\
pp'\,\,\,\,p'p &\,\,\,\,\,\,\,\,{\rm NS-NS}(-1)^F\,\,\,\,\,\,&\,\,\,\,\,\,\,\,{\rm R}\\
pp'\,\,\,\,p'p &\,\,\,\,\,\,\,\,{\rm R-R}(-1)^F\,\,\,\,\,\,&\,\,\,\,\,\,\,\,{\rm R}(-1)^F
\end{array}
\label{schema}
\end{eqnarray}
{From} this scheme, it is  easy to see that the spectrum of $pp$ and
$p'p'$ strings contains only the NS and NS$(-1)^F$ sectors
\cite{KT9811} whose massless excitations are the bosons of the gauge
theory. In the case of $D3$ branes one has:
\begin{eqnarray}
&& A^{\alpha} \equiv \left(\begin{array}{ll}
                     pp & 0\\
                      0&p'p'
                     \end{array}
                     \right)\,\otimes \psi^{\alpha}_{-1/2}|0\rangle_{-1} \qquad
                     \,\,\,\alpha=0, \dots, 3 \label{boson} \\
&& \phi^{i} \equiv \left(\begin{array}{ll}
                     pp & 0\\
                      0&p'p'
                     \end{array}
                     \right)\,\otimes \psi^{i }_{-1/2}|0\rangle_{-1} \qquad
                     \,\,\,i=4, \dots, 9~~.
\label{boson1}
\end{eqnarray}
Here $A^{\alpha}$ corresponds to the gauge field, while the
$\phi^{i}$'s represent six adjoint scalars. On the other hand $p p'$
strings have only the R spectrum \cite{KT9811} which provides
fermions to the gauge theory supported by the branes. The lowest
excitations of these strings are:
\begin{eqnarray}
\psi^{A} \equiv \left(\begin{array}{ll}
                     0 & pp'\\
                      p'p&0
                     \end{array}
                     \right)\, \otimes |A\rangle_{-\frac{1}{2}} ~~.
\label{ferm}
\end{eqnarray}
being $|A\rangle_{-\frac{1}{2}}$ a Majorana-Weyl spinor of the
ten-dimensional Lorentz group.

Finally, on a dyonic brane there are both fermionic and bosonic
degrees of freedom. Indeed a stack of $N$ dyonic D-branes of type 0
contains $N$ $p$-branes and $N$ $p'$-branes, and therefore the
massless open string states living on their world-volume are the
subset of the massless open string states living on the world-volume
of $2N$ D$p$ branes of type II theories, that are invariant under
the action of the operator $(-1)^{F_s}$, where we are looking at
type 0B as the orbifold IIB$/(-1)^{F_s}$.
In order to select the invariant open string states, one has to pay
attention to the action of the space-time fermion number operator on
the Chan-Paton factors.\cite{TZ9902,BFL9906} Indeed consistency
requirements between the two approaches to type 0B theory impose the
following non trivial action of $(-1)^{F_s}$ on the Chan-Paton
factors:
\begin{eqnarray}
(-1)^{F_s} \lambda_{ij} \equiv \left( \gamma_{(-1)^{F_s}}
\right)_{ih} \lambda_{hk} \left(
\gamma_{(-1)^{F_s}}^{-1}\right)_{kj}~~,\label{pfscp}
\end{eqnarray}
where~\cite{BFL9906}
\begin{eqnarray}
\gamma_{(-1)^{F_s}}=\left(
\begin{array}{cc}
\mathbb{ I }_{N{\times} N} & 0 \\
0 & - \mathbb{ I }_{M{\times} M}
\end{array} \right)
\label{gpfs}
\end{eqnarray}
and $N$, $M$  denote the number of $p$ and  $p'$-branes
respectively. The requirement of invariance of the physical states
under the action of $ (-1)^{F_s}$ imposes the following constraints on the Chan-Paton factors:
\begin{eqnarray}
\lambda^{\rm (NS)}= \gamma_{(-1)^{F_s}} \lambda^{\rm (NS) }
\gamma_{(-1)^{F_s}}^{-1} \qquad \lambda^{\rm (R)}=
-\gamma_{(-1)^{F_s}} \lambda^{\rm (R) } \gamma_{(-1)^{F_s}}^{-1},
\label{ccp}
\end{eqnarray}
where the minus sign is due to the action of $F_s$ on the space-time
fermion $|A \rangle$. It is easy to see that the previous equations
are satisfied  by the matrices given in Eqs. (\ref{boson}),
(\ref{boson1}) and (\ref{ferm}).

Therefore the spectrum of the open strings attached on a dyonic
brane
 can be easily derived by writing, in the NS and R sectors, the
massless states:
\begin{eqnarray}
&&A_{\alpha}\equiv \left( \begin{array}{cc}
                         A_{NN} & 0 \\
                          0     & B_{NN}
                         \end{array}
                  \right) \psi_{-1/2}^\alpha |0,k\rangle \qquad
\alpha=0,\dots, 3 \label{vecnn}\\
&&\phi^{i}\equiv \left( \begin{array}{cc}
                         A_{NN} & 0 \\
                          0     & B_{NN}
                         \end{array}
                  \right) \psi_{-1/2}^i |0,k\rangle \qquad
i=4\dots 9 \label{scalnn} \\
&&\Psi^{i}\equiv \left( \begin{array}{cc}
                           0 &   A_{NN} \\
                          B_{NN} & 0
                         \end{array}
                  \right) |s_1\, s_2\,s_3\, s_4\rangle \qquad
\sum_{i=1}^4s_i={\mbox odd}~~,\label{fernn}
\end{eqnarray}
where the last relation between the $s_i$ follows from the GSO
projection  in Eq.s (\ref{GSO}) and (\ref{fr79}).

Thus the world-volume of a dyonic D3 brane configuration
supports a $U(N){\times} U(N)$ gauge theory with six adjoint scalars
for each gauge factor and four Weyl fermions in the bifundamental
representation of the gauge group $(N\,,\,\bar{N})$ and
$(\bar{N}\,,\,N)$ (see Table 1).

The number of bosonic degrees of freedom of the open strings
attached on the $N$ dyonic branes is $8N^2 {\times}2$, and coincides
with the number of the fermionic ones. Therefore, the gauge theory
supported by these bound states, even if non-supersymmetric,
exhibits a Bose-Fermi degeneracy. Its $\beta$-function is zero at
one-loop level, and it is argued that this non-supersymmetric theory
is conformal in the large $N$ limit.\cite{KT9901,NS9902}

\subsection{{One-loop vacuum amplitude}}
In this section we compute the interaction between two branes in
type 0B theory, first in the absence of an external field and then
turning on an $SU(N)$ gauge field on one of them. As usual, this
interaction can be computed either in the open string channel or in
the closed string one.

In the open channel the interaction between two branes of the same
kind is:\cite{KT9811}
\begin{eqnarray}
Z^{o}_{pp}&=&2\int \frac{d \tau}{2 \tau} \mbox{Tr}_{\rm NS}
\left[e^{-2\pi \tau L_0}
 (-1)^{G_{bc}} P_{GSO}
\right]\nonumber \\
&=& V_{p+1}(8\pi^2\alpha')^{-\frac{p+1}{2}}\int
\frac{d\tau}{\tau^{\frac{p+3}{2} }} e^{-\frac{y^2\tau}{2\pi\alpha'}}
\frac{1}{2}\left[\left( \frac{f_3(k)}{f_1(k)}\right)^8-\left(
\frac{f_4(k)}{f_1(k)}\right)^8\right]~~,
 \label{free0pp}
\end{eqnarray}
with $P_{GSO}$ defined in Eq. (\ref{GSO}).
The interaction between a $p$ and  a $p'$-brane is obtained by
computing the trace in the R-sector:\cite{KT9811}
\begin{eqnarray}
Z^{o}_{pp'} &=&2\int \frac{d \tau}{2 \tau} \mbox{Tr}_{\rm R}
\left[e^{-2\pi \tau L_0}
 (-1)^{G_{bc}} P_{GSO}
\right] \nonumber\\
&=&- V_{p+1}(8\pi^2\alpha')^{-\frac{p+1}{2}}\int
\frac{d\tau}{\tau^{\frac{p+3}{2}}} e^{-\frac{y^2\tau}{2\pi\alpha'}}
\frac{1}{2}\left( \frac{f_2(k)}{f_1(k)}\right)^8 ~~.\label{free0pp'}
\end{eqnarray}
Thinking of type 0B as type IIB$/(-1)^{F_s}$, Eq.s (\ref{free0pp})
and (\ref{free0pp'}) can be written in a more compact form by
introducing, in the trace of the free energy, the following
projector:
\begin{eqnarray}
P_{(-1)^{F_s}} =\frac{1+(-1)^{F_s}}{2} \label{pfs}
\end{eqnarray}
that, as we have previously mentioned, eliminates all fermionic
states from the spectrum in the closed channel. The free-energy is
now written as:
\begin{eqnarray}
&&Z^{o}=2\int \frac{d \tau}{2 \tau} \mbox{Tr}_{\rm NS-R}
\left[e^{-2\pi \tau L_0} (-1)^{G_{bc}}\,P_{\rm GSO} P_{(-1)^{F_s}} \right]\nonumber \\
&&= \frac{1}{2} {\rm Tr}\left[\,\mathbb{ I }\,\right]^2
V_{p+1}(8\pi^2\alpha')^{-\frac{p+1}{2}}\int
\frac{d\tau}{\tau^{\frac{p+3}{2}}} e^{-\frac{y^2\tau}{2\pi\alpha'}}
\frac{1}{2}\left[ \left( \frac{f_3(k)}{f_1(k)}\right)^8-\left(
\frac{f_4(k)}{f_1(k)}\right)^8-
\left( \frac{f_2(k)}{f_1(k)}\right)^8\right]\nonumber\\
&& + \frac{1}{2} {\rm Tr}\left[\gamma_{(-1)^{F_s} }\right]^2\!
V_{p+1}(8\pi^2\alpha')^{-\frac{p+1}{2}}\!\int \!
\frac{d\tau}{\tau^{\frac{p+3}{2}}} e^{-\frac{y^2\tau}{2\pi\alpha'}}
\frac{1}{2}\left[ \left( \frac{f_3(k)}{f_1(k)}\right)^8\!\!-\left(
\frac{f_4(k)}{f_1(k)}\right)^8\!\!+
\left( \frac{f_2(k)}{f_1(k)}\right)^8\right]\nonumber\\
&&\equiv Z_{pp}^{o} + 2 Z_{pp'}^{o} + Z_{p'p'}^{o} ~~.\label{cfre}
\end{eqnarray}
Since the traces of the Chan-Paton factors are given by
\begin{eqnarray}
{\rm Tr}\left[\,\mathbb{ I }\,\right]^2=(N+M)^2 \qquad  {\rm Tr}
\left[\gamma_{(-1)^{F_s} }\right]^2=(N-M)^2 ,
\end{eqnarray}
we can rewrite the previous equation as follows:
\[
Z^{o} =  V_{p+1}(8\pi^2\alpha')^{-\frac{p+1}{2}}\!\int \!
\frac{d\tau}{\tau^{\frac{p+3}{2}}} e^{-\frac{y^2\tau}{2\pi\alpha'}}
\]
\begin{eqnarray}
\times \left\{  \frac{N^2 + M^2}{2} \left[   \left(
\frac{f_3(k)}{f_1(k)}\right)^8-\left( \frac{f_4(k)}{f_1(k)}\right)^8
\right] -{MN} \left( \frac{f_2(k)}{f_1(k)}\right)^8 \right\} .
\label{zop89}
\end{eqnarray}
We see that Eq.s (\ref{free0pp}) and (\ref{free0pp'}) are obtained
by putting respectively $M=0$, $N=1$ and $N=M=1$ in Eq.
(\ref{zop89}), while by taking $N=M$ we get the interaction among
$N$ dyonic branes.

Following Ref. \cite{LMP}, whose results are outlined in Sect. 2, we
compute now the one-loop vacuum amplitude of an open string
stretching between a D$3$ brane dressed with an external $SU(N)$
gauge field and a stack of $N$ D$3$ branes, located at a distance
$y$ from the first one.

In the open channel the interaction between two branes {\it of the
same kind} can be read from Eq. (\ref{zetae}) by taking only the
contribution of the spin structures NS and NS $(-1)^F$ and
multiplying it by a factor 2 (there is no orbifold projection in
this case), while the closed channel expression can be obtained from
Eq. (\ref{op2}) by considering only the spin structures NS-NS and
R-R always multiplied by the same factor. {From} Eq. (\ref{zetae}) it
follows that the contributions to the term $F^2$ coming from the NS
and NS$(-1)^{F}$ spin structures exactly cancel the one coming from
the R spin structure. Therefore, instead of considering the sum of
the spin structures NS and NS $(-1)^F$ in Eq. (\ref{zetae}), we can
consider only the spin structure R, which  is equal to the first
two, up to a sign. By using Eq.s (\ref{thex}), (\ref{th1ex}) and
(\ref{exp77})  of Appendix B and inserting them in the last term in
Eq.  (\ref{zetae}), after having changed its sign and multiplied it
by a factor 2, the coefficient of the gauge kinetic term turns out
to be:
\begin{equation}
\frac{1}{g_{YM}^{2}} = - \frac{2N}{16 \pi^2} \int_{0}^{\infty}
  \frac{d\tau}{\tau} e^{-\frac{y^2 \tau}{2\pi\alpha'}} \left(\frac{f_2
  (k)}{ f_{1} (k)}\right)^8
\left [ \frac{1}{12 \tau^2} +  k \frac{d}{dk} \log f_2
  (k) \right] .
\label{opestr88}
\end{equation}
Differently from the type IIB case, in type 0B, there are  threshold
corrections to the running coupling constant. In order to select
only the contribution of the massless states and to compare it with
the gauge theory expectation, we can perform the field theory limit
obtaining:
\begin{eqnarray}
\frac{1}{g_{YM}^{2}}=  -\frac{2N}{16 \pi^2}
  \int_{1/(\alpha' \Lambda^2 )}^{\infty}
  \frac{d\tau}{\tau} e^{-\frac{y^2 \tau}{2\pi\alpha'}} {\times}
  \frac{16}{12} = \frac{N}{(4 \pi)^2} \frac{8}{3} \log
\frac{y^2}{2 \pi (\alpha')^2 \Lambda^2} ~~,\label{ftlim85}
\end{eqnarray}
that agrees with the expected behaviour of the running coupling
constant of a gauge theory with one vector and six adjoint scalars!

In the closed string channel one can consider only the contribution
of the spin structure NS-NS $(-1)^F$ in  Eq. (\ref{op2}) which
exactly cancels the ones of the NS-NS and R-R spin structures. Here,
by making use of the expansions given in Eq.s (\ref{exf1}),
(\ref{exf2}) and (\ref{exp77}), and inserting them in the second
term in  Eq. (\ref{op2}) (after having changed its sign and
multiplied by a factor 2) we get
the following expression for the running coupling constant in the
closed channel:
\begin{equation}
\frac{1}{g_{YM}^{2}} = \frac{2N}{16 \pi^2} \int_{0}^\infty
\frac{dt}{t^3}e^{-\frac{y^{2} }{2\pi\alpha't}} \left( \frac{f_{4}
(q)}{f_{1} (q)} \right)^8 \left[ - \frac{1}{12} + q \frac{d}{dq}
\log f_4 (q) \right]~~. \label{clostr87}
\end{equation}
By using Eq. (\ref{rel98}) and the modular properties of $f_1$ and
$f_2$ (see Eq. (\ref{mtf1})) it can be seen that Eq.
(\ref{clostr87}) reduces to Eq. (\ref{opestr88}). However, by
performing the field theory limit ($\alpha'\rightarrow 0$ and
$t\rightarrow\infty$ with $\alpha't$ finite) in the closed
channel, one selects the  massless  states contribution to the one
loop running of the coupling constat:
\begin{equation}
\frac{1}{g_{YM}^{2}}= -\frac{N}{64 \pi^2} \int_{0}^\infty
\frac{dt}{t^3}e^{-\frac{y^{2} }{2\pi\alpha't}} \left( e^{\pi t}-8
\right)~~. \label{clostr87b}
\end{equation}
In this limit  the closed string tachyon gives a divergent
contribution, while the massless states give a vanishing one. In
other words, the field theory limit of Eq. (\ref{clostr87}) does not
reproduce the correct answer for the running coupling constant,
revealing that the gauge/gravity correspondence does not work in
this case.

In this section we have computed the one-loop vacuum amplitude of
an open string stretching between a stack of $N$ D3 branes and
another brane of the same kind dressed with an $SU(N)$ gauge
field. An important question  is, however, whether the stack of
$N$ D-branes would fly apart. In order to understand this point
one should look more carefully at the interaction between two
identical branes in absence of any background field. Performing
the modular transformation $\tau\rightarrow 1/t$ on  Eq.
(\ref{free0pp}) and extracting the contribution of each closed
string mass level, it turns out that for those levels
corresponding to an even power of $q$ the R-R repulsion is twice
the NS-NS attraction, while for those corresponding to odd powers
of $q$ only the NS-NS states contribute with an attractive term.
Hence, level by level, the contribution to the interaction is
always different from zero. One can conclude that, unless some
geometrical constraint arises to force the system to be stable,
there are problems in piling up $N$ identical branes on top of
each other. The authors of Ref.~\cite{KT9811,K9906} argue that a
stable bound state is formed when the tachyon mass is shifted to
remove the instability. In this review we will not try to make
this argument more quantitative, because little is known about the
tachyon condensation in closed string. However this problem can be
avoided by considering a dyonic configuration. {From} Eq.
(\ref{zop89}) one easily deduces that the zero-force condition
between two stacks made of a superposition of an equal number of
electric and magnetic D3 branes is indeed satisfied. In terms of
open string states, the interaction between two dyonic branes
vanishes because of a cancellation between the contribution of
bosonic and fermionic degrees of freedom, ensuring the stability
of the configuration. On the same footing, the one-loop vacuum
amplitude of an open string stretching between a stack of $N$
dyonic branes and a further dyonic brane dressed with an $SU(N)$
gauge field, is twice the corresponding one in type IIB theory,
and in particular, the coefficient of the gauge kinetic term turns
out to be identically zero, leading to the correct vanishing of
the one-loop beta function, both in the open and closed channel.
Hence for a dyonic brane, the gauge/gravity correspondence holds.

One can also consider orbifolds of type 0B string theory and, by
introducing dyonic fractional branes which live at orbifold
singularities, one breaks the conformal invariance of the model,
obtaining a non-supersymmetric and non-conformal gauge theory. These
are examples of the so-called {\em orbifold field
theories}~\footnote{See
  Ref.~\cite{ASV} and Ref.s therein. See also Ref.~\cite{AK9906}.}, which are
non-supersymmetric gauge theories (daughter theories) that in the
planar limit are perturbatively equivalent to some supersymmetric
gauge theories (parent theories). The gauge groups of the daughter
and parent theories are not the same: in the present case the gauge
theory living on $N$ dyonic branes has an $SU(N){\times}SU(N)$ gauge
symmetry \footnote{The factor $U(1)$ is neglected in the
decomposition $U(N)=SU(N) {\times} U(1)$.}, while the corresponding
parent theory is an $SU(2N)$ SYM, and also the field content is
actually different. However there exists a common sector in the
spectrum for which scattering amplitudes in the parent and daughter
theory are the same in the planar limit.\cite{KS,BKV9803}
The world-volume gauge theories
living on $N$ dyonic D$3$ branes in flat space and its orbifolds
 $\mathbb{C}^2/\mathbb{Z}_2$ and
 $\mathbb{C}^3/(\mathbb{Z}_2{\times}\mathbb{Z}_2)$ are listed in
Table 1.

\begin{table}[h]
\caption{Spectrum of the $SU(N)$ world-volume gauge theory
living on $N$ dyonic D$3$ branes.}
\vskip 0.5cm
{\begin{tabular}
{@{}c|c c c@{}}
{\,\,\,\,\,\,\,\,\,\,\,\,\,\,\,\,\,\,\,\,
} & { 0B {\it on flat space}} & {0B {\it on} $\mathbb{C}^2/\mathbb{Z}_2$}
  & { 0B {\it on} $\mathbb{C}^3/(\mathbb{Z}_2{\times}\mathbb{Z}_2)$}
\\
\hline
     {\it Gauge vector}   & {(Adj,1)+(1,Adj) }               &
{(Adj,1)+(1,Adj) } & {(Adj,1)+(1,Adj)}
\\ \hline
{ \it Scalars}     &{6[(Adj,1)+(1,Adj)] }     & {2[(Adj,1)+(1,Adj)]
}  &{--}
\\
\hline {\it Weyl fermions} & {$4[(\funda\,,\bfunda\,)+(\bfunda\,,\funda\,)]$
}& { $2[(\funda\,, \bfunda\,)+(\bfunda\,, \funda\,)]$  }
&{$(\funda\,,\bfunda\,)+(\bfunda\,,\funda\,)$}
\\ \hline
{\it 1-loop $\beta$-function}  & {$0$ }&{
$-\frac{N}{8\pi^2}g^3_{YM}$ }&{  $-\frac{3N}{16\pi^2}g^3_{YM}$}
\\\\ &{(${\cal N}=4$ SYM) }&{ (${\cal N}=2$ SYM)}&{(${\cal N}=1$ SYM) }\\
\hline
\end{tabular}}
\end{table}

Let us first analyse the orbifold $\mathbb{C}^2/\mathbb{Z}_2$. This
orbifold, in the case of fractional branes, acts only on the
oscillator part of the states in Eq.s (\ref{vecnn}), (\ref{scalnn})
and (\ref{fernn}). The action on the bosonic and fermionic
coordinates is given in Eq. (\ref{z2z3}), while that on the R-R
vacuum is given by:
\begin{eqnarray} h:|s_1\, s_2\,s_3\,
s_4\rangle \rightarrow e^{i \pi (s_3+s_4)}|s_1\, s_2\,s_3\,
s_4\rangle~~. \label{orbf}
\end{eqnarray}
The states left invariant are, for each gauge group, one gauge
vector, two scalars in the adjoint,  two Weyl fermions in the
$(N\,,\bar{N})$ representation and two Weyl fermions in the
$(\bar{N}\,,N)$ one. It is simple to check that the $\beta$-function
of the daughter theory relative to each gauge group is, at one-loop
level, the same as the pure ${\cal N}=2$ super Yang-Mills with gauge
group $SU(2N)$, if the gauge coupling of the latter $g_P$ is related
to the one of the former $g_D$ by $g^2_D=2g^2_P$.\cite{ASV}

Analogously one can  analyse the orbifold
$\mathbb{C}^3/(\mathbb{Z}_2{\times} \mathbb{Z}_2)$ acting  on the
coordinates according to Eq. (\ref{erre}) and on the fermions as
follows:
\begin{eqnarray}
&&h_1|s_1\, s_2\,s_3\, s_4\rangle=
e^{i \pi (s_3+s_4)}|s_1\, s_2\,s_3\, s_4\rangle \nonumber\\
&&h_2|s_1\, s_2\,s_3\, s_4\rangle=
e^{i \pi (s_2+s_4)}|s_1\, s_2\,s_3\, s_4\rangle  \nonumber\\
&&h_3|s_1\, s_2\,s_4\, s_4\rangle= -e^{i \pi (s_2+s_3)}|s_1\,
s_2\,s_3\, s_4\rangle~~,\label{orbf22}
\end{eqnarray}
where the sign in front of the last equation is required by the
group properties. The states left invariant are, for each gauge
group, one gauge vector,  one Weyl fermion in $(N\,,\bar{N})$ and
one Weyl fermion in $(\bar{N}\,,N)$. It is simple to check that the
$\beta$-function of each gauge factor of the daughter theory is, at
one-loop, the one of  ${\cal N}=1$ super Yang-Mills with gauge group
$SU(2N)$, with the appropriate rescaling of the running coupling
constant. Notice that for $N=3$ , if one of the two $SU(N)$ factors
is interpreted as a colour index and the other as a flavour index,
this model provides an example of a three-colour/three-flavour QCD
.\cite{ASV}

The interesting aspect of these orbifold theories is that the
expression of the one-loop vacuum amplitude of an open string
stretching between a stack of $N$ dyonic branes and one electric or
magnetic brane dressed with an $SU(N)$ gauge field coincides with
the corresponding one computed in type IIB and studied in
Ref.~\cite{LMP}. This follows from the fact that, because of Eq.s
(\ref{btipoo}) and (\ref{tens}), the boundary state of $N$ dyonic branes is
the same as the one of type IIB and therefore, when we multiply it
by a closed string propagator and sandwich it with the boundary
state of an electric or magnetic brane, we get exactly the same
result as in type IIB theory.
Therefore, all the features discussed in Sect. 3 are also shared by
these non-supersymmetric theories. In particular the gauge/theory
parameters do not admit threshold corrections either in the open or
in the closed channel and the gauge/gravity correspondence perfectly
holds.

\section{Gauge/Gravity  Correspondence in Type $0'$ Theories}
\label{0'}

In this section we are going to study the gauge/gravity
correspondence for type $0'$ theories.\cite{SAGNOTTI} These are
unoriented, non-supersymmetric string models that can be constructed
as orientifolds of type 0 theories by taking  the quotient 0B
/$\Omega',$ where $\Omega'$ is a suitable world-sheet parity
operator which imposes the existence of a non
trivial background made of an orientifold $9$-plane (O9) and $32$
D$9$ branes necessary to ensure the R-R tadpole cancellation, as we
will discuss in detail later.

\subsection{World-sheet parity definitions}
\label{subsec1} In the literature one can find two different definitions
of the world-sheet parity operator $\Omega$. In particular, the
following definition is given in the closed string sector in Ref.s
\cite{GP9601} and \cite{BG9701}:
\begin{eqnarray}
\tilde{\Omega} \alpha^{\mu}_n \tilde{\Omega}^{-1} & = &
\tilde{\alpha}^{\mu}_n
\nonumber \\
\tilde{\Omega} \psi^{\mu}_r \tilde{\Omega}^{-1} & = &
\tilde{\psi}^{\mu}_r
\nonumber \\
\tilde{\Omega} \tilde{\psi}^{\mu}_r \tilde{\Omega}^{-1} & = & -
\psi^{\mu}_r
\end{eqnarray}
and
\begin{eqnarray}
\tilde{\Omega}\left(|0\rangle_{-1}\otimes |\tilde{0}
\rangle_{-1}\right)& = &
|0 \rangle_{-1} \otimes |\tilde{0} \rangle_{-1} \nonumber \\
\tilde{\Omega} \left(|A \rangle_{-1/2} \otimes | \tilde{B}
\rangle_{-1/2}\right) & = & - | B \rangle_{-1/2} \otimes |
\tilde{A}\rangle_{-1/2}~~, \label{ovacc}
\end{eqnarray}
where the NS-NS and the R-R vacua are taken respectively in the
$(-1,-1)$ and $(-\frac{1}{2},-\frac{1}{2})$ symmetric pictures.
Notice that the operator $\tilde\Omega$ satisfies the condition
${\tilde\Omega}^2=(-1)^{F+\tilde F}$, where $(-1)^{F}$ is  defined
as in Eq.s (\ref{fns78}) and (\ref{fr79}) with the analogous
definition for $(-1)^{\tilde F}$. In the open string sector the
world-sheet parity operator $\tilde{\Omega}$ acts as follows:
\begin{eqnarray}
\tilde{\Omega} \alpha_{m} \tilde{\Omega}^{-1} = \pm e^{i \pi m}
\alpha_m \qquad
  \tilde{\Omega}   \psi_r \tilde{\Omega}^{-1} = \pm e^{i \pi r} \psi_r
\label{omegaos}
\end{eqnarray}
for integer and half-integer $r$ and the signs $\pm$ refer
respectively to NN and DD boundary conditions.

{From} Eq.s (3.11) and (3.12) of Ref. \cite{GP9601} it is possible to
get the action of $\tilde{\Omega}$ on the NS vacuum:
\begin{eqnarray}
\tilde{\Omega}\, |0 \rangle_{-1} = -i \,|0 \rangle_{-1}
\label{omegavac}
\end{eqnarray}
and, in order to get a supersymmetric theory, the following action
must hold on the R vacuum:
\begin{eqnarray}
\tilde{\Omega} |A \rangle_{-1/2} = - \left(\Gamma^{p+1} \dots
\Gamma^{9}\right)^A_{\,\,B}
|B \rangle_{-1/2}~~,\label{ramo23}
\end{eqnarray}
where we have NN boundary conditions along the world-volume of D$p$
brane corresponding to the directions $0, 1, \dots, p$ and  DD
boundary conditions along the remaining ones. In Ref.s \cite{FGLS},
\cite{tesia} and \cite{LR9905} a different definition of the
world-sheet parity operator $\Omega$ is used.

In the closed string channel one has:
 \begin{eqnarray}
\Omega \alpha^{\mu}_n \Omega^{-1} & = & \tilde{\alpha}^{\mu}_n \nonumber \\
\Omega \psi^{\mu}_r \Omega^{-1} & = & \tilde{\psi}^{\mu}_r \nonumber \\
\Omega \tilde{\psi}^{\mu}_r \Omega^{-1} & = &  \psi^{\mu}_r
\end{eqnarray}
with $\Omega^2=1$ and
\begin{eqnarray}
\Omega \left(|0 \rangle_{-1} \otimes |\tilde{0}\rangle_{-1}\right) &
= &
- |0 \rangle_{-1} \otimes |\tilde{0} \rangle_{-1} \nonumber\\
\Omega \left(|A \rangle_{-1/2} \otimes | \tilde{B}
\rangle_{-1/2}\right) & = & - | B \rangle_{-1/2} \otimes | \tilde{A}
\rangle_{-1/2}~~,\label{ovacc1}
\end{eqnarray}
while in the open string channel the definition is like the
corresponding one given by $\tilde{\Omega}$. It is straightforward
to see that for type IIB theory, the two definitions are equivalent
and both of them yield type I theory, i.e.:
\begin{eqnarray}
\mbox{type I} = \mbox{ type IIB}/ \Omega=  \mbox{ type IIB}/
\tilde{\Omega}~~. \label{typeI}
\end{eqnarray}
It is important here to observe that one can define the world-sheet
parity action on the closed string R-R vacuum, given in Eq.s
(\ref{ovacc}) and (\ref{ovacc1}), only in type II/0B theories where
the GSO projection imposes, in the symmetric picture, that the
spinors have the same chirality in both left and right sectors. As a
consequence, Eq.s (\ref{ovacc}) and (\ref{ovacc1}) define the
world-sheet parity action only on the following combination of
states:
\begin{eqnarray}
\tilde{\Omega} \,\left( |\alpha\rangle \otimes |\tilde{\beta}
\rangle\,,\,|\dot{\alpha}\rangle \otimes |\dot{\tilde{\beta}}\rangle
\right)=-\left( |\beta\rangle \otimes |\tilde{\alpha}
\rangle\,,\,|\dot{\beta}\rangle \otimes |\dot{\tilde{\alpha}}
\rangle\right)~~,\label{ova16}
\end{eqnarray}
with an analogous expression for $\Omega$, after having used the
$\Gamma$-matrices in the chiral base, given in Ref.s
\cite{DFPSLR9702} and \cite{DLII}, to decompose the $32$-dimensional
spinor $|A \rangle$ into two $16$-dimensional spinors with opposite
chirality: $ |A \rangle \equiv | \alpha , {\dot{\alpha}} \rangle$,
This remark will be crucial when we analyze the world-sheet parity
action on the boundary state. The boundary state, in the R-R sector,
is naturally written in the asymmetric picture $(-1/2\,,\,-3/2)$, in
which the GSO projected states of the R-R sector are constructed
starting from 16-dimensional spinors having opposite
chirality.\cite{BDFLPRS} The world-sheet parity action on these
spinors, for the reasons explained before, is not well-defined and,
as we will see next, its definition has to be given in a consistent
way.

When used to construct an orientifold of type 0 theory, $\Omega$ and
$\tilde{\Omega}$ do not yield any interesting theory. Indeed the
former gives a tachyon-free theory having the same kinds of branes
as type I (hence there are no D3 branes) while the latter gives a
theory with tachyons. An alternative definition of the world-sheet
parity operator allows to get a tachyon-free orientifold theory of
type 0B having D3 branes in the spectrum. It is given by the
following operator:
\begin{equation}
\Omega' = \tilde{\Omega} (-1)^{\tilde{F}} \,\,\,\,\,\, {\rm
with}\,\,\,\,\,\, \Omega'^2=1~~.\label{ome'}
\end{equation}
The action of $\Omega'$ in the closed string sector is:
 \begin{eqnarray}
\Omega' \alpha^{\mu}_n \Omega^{'-1} & = & \tilde{\alpha}^{\mu}_n \nonumber \\
\Omega' \psi^{\mu}_r \Omega^{'-1} & = & \tilde{\psi}^{\mu}_r \nonumber \\
\Omega' \tilde{\psi}^{\mu}_r \Omega^{'-1} & = &  \psi^{\mu}_r
\label{ome65}
\end{eqnarray}
and
\begin{eqnarray}
\Omega' \left(|0 \rangle_{-1} \otimes |\tilde{0} \rangle_{-1}
\right)& = &
- |0 \rangle_{-1} \otimes |\tilde{0} \rangle_{-1} \label{ome66} \\
\Omega' \left(|A \rangle_{-1/2} \otimes | \tilde{B}
\rangle_{-1/2}\right) & = & \left( \Gamma^{11} \right)^{B}_{\,\,\,D}
| D \rangle_{-1/2} \otimes | \tilde{A} \rangle_{-1/2}~~,
\label{omegavuo}
\end{eqnarray}
where in the last equality we have used the following identity:
\begin{eqnarray}
(-1)^{\tilde{F}} \left(|A \rangle_{-1/2} \otimes | \tilde{B}
\rangle_{-1/2} \right)= - \left( \Gamma^{11} \right)^{B}_{\,\,\,D}|A
\rangle_{-1/2} \otimes | \tilde{D} \rangle_{-1/2}~~.
\end{eqnarray}
In particular, for the same reasons explained soon after Eq.
(\ref{typeI}), Eq. (\ref{omegavuo}) defines the world-sheet parity
action only on the spinors having the same chirality in both
sectors. This means that, introducing the 16-dimensional basis for
the left and right spinors, we have:
\begin{eqnarray}
\Omega' \,\left( |\alpha\rangle \otimes |\tilde{\beta}
\rangle\,,\,|\dot{\alpha}\rangle \otimes |\dot{\tilde{\beta}}\rangle
\right)=\left( |\beta\rangle \otimes |\tilde{\alpha}
\rangle\,,\,-|\dot{\beta}\rangle \otimes |\dot{\tilde{\alpha}}
\rangle\right) \label{o'va16}~~.
\end{eqnarray}
It can be seen from the definition of $\Omega'$ that it acts, in the
open string channel, like both $\Omega$ and $\tilde{\Omega}$ do.

Notice that,  being the action of $\Omega'$, $\Omega$ and
$\tilde{\Omega}$ on physical states identical in type IIB theory,
one can also define  type I as:
\begin{eqnarray}
\mbox{ type I}= \mbox{type IIB}/\Omega' ~~.\label{Io'}
\end{eqnarray}

\subsection{Closed string spectrum}

The spectrum of the type $0'$ theory is obtained from that of the
type $0$B one by selecting the $\Omega'$ invariant states.

Let us start from the closed string spectrum. In the NS-NS sector,
from Eq. (\ref{ome66}) it follows that the closed string tachyon is
projected out, while at the massless level only the graviton and
dilaton fields belong to the spectrum, being the Kalb-Ramond field
projected out. For what concerns the R-R spectrum, the action of
$\Omega '$ on the state in Eq. (\ref{rrsta7}) is obtained by using
Eq. (\ref{omegavuo}). One gets:
\begin{equation}
\Omega ' \left[ (  u_A (k) {\tilde{u}}_{B} (k) |A \rangle
|{\widetilde {B }} \rangle ) \right] = - (\pm)  {\tilde{u}}_A (k)
{{u}}_{B} (k) |A  \rangle |{\widetilde{B }} \rangle~~,\label{ome'8}
\end{equation}
where the upper [lower] sign is valid for the upper sign in
Eq. (\ref{pro74}) [(\ref{pro75})] which corresponds to type 0B theory. By
expanding in terms of a complete system of $\Gamma$-matrices, one
can see that the R-R field is invariant under $\Omega'$ if the
following condition is satisfied:
\begin{equation}
u_A ( \Gamma_{\mu_1 \dots \mu_n } C^{-1} )^{AB} {\tilde{u}}_B = - (
\pm) u_A [( \Gamma_{\mu_1 \dots \mu_n } C^{-1} )^{T}]^{AB}
{\tilde{u}}_B~~,\label{rela529}
\end{equation}
where the upper index $T$ indicates the transposed matrix. But since
one has, for odd $n$:
\begin{equation}
( \Gamma_{\mu_1 \dots \mu_n }C^{-1} )^T = (-1)^{n(n-1)/2} (
\Gamma_{\mu_1 \dots \mu_n }C^{-1} ) ~~,\label{rel87}
\end{equation}
then Eq. (\ref{rela529}) implies
\begin{equation}
 1 = -  (\pm)
(-1)^{n(n-1)/2}~~. \label{rel88}
\end{equation}
If we take the upper sign, the previous relation is satisfied for
$n=3$ corresponding to a R-R potential $C_2$, while if we take the
lower sign it is satisfied for $n=1,5$ corresponding to the R-R
potentials $\bar{C}_0$ and $\bar{C}_4$. This implies that the R-R
massless bosonic spectrum of $0'$ theory is the same as IIB.

Moreover notice that because of Eq. (\ref{omegavuo}),  when acting
on the original ${\rm (R+,R+)}$ sector of type $0$B (where
$\Gamma^{11}=1$) $\Omega '$ leaves $C_2$ invariant and changes the
sign of $C_0$ and $C_4$. On the contrary in the sector ${\rm
(R-,R-)}$ (where $\Gamma^{11} =-1$) $\Omega'$ leaves $\bar C_{0}$
and $\bar C_{4}$ invariant and changes the sign of $\bar C_{2}$. In
particular this implies that the action of $\Omega'$ on the
combination $C_4^{\pm}$ defined in Eq. (\ref{elmag}) is to send
$C_4^\pm\rightarrow C_4^{\mp}$, up to a phase, namely it exchanges
the role of the electric and magnetic branes. This suggests that the
only $\Omega'$-invariant D$p$ brane of type $0'$ would be a
combination of the electric and magnetic branes of type $0$B, as we
are going to discuss in detail in the next subsection.

\subsection{Boundary state}
\label{Boundary state}

Let us now consider the brane content of type $0'$ theory, looking
at the action of $\Omega'$ on the type $0$ boundary states. The
expression of the boundary state is given by:
\begin{eqnarray}
|B\,,\eta\rangle= |B_X \rangle |B_{\psi}\,,\eta \rangle_{\rm
NS-NS,\,R-R} |B_{\rm gh.} \rangle |B_{\rm sgh.}\,,\eta\rangle_{\rm
NS-NS,\,R-R}~~,\label{boundary}
\end{eqnarray}
where the various factors in Eq. (\ref{boundary}) can be found, for
example, in Ref. \cite{DL0307}. In order to determine the
world-sheet parity action on it, we first remind how such an
operator acts on the  ghost and superghost oscillators:
\begin{eqnarray}
&&\!\!\Omega' b_n \Omega'^{-1}=\tilde{b}_n \qquad \Omega'
\tilde{b}_n \Omega'^{-1}=b_n\qquad \Omega' c_n
\Omega'^{-1}=\tilde{c}_n \qquad \Omega' \tilde{c}_n
\Omega'^{-1}=c_n
\label{ogh}\\
&&\!\! \Omega' \beta_t \Omega'^{-1}=\tilde{\beta}_t \qquad \Omega'
   \tilde{\beta}_t \Omega'^{-1}=\beta_t\qquad \Omega' \gamma_t \Omega'^{-1}=
\tilde{\gamma}_t \qquad \Omega' \tilde{\gamma}_t
\Omega'^{-1}=\gamma_t~,\label{osgh}
\end{eqnarray}
while the action on the vacuum states, in the NS-NS sector, can be
determined by observing that $\Omega'$, by definition, exchanges
  left and right sectors, i.e:
\begin{eqnarray}
&&\Omega' \left[|0\rangle_{-1} \otimes |\tilde{0}
\rangle_{-1}\right]= |\tilde{0}\rangle_{-1} \otimes |0
\rangle_{-1} = |0\rangle_{-1} \otimes
|\tilde{0} \rangle_{-1} \label{ovacsg}\\
&& \Omega' \left[|q=1\rangle \otimes |\tilde{q}=1 \rangle\right]=
\tilde{c}_{1} |\tilde{0}\rangle\otimes c_1|0\rangle= |q=1\rangle
\otimes |\tilde{q}=1\rangle~~. \label{ovgh}
\end{eqnarray}
By using the previous transformations for the ghost and superghost
degrees of freedom and Eq.s (\ref{ome65}) and (\ref{ome66}) we
obtain:
\begin{eqnarray}
\Omega' |B_X \rangle = |B_X \rangle && \Omega' |B_{\psi}, \eta
\rangle_{\rm NS-NS}
= - |B_{\psi}, - \eta \rangle_{\rm NS-NS} \nonumber \\
\Omega' |B_{\rm gh.} \rangle = |B_{\rm gh.} \rangle && \Omega'
|B_{\rm sgh.}\, \eta \rangle_{\rm NS-NS} =  |B_{\rm sgh.}\, -\eta
\rangle_{\rm NS-NS}~~. \label{action}
\end{eqnarray}
These actions are compatible with the overlap conditions. For
instance, the fermionic part of the boundary state satisfies the
following overlap condition:
\begin{eqnarray}
\Omega' \left( \psi^\mu_t -i\eta S^\mu_{\,\nu}
\tilde{\psi}_{-t}^\nu\right)|B_\psi,\,\eta \rangle=-i \eta S^\mu_{\,
\nu} \left( \psi^\nu_{-t} +i\eta S^\nu_{\,\rho}
\tilde{\psi}_t^\rho\right) \Omega' |B_\psi,\,\eta \rangle=0~,
\label{ovl}
\end{eqnarray}
where the index $t$ is integer [half-integer] in the R-R [NS-NS]
sector. This condition clearly shows that $\Omega' |B_\psi,\,\eta
\rangle$ satisfies the same overlap as $|B_\psi,\,-\eta \rangle$ and
therefore, apart from an overall factor, they can be identified.

At this point we are able to give the world-sheet parity action on
the whole boundary state in the NS-NS sector:
\begin{eqnarray}
\Omega' |Bp,\eta\rangle_{\rm NS-NS}=-|Bp,-\eta\rangle_{\rm NS-NS}~~.
\label{omegnzm}
\end{eqnarray}
Like all physical states, also the boundary state must be invariant
under the action of $\Omega'$. The invariant boundary state has the
following form:
\begin{eqnarray}
|B p \rangle_{\rm NS-NS} = \frac{1+ \Omega'}{2}|B p, \, +
\rangle_{\rm NS-NS}
 = \frac{1}{2} \left[ |B p, \, + \rangle_{\rm NS-NS} -
|B p, \, - \rangle_{\rm NS-NS} \right]~.\label{Bp}
\end{eqnarray}
By comparing this equation with Eq. (\ref{btipoo}), one can easily
see that, as far as the NS-NS sector is concerned, the boundary
state describing a D$p$ brane in type $0'$ is the sum of the
boundary states of the electric and magnetic branes of type $0$B.
This is consistent also with the comment at the end of the previous
subsection.

Let us now consider the R-R sector where we have only to analyse the
transformation properties of the fermionic and superghost part of
the boundary states under the world-sheet parity. In particular, it
is convenient to separate the zero-modes from the tower of
oscillators (nzm), i.e.:
\begin{eqnarray}
|B_{\psi} , \eta\rangle_{\rm R-R} &=& |B_{\psi} , \eta\rangle_{\rm
R-R}^{(0)} |B_{\psi},  \eta\rangle_{\rm R-R}^{\rm nzm} \nonumber\\
|B_{\rm sgh}, \eta\rangle_{\rm R-R}&=&|B_{\rm sgh}, \eta\rangle_{\rm
R-R}^{(0)} |B_{\rm sgh}, \eta\rangle_{\rm R-R}^{\rm nzm}~~.\label{bfs}
\end{eqnarray}
It is easy to verify that:
\begin{eqnarray}
\Omega'| B_{\psi}, \eta\rangle_{\rm R-R}^{\rm nzm} = |B_{\psi},
-\eta\rangle_{\rm R-R}^{\rm nzm} \qquad \Omega' |B_{\rm sgh},
\eta\rangle_{\rm R-R}^{\rm nzm} = |B_{\rm sgh}, -\eta\rangle_{\rm
R-R}^{\rm nzm}~~. \label{onzmfg}
\end{eqnarray}
What is less trivial  is to determine the action of $\Omega'$ on the
zero modes. Indeed the vacuum in the R-R sector is written in a
picture which is not symmetric under the exchange of the right and
left sectors. Furthermore, the action of $\Omega'$  must be defined
in such a way to be  compatible with the following definitions:
\begin{eqnarray}
\psi_0^\mu |A \rangle \otimes |\tilde{B} \rangle &=&
\frac{1}{\sqrt{2}} \left(\Gamma^\mu \right)^A_{\,\,C}
|C\rangle\otimes|\tilde{B}\rangle \nonumber \\
\tilde{\psi}_0^\mu |A\rangle\otimes|\tilde{B}\rangle &=&
\frac{1}{\sqrt{2}} \left(\Gamma^{11} \right)^A_{\,\,C}
  \left(\Gamma^\mu \right)^B_{\,\,D}
|C\rangle\otimes|\tilde{D}\rangle~~,\label{gammare}
\end{eqnarray}
where $|A \rangle$ and $ |\tilde{B} \rangle$ are 32-dimensional
Majorana spinors.

It turns out that it is not possible to give a well-defined action
of $\Omega'$ on such a kind of spinors that is compatible with Eq.
(\ref{gammare}). It is, however,  possible to overcome this
difficulty by giving the action of $\Omega'$ on the 16-dimensional
chiral representation of the $32$-dimensional spinors. In Eq.
(\ref{o'va16}) this action is already given on the 16-dimensional
spinors having the same chirality in the left and right sectors.
Therefore, we have only to define as the world-sheet parity acts on
spinors having opposite chiralities and it has to be defined in such
a way to be compatible with Eq. (\ref{gammare}). This requirement
leads to:
\begin{eqnarray}
\Omega' \left( \begin{array}{cc}
                |\alpha \rangle \otimes \tilde{|\beta \rangle} &
                |\dot{\alpha} \rangle \otimes \tilde{|\beta \rangle} \\
                |\alpha \rangle \otimes |\dot{\tilde{\beta}} \rangle &
                |\dot{\alpha} \rangle \otimes |\dot{\tilde{\beta}} \rangle
               \end{array} \right) =
        \left( \begin{array}{cc}
                |\beta \rangle \otimes \tilde{|\alpha \rangle} &
                |\beta \rangle \otimes |\dot{\tilde{\alpha}} \rangle \\
                |\dot{\beta} \rangle \otimes \tilde{|\alpha \rangle} &
               -|\dot{\beta} \rangle \otimes |\dot{\tilde{\alpha}} \rangle
                \end{array} \right)~~.
                \label{016}
\end{eqnarray}
Now we have all the ingredients to analyse how the zero-mode part of
the boundary state transforms under $\Omega'$.
We remind that:\cite{DL0307}
\begin{eqnarray}
|B_{\psi}, \, \eta \rangle^{(0)}_{\rm R-R} = \left[ C \Gamma^{0}
\dots \Gamma^{p} \left( 1 + i \eta \Gamma^{11} \right) \right]_{AB}
|A \rangle |\tilde{B} \rangle~~,\label{Bpsi}
\end{eqnarray}
where $C$ is the charge conjugation matrix and the usual factor
$(1+i \eta)^{-1}$ has been omitted here and will be included, for
future convenience, in the superghost part of the boundary state.

Since $p$ is odd then Eq. (\ref{Bpsi}) contains only an even number
of $\Gamma$-matrices that in the chiral basis are all anti-diagonal.
This implies that the $32$-dimensional spinors appearing in the
vacuum of Eq. (\ref{Bpsi}) can be decomposed in 16-dimensional
chiral spinors, always having opposite chirality, i.e.
\begin{eqnarray}
| A \rangle \otimes |\tilde{B} \rangle \equiv \left( |\alpha \rangle
\otimes | \dot{\tilde{\beta}} \rangle, \,  | \dot{\alpha} \rangle
\otimes |\tilde{\beta} \rangle \right).
\end{eqnarray}
In conclusion, for $p$ odd, $\Omega'$ acts as follows on the
32-dimensional Majorana spinors:
\begin{eqnarray}
\Omega' | A \rangle \otimes |\tilde{B} \rangle = | B \rangle \otimes
|\tilde{A} \rangle
\end{eqnarray}
that implies the following action on the boundary state:
\begin{eqnarray}
\Omega' |B_{\psi}, \, \eta \rangle^{(0)}_{\rm R-R} =
(-1)^{\frac{5-p}{2}} |B_{\psi}, \, -\eta \rangle^{(0)}_{\rm R-R}~~,
\label{omega'}
\end{eqnarray}
where we have used the identity
\[
\left[ C \Gamma^{0} \dots \Gamma^{p} \left( 1 + \eta \Gamma^{11}
\right) \right]^{T} = (-1)^{\frac{5-p}{2}} \left[ C \Gamma^{0} \dots
\Gamma^{p} \left( 1 - \eta \Gamma^{11} \right) \right]~~.
\]
The same result can be obtained by following a different strategy.
The zero modes of the R-R boundary states can be written in the
following form:
\begin{eqnarray}
&&|Bp,\eta\rangle^{(0)}_{\rm R-R}=2^{\frac{p-9}{2}}\left( \psi_0^9 +
i\eta \tilde{\psi}_0^9\right) \dots\left( \psi_0^{p+1} + i\eta
\tilde{\psi}_0^{p+1}\right)|B9,\eta\rangle^{(0)}_{\rm R-R}~,
\label{zmp}
\end{eqnarray}
with
\begin{eqnarray}
|B9,\eta\rangle_{\rm R-R}^{(0)}=\frac{1}{2^4}\left(\psi^0_0-
i\eta\tilde{\psi}^0_0\right)\dots\left(\psi^9_0-i\eta\tilde{\psi}^9_0\right)
C_{AB}|A\rangle \otimes |\tilde{B}\rangle~~. \label{d9o'}
\end{eqnarray}
If we observe that
\begin{eqnarray}
\Omega' |B9,\, \eta\rangle_{\rm
R-R}^{(0)}=-(-i\eta)^{10}|B9,\,-\eta\rangle_{\rm
R-R}^{(0)}=|B9,\,-\eta\rangle_{\rm R-R}^{(0)}~~,\label{omd90'}
\end{eqnarray}
we can write:
\begin{eqnarray}
\Omega' |Bp,\eta\rangle^{(0)}_{\rm R-R}=(i\eta)^{9-p}
|Bp,-\eta\rangle^{(0)}_{\rm R-R}=-(-i\eta)^{7-p}
|Bp,-\eta\rangle^{(0)}_{\rm R-R}~~,
\end{eqnarray}
that coincides with Eq. (\ref{omega'}) if $p$ is odd.

Finally let us analyze the action of $\Omega'$ on the superghost
boundary state. {From} the overlap conditions
\begin{eqnarray}
(\gamma_t+i\eta\tilde\gamma_{-t})|B_{\rm sgh},\,\eta\rangle=0
\qquad(\beta_t+i\eta\tilde\beta_{-t})|B_{\rm sgh},\,\eta
\rangle=0~~,\label{ovsgh0}
\end{eqnarray}
which hold for both  NS-NS ($t$ is half-integer) and R-R ($t$ is an
integer) sector, we deduce that $\Omega' |B_{sgh},\eta\rangle $
satisfies the same overlap condition as $ |B_{sgh},-\eta\rangle$,
therefore we conclude:
\begin{eqnarray}
\Omega' |B_{sgh},\eta\rangle= k \, |B_{sgh},-\eta\rangle  \qquad
k^2=1~~,\label{ospghrr}
\end{eqnarray}
together with
\begin{eqnarray}
\Omega' |B_{sgh},\eta\rangle_{\rm R-R}=  |B_{sgh},-\eta\rangle_{\rm
R-R}~~. \label{ospghrr1}
\end{eqnarray}

In order to find an expression of the boundary state
$|B_{sgh},\,\eta\rangle_{\rm R-R}$ satisfying the conditions
(\ref{ovsgh0}) for integer values of $t$, we notice that the overlap
equations can be alternatively written in a different way by
factorizing $i\eta$ and sending $t \rightarrow -t$.
\begin{eqnarray}
(\tilde\gamma_t-i\eta\gamma_{-t})|B_{\rm sgh},\,\eta\rangle_{\rm
R-R}=0 \qquad(\tilde\beta_t-i\eta\beta_{-t})|B_{\rm
sgh},\,\eta\rangle_{\rm R-R}=0~~. \label{ovsgh01}
\end{eqnarray}
The boundary state $|B_{sgh},\,\eta\rangle_{\rm R-R}$ satisfying
both conditions given in Eq.s (\ref{ovsgh0}) and (\ref{ovsgh01}) has
to be symmetric under the exchange of $\gamma$ with $\tilde\gamma$,
$\beta$ with $\tilde\beta$ and $\eta$ with $-\eta$. The usual
expression of the non-zero mode part of the boundary state, given
for example in Ref. \cite{DL0307}, trivially satisfies this latter
symmetry. The zero-mode part, instead, has to be changed in order to
be invariant under the world-sheet parity and it is natural to
define it as follows:
\begin{eqnarray}
|B_{\rm sgh},\eta\rangle_{\rm
R-R}^{(0)}=\frac{1}{\sqrt{2}}\left[
 \frac{ e^{i\eta \gamma_0
\tilde\beta_0}}{1+i\eta}|0\rangle_{-\frac{1}{2}}\otimes |\tilde{0}
\rangle_{-\frac{3}{2}} +
\frac{ e^{-i\eta \tilde\gamma_0
\beta_0}}{1-i\eta} |\tilde{0}\rangle_{-\frac{1}{2}}\otimes
|0\rangle_{-\frac{3}{2}}\right],
 \label{nbsgzm}
\end{eqnarray}
where we have chosen for the  phase factor the value $k=1$ and, as
previously said, we have included the factors $(1 \pm i
\eta)^{-1}$. By using the well-known identity:
\begin{eqnarray}
e^A\,e^B=e^B\,e^A\,e^{[A,\,B]}
\end{eqnarray}
valid when $[A,\,B]$ is a $c$-number and using the formulas:
\begin{eqnarray}
\tilde\gamma_0|\tilde{0}\rangle_{-\frac{3}{2}}=0 \qquad
\beta_0|0\rangle_{-\frac{1}{2}}=0
\end{eqnarray}
one can get the following  relations~\footnote{Remember
  that $[ \gamma_0 , \beta_0 ] =1$.}:
\begin{eqnarray}
&&\gamma_0{\rm exp}[-i\eta \tilde\gamma_0 \beta_0]={\rm exp}[-i\eta
\tilde\gamma_0 \beta_0](\gamma_0-i\eta\tilde\gamma_0)
\qquad\tilde\gamma_0{\rm exp}[-i\eta \tilde\gamma_0 \beta_0]={\rm
exp}[-i\eta \tilde\gamma_0 \beta_0]\tilde\gamma_0 \nonumber\\
&&\tilde\beta_0{\rm exp}[-i\eta \tilde\gamma_0 \beta_0]={\rm
exp}[-i\eta \tilde\gamma_0 \beta_0](\tilde\beta_0+i\eta\beta_0)
\qquad\beta_0{\rm exp}[-i\eta \tilde\gamma_0 \beta_0]={\rm
exp}[-i\eta \tilde\gamma_0 \beta_0]\beta_0.\nonumber
\end{eqnarray}
These equations allow us to prove that the boundary state in Eq. (\ref{nbsgzm})
satisfies the required overlap conditions.
Finally, after collecting Eq.s (\ref{onzmfg}), (\ref{omega'}) and
(\ref{ospghrr1}), we can write:
\begin{eqnarray}
\Omega' |Bp, \, \eta \rangle_{\rm R-R} = (-1)^{\frac{9-p}{2}} |Bp,
\, -\eta \rangle_{\rm R-R} ~~.\label{complete}
\end{eqnarray}
This means that the boundary state invariant under  $\Omega'$ is the
following:
\begin{eqnarray}
|Bp \rangle_{\rm R-R} = \frac{1+\Omega'}{2} |Bp, \, + \rangle_{\rm
R-R} = \frac{1}{2} \left[ |Bp, \, + \rangle_{\rm R-R} +
(-1)^{\frac{9-p}{2}} |Bp, \, -
  \rangle_{\rm R-R} \right].
\label{BpR}
\end{eqnarray}
Hence in type $0'$ theory one has:
\begin{eqnarray}
|Bp \rangle = |Bp \rangle_{\rm NS-NS} +|Bp \rangle_{\rm R-R}~~,
\label{bound}
\end{eqnarray}
with $p$ odd, in agreement with the result of Ref.~\cite{BFL9904},
where the first term is given in Eq. (\ref{Bp}) and the second one
in Eq. (\ref{BpR}).

Eq.s (\ref{omegnzm}) and (\ref{complete}) imply that the boundary
states of type IIB theory, given in Eq. (\ref{bs22ab}), is invariant
under $\Omega'$ only for the values of $p=1,5,9$. As a consequence,
the only boundaries that we can have in type I theory are the ones
associated with the D1, D5 and D9 branes. This remark can be seen as
a check on the validity of our construction.

Let us notice that the boundary state given in Eq. (\ref{bound})
differs from the standard one given in the literature, only for the
part regarding the superghost zero modes. However, it is simple to
check that, in computing the interaction between branes, the
contribution due to such modified boundary remains unchanged.

As previously mentioned in type $0'$ there is a non trivial
background made of the O9-plane and a stack of $32$ D$9$ branes. These
latter are described by the boundary states with $p=9$ that we
have constructed in this subsection.  For what concerns the
O9-plane, it also admits a microscopic description in the language
of perturbative closed string theory, which is given by the crosscap
state that we explicitly construct in Subsect. 6.5.

\subsection{Open string spectrum}
\label{openstringspectrum} In this subsection we determine the
spectrum of the massless states living on the world-volume of $N$ D3
branes of type $0'$ theory. We denote the generic open string state
living  in the world-volume of a D$p$ brane as:
\begin{eqnarray}
|\chi,\,ij\rangle\equiv \lambda_{ij}|\chi\rangle \qquad i,j=1\dots N~~,
\label{ost1}
\end{eqnarray}
where $\lambda$ is an hermitian matrix~\cite{D9804} corresponding to
the Chan-Paton factors that describe the gauge degrees of
freedom and  $\chi$ is the state made by the string oscillators. We
denote by  $\tilde{T}^{\tilde{a}}, \,\,\,\tilde{a}=1,\dots, N^2-1$
the generators of the group $SU(N)$  normalized as follows:
\begin{eqnarray}
\mbox{Tr}\,\left[
  \tilde{T}^{\tilde{a}} \,\tilde{T}^{\tilde{b}}\right]=
\frac{1}{2}\delta^{{\tilde{a}}{\tilde{b}}}~~.\label{sung1}
\end{eqnarray}
By adding the properly normalized identity generator we obtain the
generators of the group $U(N)$ given by
$T^a\equiv\left(\tilde{T}^{\tilde{a}} \,,\frac{1}{\sqrt{2\,N}}{\rm
I\!l} \right)$ with $a=1, \dots, N^2$. They satisfy the relations:
\begin{equation}
\mbox {Tr} \,\, \left[ T^a T^b \right] = \frac{1}{2} \delta^{ab}
~~~;~~~
\sum_{a=1}^{N^2}T^a_{ij}\,T^a_{kl}=\frac{1}{2}\delta_{il}\,\delta_{jk}~~.
\label{ihm1}
\end{equation}
Since any hermitian matrix can be expanded in terms of the $U(N)$
generators we can write $\lambda$ as follows:
\begin{eqnarray}
\lambda_{ij}|\chi\rangle  =\sum_{a=1}^{N^2}c_a (T^a)_{ij} |\chi
\rangle~~,
\end{eqnarray}
where the $c_a$'s are arbitrary real coefficients. By considering:
\begin{eqnarray}
\sum_{i,j=1}^N(T^b)_{ji}\lambda_{ij}|\chi\rangle  =\sum_{a=1}^{N^2}
c_a(T^b)_{ji} (T^a)_{ij} |\chi \rangle= \nonumber \\
= \sum_{a=1}^{N^2}c_a {\rm
Tr}[T^b\, T^a] |\chi \rangle= \frac{c_b}{2}|\chi \rangle \equiv |
\chi, b \rangle
\end{eqnarray}
and using Eq. (\ref{ost1}), we can write:
\begin{eqnarray}
|\chi, b\rangle= \sum_{i,j=1}^N(T^b)_{ji}|\chi,ij\rangle \qquad b=1,
\dots, N^2~~.\label{state2}
\end{eqnarray}
Eq.s (\ref{ost1}) and (\ref{state2}) give two different
representations of the same open string state. In the following we
are going to use the basis defined in Eq. (\ref{ost1}). The two
basis correspond to the two ways of representing a gauge field in
field theory. Indeed we can represent it by a matrix $A^{\mu}_{ij} =
\sum_{a=1}^{N^2} T^{a}_{ij} A^{\mu}_{a}$ or simply by $A^{\mu}_{a}$.
In particular, using the notation in Eq. (\ref{ost1}), a massless
gluon state of the open string can be written as follows:
\begin{equation}
c_a T^{a}_{ij} \epsilon_{\mu} \psi_{-1/2}^{\mu} | 0 \rangle~~,
\label{pola45}
\end{equation}
where $c$ and $\epsilon$ are the two ``polarization'' vectors in the
gauge and Minkowski space respectively.

As usual, physical states are left invariant by the world-sheet
parity. In order to determine such states we also have  to define
the combined action of $\Omega'$ on the Chan-Paton factors and on the
oscillators, given by:\cite{GP9601}
\begin{eqnarray}
&&\Omega' |\chi,\,ij\rangle= \left( \gamma_{\Omega'_p} \right)_{im}
|\Omega'\chi,\,nm\rangle
\left(\gamma_{\Omega'_p}^{-1}\right)_{nj}\label{oaccp}\\
&&{\Omega'}^2 |\chi,\,ij\rangle=
\left(\gamma_{\Omega'_p}^T\gamma_{\Omega'_p}^{-1}\right)_{sj}
\left(\gamma_{\Omega'_p}^T \gamma_{\Omega'_p}^{-1} \right)^{-1}_{ir}
|{\Omega'}^2\chi,\,rs\rangle ~~.\label{oacp}
\end{eqnarray}
The explicit form of the matrix $\gamma_{\Omega'_p} $ is obtained by
imposing the constraint:
\begin{eqnarray}
{\Omega'}^2 |\chi,\,ij\rangle = |\chi,\,ij\rangle~~,
\end{eqnarray}
with  ${\Omega '}^2=1$ that is a consequence of
the fact that physics is invariant under a double inversion of the
string endpoints.

The massless states living on the world-volume of a D3 brane are
given by:
\begin{eqnarray}
\lambda_A \psi^{\alpha}_{-1/2} | 0 \rangle_{-1}~~~,~~~ \lambda_\phi
\psi^{i}_{-1/2} | 0 \rangle_{-1} \label{boson2}
\end{eqnarray}
in the NS sector with $\alpha=0,1,2,3$ and $i = 4 \dots 9$ and by
\begin{eqnarray}
\lambda_\psi  | A \rangle_{-1/2} \label{ferm2}
\end{eqnarray}
in the R sector. The generalization of the last two equations to a
generic $p$ brane is straightforward: one has just to notice that in
this case $\alpha=0,\dots, p$ and $i = p+1, \dots, 9$. In the
following,  by imposing the invariance under the action of
$\Omega'$, we determine the precise structure of the Chan-Paton
factors and show that it is the same as in Eq.s (\ref{boson}),
(\ref{boson1}) and (\ref{ferm}).

In the NS sector it is easy   to verify that $|{\Omega'}^2 \chi
\rangle = | \chi \rangle$ and therefore the Chan-Paton factors
satisfy the condition:
\begin{eqnarray}
\left(\gamma_{\Omega'_p}^T\gamma_{\Omega'_p}^{-1}\right)^{-1}
\lambda_{A,\,\phi}\left(\gamma_{\Omega'_p}^T \gamma_{\Omega'_p}^{-1}
\right)=\lambda_{A,\,\phi}~~.\label{nscpc}
\end{eqnarray}
In the R-sector instead we have, for odd values of $p$:
\begin{eqnarray}
|{\Omega'}^2\chi\rangle=\left(\Gamma^{p+1}\dots\Gamma^9\right)^2
 |A\rangle =
(-1)^{\frac{5-p}{2}}  |A\rangle \label{rcpc}
\end{eqnarray}
and therefore the following condition has to be satisfied:
\begin{eqnarray}
\left(\gamma_{\Omega'_p}^T\gamma_{\Omega'_p}^{-1}\right)^{-1}
\lambda_\psi\left(\gamma_{\Omega'_p}^T \gamma_{\Omega'_p}^{-1}
\right) (-1)^{\frac{5-p}{2}}=\lambda_\psi ~~.\label{rcpc1}
\end{eqnarray}
The constraints given in Eq.s (\ref{nscpc}) and (\ref{rcpc1}) seem
to be uncompatible. But we have to remember that the matrix
$\lambda$ is diagonal in the NS sector (see Eq.s (\ref{boson}) and
(\ref{boson1})) and off-diagonal in the R sector (see Eq.
(\ref{ferm})) and therefore there is no contradiction.

In order to find an explicit expression of the world-sheet parity
action on the Chan-Paton factors, satisfying both Eq.s (\ref{nscpc})
and (\ref{rcpc1}), let us write, in the case of a single brane:
\begin{eqnarray}
\gamma_{\Omega'_p}^T\gamma_{\Omega'_p}^{-1} =\left(\begin{array}{cc}
                           a& b \\
                                           c   & d
                                          \end{array}\right),
\label{gamma1}
\end{eqnarray}
where of course $ \mbox{det} [
\gamma_{\Omega'_p}^T\gamma_{\Omega'_p}^{-1} ] = 1$ that implies $
ad -bc =1$ and therefore:
\begin{eqnarray}
\left(\gamma_{\Omega'_p}^T\gamma_{\Omega'_p}^{-1}\right)^{-1} =
\left(\begin{array}{cc}
                           d& -b \\
                          - c   & a
                                          \end{array}\right)~~.
\label{gamma2}
\end{eqnarray}
Eq. (\ref{nscpc}) imposes the following constraints:
\begin{eqnarray}
b=0 \qquad c=0 \qquad a\,d=1 ~~,\label{co1}
\end{eqnarray}
while Eq. (\ref{rcpc1}) yields:
\begin{eqnarray}
&&
\gamma_{\Omega'_p}^T\gamma_{\Omega'_p}^{-1}=\frac{1}{a}\left(\begin{array}{cc}
                                           a^2   &0\\
                                            0 &1
                                          \end{array}\right)
 \qquad\mbox{with} \qquad a^2= (-1)^{\frac{5-p}{2}}~~.
\label{co2}
\end{eqnarray}
{From} the previous equation we see that  $\gamma_{\Omega'_p}$ is
symmetric or antisymmetric if  $p=1,5,9$ and that ${\rm Tr}
[\gamma_{\Omega'_p}^T\gamma_{\Omega'_p}^{-1}] =0$ for the D3 and D7
branes, in agreement with the results of Ref. \cite{DM0010}. By
collecting the previous results we can write:
\begin{eqnarray}
\gamma_{\Omega'_p}^T\gamma_{\Omega'_p}^{-1} = \epsilon' e^{\epsilon
i \pi \left(\frac{5-p}{4}\right)} \left(\begin{array}{cc}
                            (-1)^{\frac{5-p}{2}}&  0 \\
                          0  & 1
                                          \end{array}\right)~~,
\label{gammagamma}
\end{eqnarray}
where $\epsilon$ and $\epsilon'$ are equal to $\pm 1$. In general
one should consider $\epsilon$ and $\epsilon'$ dependent on $p$.
However, as it will be shown in a while, the assumption that they
are independent on $p$ gives the right result when we compare the
one-loop open string diagrams, which explicitly depend on the
structure of the Chan-Paton factors, to the corresponding
calculation in the closed string channel where instead we use the
boundary state. The cancellation of the R-R tadpole in the case of
the D9 brane implies that $\gamma_{\Omega'_9}$ is a symmetric matrix,
\cite{BFL9904} i.e.
$\gamma_{\Omega'_9}^T\gamma_{\Omega'_9}^{-1}=+1$, which fixes
$\epsilon'=-1$. This yields:
\begin{eqnarray}
 {\rm Tr}
[\gamma_{\Omega'_p}^T\gamma_{\Omega'_p}^{-1}] =-2 e^{i\pi
\left(\frac{5-p}{4}\right)} \delta_{p,\,(1\,,5\,,9)}~~. \label{tgg}
\end{eqnarray}
Eq. (\ref{gammagamma}) implies that the matrices
$\gamma_{\Omega'_{9,1}}$ are symmetric, while the matrix
$\gamma_{\Omega'_5}$  is antisymmetric. In order to find the
explicit expression of $\gamma_{\Omega'_p}$ we have to impose Eq.
(\ref{co2}). In so doing one gets:
\begin{eqnarray}
\gamma_{\Omega'_p} =  \left( \begin{array}{cc} 0 &
 \beta \\ \gamma & 0 \end{array} \right)\qquad {\rm with}\,\,
{\gamma}=a {\beta}~~.\label{gammaomegap}
\end{eqnarray}
However Eq. (\ref{co2}) does not fix either the determinant or the
phase of $\gamma_{\Omega'_p}$. We can choose the determinant to be
equal to $(\pm 1)$ without changing the spectrum of the open strings
attached to a stack of branes of the same kind. For $p=3,7$, by
choosing $\mbox{det} \gamma_{\Omega'_p}= - \beta \gamma= -1$ and
fixing the phase of  $\gamma_{\Omega'_p}$ to be equal $1$, one
reproduces Eq. (71) of Ref. \cite{DM0010}, i.e.:
\begin{eqnarray}
\gamma_{\Omega'_p} = \left(\begin{array}{cc}
                           0&  e^{\epsilon i \pi \left(\frac{5-p}{8}\right)} \\
                           e^{\epsilon i \pi\left( \frac{p-5}{8}\right)}& 0
                                          \end{array}\right)\qquad p=3,7~~.
\label{gamma}
\end{eqnarray}
By imposing the invariance of the states (\ref{boson2})
 under the action of $\Omega'$ and using Eq.
(\ref{oaccp}) together with Eq.s (\ref{omegaos}) and
(\ref{omegavac}), one finds the following constraints on the
Chan-Paton factors:
\begin{eqnarray}
\lambda_A=-\gamma_{\Omega' }\lambda_A^T{\gamma_{\Omega'}^{-1}}
\qquad\lambda_\phi=\gamma_{\Omega'
}\lambda_\phi^T{\gamma_{\Omega'}^{-1}}~~,\label{chcon}
\end{eqnarray}
that are  satisfied by taking
\begin{eqnarray}
\lambda_A=\left(\begin{array}{ll}
                     A & \,\,\,\,\,\,0\\
                      0&-A^T
                     \end{array}
                     \right)\qquad \lambda_\phi=\left(\begin{array}{ll}
                     A & \,\,\,0\\
                      0&A^T
                     \end{array}
                     \right) ,
\label{cpbos}
\end{eqnarray}
being $A$ an $N{\times} N$ matrix. The last condition to be imposed
on the Chan-Paton factors is $\lambda^\dag=\lambda$ which fixes $A$
to be an hermitian matrix, i.e. $A=A^\dag$. The gauge group $U(N)$
acts on the Chan-Paton  as follows:
\begin{eqnarray}
\lambda \rightarrow \tilde{U}\lambda \tilde{U}^{\dag} ~~,\label{ggac}
\end{eqnarray}
where
\begin{eqnarray}
\tilde{U}=\left(\begin{array}{ll}
                     U & 0\\
                      0&\bar{U}
                     \end{array}
                     \right),
\label{grel}
\end{eqnarray}
with $U$ being an $N{\times} N$ matrix of the $U(N)$ group, and
$\bar{U}$ being its complex conjugate.  {From} Eq. (\ref{ggac}) we get
the following conditions:
\begin{eqnarray}
A'=UAU^{\dag} \qquad A^{'T}=\bar{U}A^T \bar{U}^{\dag}~~,\label{Agt}
\end{eqnarray}
that are compatible since $A$ is an hermitian matrix. The states
given in Eq. (\ref{boson2}) describe a vector boson and six adjoint
scalars of the gauge group $U(N)$.

Let us now consider the Ramond sector of the open string attached to
the D3 branes. The world-sheet parity $\Omega'$ acts on the massless
states as follows ($s_i = \pm \frac{1}{2}$):
\begin{eqnarray}
\Omega'  |s_1,\,s_2,\,s_3,\,s_4;\,ij\rangle= - \left(
\gamma_{\Omega'_{3}} \right)_{im} \left( \gamma_{\Omega'_3}^{-1}
\right)_{nj} \left(\Gamma^4\dots\Gamma^9\right)
|s_1,\,s_2,\,s_3,\,s_4;\,nm\rangle~,\label{rst}
\end{eqnarray}
where we have used the second equation in (\ref{ramo23}) and the
fact that the action of $\Omega '$ is the same as that of
${\tilde{\Omega}}$. Remember that $s_0=1/2$ and  the GSO projection
imposes $\sum_{i=1}^4s_i={\rm odd}$. Using the previous equations we
get:
\[
\left(\Gamma^4\dots\Gamma^9\right)
|s_1,\,s_2,\,s_3,\,s_4;\,nm\rangle
 = -i (2s_2 ) {\cdot} (2 s_3 ) {\cdot} (2 s_4 )
|s_1,\,s_2,\,s_3,\,s_4;\,nm\rangle =
\]
\begin{equation}
= 2 i s_1 |s_1,\,s_2,\,s_3,\,s_4;\,nm\rangle~~,\label{eq786}
\end{equation}
where  the GSO condition has been imposed. Then Eq. (\ref{rst})
becomes:
\begin{eqnarray}
\Omega'  |s_1,\,s_2,\,s_3,\,s_4;\,ij\rangle= - 2i s_1 \left(
\gamma_{\Omega'_3} \right)_{im} \left( \gamma_{\Omega'_3}^{-1}
\right)_{nj} |s_1,\,s_2,\,s_3,\,s_4;\,nm\rangle~~.\label{rst1}
\end{eqnarray}
Notice that $(2 s_1 )$ coincides with the eigenvalue of
four-dimensional chirality defined as:
\begin{equation}
\Gamma_5=i\, \Gamma^0\Gamma^1\Gamma^2\Gamma^3= 4\,N_0 N_1 = 2 \, N_1
\ ~~.\label{ventidue}
\end{equation}
Then Eq.. (\ref{rst1}) becomes
\begin{eqnarray}
\Omega'  |s_1,\,s_2,\,s_3,\,s_4;\,ij\rangle= - i \Gamma_5 \left(
\gamma_{\Omega'_3} \right)_{im} \left( \gamma_{\Omega'_3}^{-1}
\right)_{nj} |s_1,\,s_2,\,s_3,\,s_4;\,nm\rangle~~. \label{rst12}
\end{eqnarray}
The invariance of the state under $\Omega'$ yields the
 condition:
\begin{eqnarray}
\lambda_\psi= -i\,\Gamma_5 \gamma_{\Omega'_3}\lambda_\psi^T
\gamma_{\Omega'_3}^{-1}~~,\label{rcon}
\end{eqnarray}
where we have denoted by $\Gamma_5$ the four-dimensional chirality
of the spinor.

Furthermore, writing
\begin{eqnarray}
\lambda_\psi=\left(\begin{array}{ll}
                     0 & B\\
                     C&0
                     \end{array}
                     \right)
\label{cpfer}
\end{eqnarray}
and imposing the hermiticity condition
$\lambda_{\psi}=\lambda^{\dag}_{\psi}$, we get $C=B^{\dag}$.

Eq. (\ref{rcon}), together with the explicit expression of the
action of the world-sheet parity on the Chan-Paton factors given in
Eq. (\ref{gamma}), gives:
\begin{eqnarray}
B= \epsilon \Gamma_5 B^T \qquad B^{\dag}= - \epsilon \Gamma_5
\bar{B}~~,\label{cobc}
\end{eqnarray}
where $\bar{B}$ denotes the complex conjugate of $B$. The last two
conditions are compatible if one remembers that the four-dimensional
chirality changes its sign under the complex conjugate operation.

The matrix $B$ is antisymmetric  if $\Gamma_5 = - \epsilon $, while
is symmetric if  $\Gamma_5 = \epsilon$. These conditions, and hence
the open string spectrum, are exactly the same as the ones given in
Ref.~\cite{DM0010}. Consistently with that,  Eq. (\ref{ggac})
implies $B$ to transform under a gauge transformation as follows
\begin{eqnarray}
B'=UBU^T~~, \label{ftra}
\end{eqnarray}
which is the appropriate transformation property for the two-index
symmetric and antisymmetric representation. In conclusion the
spectrum of the massless states of type $0 '$ theory consists of one
gluon, six real scalars, all in the adjoint representation of the
gauge group $SU(N)$, two Dirac fermions in the two-index
antisymmetric and two Dirac fermions in the two-index symmetric
representation of the gauge group. We have two Dirac fermions for
instance in the antisymmetric representation of the gauge group
$SU(N)$ because the state in Eq. (\ref{rst1}), after taking into
account the GSO projection, has four degrees of freedom with
four-dimensional chirality $+1$ and four degrees of freedom with
four-dimensional chirality $-1$. Notice that, even if the
world-volume gauge theory is non-supersymmetric, its spectrum
satisfies the Bose-Fermi degeneracy condition. Finally the
coefficient of one-loop $\beta$-function is given by:
\begin{eqnarray}
\beta(g) = \frac{g^{3}}{(4 \pi)^2}\left[- \frac{11}{3} N + 6
\frac{N}{6} + 2 \frac{4}{3} \left(
  \frac{N-2}{2} + \frac{N+2}{2} \right) \right] =0
\label{beta67}
\end{eqnarray}
and therefore the gauge theory living on a D3 brane of type $0'$
theory is conformal invariant at least at one-loop.

\subsection{One-loop vacuum amplitude}
\label{1loop} In type $0'$ string theory
the free energy is simply:
\begin{eqnarray}
Z^{o}= Z^{o}_{1}+Z^{o}_{\Omega'}=  2\!\!\int_{0}^{\infty}\!\! \frac{d
\tau}{2 \tau} \mbox{Tr}_{\rm NS,\, R} \left[e^{-2\pi \tau L_0 }
(-1)^{G_{bc}}P_{\rm GSO} P_{(-1)^{F_s}}
\left( \frac{1+\Omega'}{2}\right) \right],\label{ppo'}
\end{eqnarray}
where the factor 2 in front of the last equation takes into account
the two orientations of the open string exchanged in the loop. The
annulus contribution corresponds to the interaction between two D$p$
branes while the M\"{o}bius strip describes the interaction between
the D$p$ brane and the O9-plane.

Let us  compute it explicitly. $Z^{o}_1$ is obtained by multiplying
the expression in Eq.
(\ref{zop89}) with $N=M$ by a factor 1/2 due to the orientifold
projection, obtaining
\begin{eqnarray}
Z^{o}_1&=&2\int_{0}^{\infty} \frac{d \tau}{2 \tau}
\mbox{Tr}_{\rm NS, \,R } \left[e^{-2\pi \tau L_0} P_{\rm GSO}
  (-1)^{G_{
bc}} P_{(-1)^{F_s}}
{\frac{1}{2}}\right]\nonumber\\
&=& N^2 V_{p+1}(8\pi^2\alpha')^{-\frac{p+1}{2}}\!\int_{0}^{\infty}
\frac{d\tau}{\tau^{\frac{p+3}{2}}} e^{-\frac{y^2\tau}{2\pi\alpha'}} \nonumber\\
&{\times}&   \frac{1}{2} \left[   \left(
\frac{f_3(k)}{f_1(k)}\right)^8-\left(
\frac{f_4(k)}{f_1(k)}\right)^8  - \left(
\frac{f_2(k)}{f_1(k)}\right)^8 \right]~~, \label{182}
\end{eqnarray}
that in the closed string channel $(\tau = 1/t)$ becomes:
\begin{eqnarray}
\!\!\!\!Z^{c}_1= N^2
V_{p+1}(8\pi^2\alpha')^{-\frac{p+1}{2}}\!\!\!\!\int_{0}^{\infty} \!\!\!
\frac{dt}{t^{\frac{9-p}{2}}} e^{-\frac{y^2 }{2\pi\alpha' t}}
\frac{1}{2} \left[   \left(
\frac{f_3(q)}{f_1(q)}\right)^8\!\!\!-\!\!\left(
\frac{f_2(q)}{f_1(q)}\right)^8\!\! \! -\!\! \left(
\frac{f_4(q)}{f_1(q)}\right)^8 \right] . \label{182clo}
\end{eqnarray}
In order to compute $Z^{o}_{\Omega'}$ let us observe that:
\begin{eqnarray}
{\rm Tr} \left[ e^{-2\pi\tau (N_{X}+N_{bc})} \Omega'\right]&=&
(ik)^{2/3} 2^{\frac{9-p}{2}}f_2^{p-9}(ik) f_1^{1-p}(ik) {\cdot}
k^{-2}
\label{tracea}\\
{\rm Tr}_{\rm NS} \left[e^{-2\pi\tau (N_{\psi}+
N_{\beta\gamma})}\Omega'\right]&=&-i (ik)^{1/3}
f_3^{9-p}(ik)f_4^{p-1}(ik) {\cdot} k \label{tracefns}\\ {\rm
Tr}_{\rm NS} \left[e^{-2\pi\tau (N_{\psi}+ N_{\beta
\gamma})}\Omega'(-1)^F\right]&=&i (ik)^{1/3}
f_3^{p-1}(ik)f_4^{9-p}(ik) {\cdot} k~~,\label{tracefnsf}
\end{eqnarray}
where $N$ is the world-sheet number operator defined in each sector
of the string Fock spaces and the contribution to the traces coming
from the ghost and superghost is already included. Instead the trace
over the Chan-Paton factors gives:
\begin{eqnarray}
\mbox{Tr}^{\rm C.P.}\left[  \langle hk |\Omega' |ij \rangle
\right] & = & \delta_{ik}\delta_{hj}\left( \gamma_{\Omega'_p}
\right)_{im} \langle hk |nm\rangle
\left(\gamma_{\Omega'_p}^{-1}\right)_{nj} \nonumber \\
& = & \mbox{Tr}\left[\gamma_{\Omega'_p}^T
\gamma_{\Omega'_p}^{-1}\right] = -2\,N\, e^{i\pi
\left(\frac{5-p}{4}\right)} \delta_{p,\,(1\,,5\,,9)}~~,\label{trcp}
\end{eqnarray}
where we have used Eq.s (\ref{oaccp}) and (\ref{tgg}) and the
normalization $\langle hk |nm\rangle =\delta_{kn}\delta_{hm}$.
Notice that this definition of the trace gives also the correct
result for the planar diagram. In fact we get:
\begin{eqnarray}
\mbox{Tr}^{\rm C.P.}\left[  \langle hk | ij \rangle   \right] =
\mbox{Tr} \left[ \lambda \lambda \right] = \delta_{ki} \delta_{hj}
\langle hk | ij \rangle =\delta_{ii} \delta_{jj}=(2 N)^2~~.
\label{planar23}
\end{eqnarray}
Furthermore, remembering the action of the space-time fermion number
operator on the Chan-Paton factors, given in Eq.s (\ref{pfscp}) and
(\ref{gpfs}),
 we can write:
\begin{eqnarray}
{\rm Tr}^{\rm C.P.} \left \langle hk|(-1)^{F_s} \Omega'|ij\rangle
\right]= {\rm Tr}\left[\gamma_{\Omega'_{p}}^{-1}
\gamma_{(-1)^{F_s}}^{-1} \gamma_{\Omega'_{p}}^{T}
\gamma_{(-1)^{F_s}}^{T} \right]=- {\rm Tr}\left[
\gamma_{\Omega'_{p}}^{-1} \gamma_{\Omega'_{p}}^{T}\right]~.
\label{ttrsf}
\end{eqnarray}
For the M{\"{o}}bius diagram, in the NS sector, we get the following
expression:
\begin{eqnarray}
Z^{o}_{NS;\,\Omega'}\!\!&=&\!\!\int_{0}^{\infty} \frac{d \tau}{\tau}
\mbox{Tr}_{\rm NS} \left[e^{-2\pi \tau L_0}  (-1)^{G_{bc}} P_{\rm
GSO} P_{(-1)^{F_{s}}}{\frac{\Omega'}{2}}\right]\nonumber\\
\!\!&=&\!\! \left(- 2\,N e^{i\pi\left(\frac{5-p}{4}\right)}
\delta_{p,\,(1,5,9)} \right)  \frac{V_{p+1}}{8}
 (8\pi^2 \alpha')^{-\frac{p+1}{2}}\left[
\tilde{Z}^{op.}_{\Omega'}+\tilde{Z}^{op.}_{\Omega'(-1)^{F_s}}\right],
\label{freeo}
\end{eqnarray}
with:
\begin{eqnarray}
\tilde{Z}^{o}_{\Omega'}&=& -
\tilde{Z}^{o}_{\Omega'(-1)^{F_s}}
= 2^{\frac{9-p}{2}}\int_0^\infty
\frac{d\tau}{\tau^{(p+3)/2}}
 e^{-\frac{y^2\tau}{2\pi\alpha'}}\nonumber\\
&\times&\left[\left(\frac{ f_3(ik)}{f_2(ik)}\right)^{9-p} \left(\frac{
f_4(ik)}{f_1(ik)}\right)^{p-1}- \left(\frac{
f_4(ik)}{f_2(ik)}\right)^{9-p}
\left(\frac{ f_3(ik)}{f_1(ik)}\right)^{p-1}\right]~,\nonumber\\
&&\label{znsom}
\end{eqnarray}
where Eq.s  (\ref{tracea}), (\ref{tracefns}), (\ref{tracefnsf}),
(\ref{trcp}) and (\ref{ttrsf}) have been taken into account. By
using Eq. (\ref{znsom}) in Eq. (\ref{freeo}) we get:
\begin{eqnarray}
 Z^{o}_{NS;\,\Omega'} = 0 ~~.
\label{185}
\end{eqnarray}
It may be helpful to give Eq. (\ref{znsom}) also in the closed
string channel by performing the modular transformation
$\tau=1/(4\,t)$, obtaining:
\begin{eqnarray}
&&{\tilde{Z}}^{c}_{\Omega'}= 2^6 e^{i\pi\left(\frac{5-p}{4 }\right)}
\int_{0}^{\infty} dt e^{-\frac{y^2}{8\pi \alpha't}}
\nonumber\\
&&\,\times \left[ \left(\frac{ f_3(iq)}{f_1(iq)}\right)^{p-1} \left(\frac{
f_4(iq)}{f_2(iq)}\right)^{9-p}- \left(\frac{
f_4(iq)}{f_1(iq)}\right)^{p-1}
\left(\frac{ f_3(iq)}{f_2(iq)}\right)^{9-p}\right]~,\nonumber\\
&&\label{zcnsom}
\end{eqnarray}
where $q = e^{- \pi t}$ and the identities in Eq.s (\ref{f1f2}) and
(\ref{f3f4}) have been used.

Because of Eq. (\ref{185}), the total bosonic contribution to the
free energy simply reduces to the first two terms of Eq.
(\ref{182}), and in particular the contribution of the bosonic
massless states turns out to be:
\begin{eqnarray}
Z^{o}_1({\rm bosonic\,\, massless})&=& 8N^2
V_{p+1}(8\pi^2\alpha')^{-\frac{p+1}{2}}\!\int_{0}^{\infty} \!
\frac{d\tau}{\tau^{\frac{p+3}{2}}} e^{-\frac{y^2\tau}{2\pi\alpha'}}~,
\label{bs1}
\end{eqnarray}
where the factor $8N^2$ in front of the previous expression counts
the number of bosonic degrees of freedom of the gauge theory living
on the brane.

In the Ramond sector, the term corresponding to the M{\"{o}}bius
diagram is given by:
\begin{eqnarray}
Z^{o}_{R;\Omega'}=-\int_{0}^{\infty} \frac{d\tau}{\tau} {\rm
Tr}_R\left[ e^{-2\pi\tau L_0}(-1)^{G_{bc}} P_{\rm GSO}
P_{(-1)^{F_{s}}} \frac{\Omega'}{2}
\right]~~.\label{frer}
\end{eqnarray}
It is easy to compute the traces over the non-zero modes, getting:
\begin{eqnarray}
&&{\rm Tr}_R^{nzm}\left[e^{-2\pi\tau (N_\psi + N_{\beta \gamma})}
\Omega'(-1)^{G_{\beta \gamma}}\right] = \frac{(ik)^{-2/3}}{
2^{(p-1)/2}} f_2(ik)^{p-1}f_1(ik)^{9-p}{\cdot} k^2\label{tro}\\
&&{\rm Tr}_R^{nzm}\left[e^{-2\pi\tau (N_\psi + N_{\beta \gamma})}
\Omega'(-1)^F\right] = \frac{(ik)^{-2/3}}{ 2^{(9-p)/2}}
f_1(ik)^{p-1}f_2(ik)^{9-p}{\cdot} k^2~~,\label{trof}
\end{eqnarray}
where $N$ are the world-sheet number operators and the ghost and
superghost non zero-mode contributions to the traces have been
already taken into account. The trace over the zero modes is instead
given by:
\begin{eqnarray}
&&{\rm Tr}_R^{z.m.}\left[\Omega'(-1)^{G^{0}_{\beta \gamma}}\right]
= -2^4\delta_{9,p}\label{trozm}\\
&&{\rm Tr}_R^{z.m.}\left[\Omega'(-1)^{F_0}\right] = -2^4 \delta_{p,1}
~~.\label{trofzm}
\end{eqnarray}
By collecting Eq.s (\ref{tracea}), (\ref{tro}), (\ref{trof}),
(\ref{trozm}) and (\ref{trofzm}) we get:
\begin{eqnarray}
Z_{R;\Omega'}^{o}= &&\frac{V_{p+1}}{4}
(8\pi^2\alpha')^{-\frac{p+1}{2}} \left(-2\,N e^{i\pi
\left(\frac{5-p}{4}\right)} \delta_{p,\,(1\,,5\,,9)}
\right)\int_{0}^{\infty} \frac{d\tau}{\tau}\tau^{-\frac{p+1}{2}}
e^{-\frac{y^2\tau}{2\pi\alpha'}}\nonumber\\
&&{\times}\left[\delta_{9,p}\left(\frac{ f_2(ik)}{f_1(ik)}\right)^8
+2^4\delta_{p,1} \right]~~. \label{freeom}
\end{eqnarray}
The previous expression can be written also in the closed channel
as:
\begin{eqnarray}
Z_{R;\Omega'}^{c}=&& 2^4\, V_{p+1} (8\pi^2\alpha')^{-\frac{p+1}{2}}
\left( -2 N e^{i\pi \left(\frac{5-p}{4}\right)}
\delta_{p,\,(1\,,5\,,9)}  \right)
\int_{0}^{\infty}  dt\,e^{-\frac{y^2}{8\pi\alpha't}}\nonumber\\
&&{\times}\left[\delta_{9,p}\left(\frac{ f_2(iq)}{f_1(iq)}\right)^8
+ \delta_{p,1} \right]~~. \label{freeomf}
\end{eqnarray}
For $p=3$ the fermionic free energy contribution reduces to the
third term of Eq. (\ref{182}) that in the field theory limit gives:
\begin{eqnarray}
Z^{o}_1({\rm fermionic\,\, massless})&=& -8N^2
V_{p+1}(8\pi^2\alpha')^{-\frac{p+1}{2}}\!\int_{0}^{\infty} \!
\frac{d\tau}{\tau^{\frac{p+3}{2}}}
e^{-\frac{y^2\tau}{2\pi\alpha'}}~~. \label{bf1}
\end{eqnarray}
The number $8N^2$ gives exactly the number of fermionic degrees of freedom
of the world-volume gauge theory which indeed consist of two Dirac
spinors in the two-index symmetric representation of the gauge group
and two Dirac spinors in the two-index antisymmetric one.

For $p=9$ Eq. (\ref{freeomf})  is divergent in the infrared limit
$t\rightarrow \infty$. This divergence signals, as in type I string
theory, the presence of a R-R tadpole which must be cancelled by
introducing a suitable background of D9 branes. In order to
determine it we compute the Klein-Bottle amplitude for this model.
It is given by:
\begin{eqnarray}
Z_{\rm KB}=\int_0^\infty \frac{d\tau}{\tau}{\rm Tr} \left[
\frac{\Omega'}{2} e^{-2\pi\tau( L_0+\tilde{L}_0) } P_{\rm
GSO}(-1)^{G_{bc}+{\tilde G}_{bc}+G_{\beta\gamma}+{\tilde
G}_{\beta\gamma}}\right]~~.\label{kbo'}
\end{eqnarray}
where $P_{GSO}$ has been defined in Eq.(\ref{gso}).

 By computing the trace over the oscillators one gets
($k=e^{-\pi\tau}$)
\begin{eqnarray}
{\rm Tr} \left[ \Omega'e^{-2\pi\tau(N_{X} +\tilde{N}_{X} +N_{bc} +
\tilde{N}_{bc})}\right] &=&\frac{k^{\frac{4}{3}}}{
\left[f_1(k^{2})\right]^{8}}{\cdot} k^{-4}
\nonumber\\
{\rm Tr}_{\rm NS-NS} \left[ \Omega'e^{-2\pi\tau(N_{\psi}
+\tilde{N}_{\psi} + N_{\beta \gamma} + \tilde{N}_{\beta \gamma} )
}\right]&=&-k^{\frac{2}{3}}\left[ f_4(k^{2})\right]^{8} {\cdot}
k^{2} \nonumber\\
{\rm Tr}_{\rm NS-NS}\left[ \Omega'e^{-2\pi\tau(N_{\psi}
+ \tilde{N}_{\psi}+ N_{\beta \gamma} + \tilde{N}_{\beta \gamma}
)}(-1)^{F+\tilde{F}}
\right]
&=&-k^{\frac{2}{3}}\left[ f_4(k^{2})\right]^{8} {\cdot}
k^{2}\nonumber\\
{\rm Tr}_{\rm R-R}\left[
\Omega'e^{-2\pi\tau(N_{\psi} +\tilde{N}_{\psi} + N_{\beta \gamma} +
\tilde{N}_{\beta \gamma}   )}\right] &=&
k^{-\frac{4}{3}}f_1^8(k^{2}) {\cdot} k^{4}\nonumber\\
{\rm Tr}_{\rm R-R}\left[
\Omega'e^{-2\pi\tau(N_{\psi} + \tilde{N}_{\psi}  N_{\beta \gamma} +
\tilde{N}_{\beta \gamma} )}(-1)^{F+\tilde{F}}
\right]
 &=&k^{-\frac{4}{3}}f_1^8(k^{2}) {\cdot} k^{4}~,
\label{traceos}
\end{eqnarray}
where we have already added the contribution coming from ghosts and
superghosts. Let us consider now the zero modes contributions.  In
the bosonic sector one has:
\begin{eqnarray}
V_d \int \frac{d^dp}{(2\pi)^d} e^{-\alpha'\pi\tau p^2}=\frac{2^{d/2}\,V_d}
{(8\pi^2\tau\alpha')^{d/2}}~~.\label{tracemo}
\end{eqnarray}
By using Eq. (\ref{omegavuo}), one can easily check that the R-R
zero modes give a vanishing contribution to the trace in the
partition function:
\begin{eqnarray}
{\rm Tr}_{z.m.\, R-R}[\Omega']&=& \sum_{A,B}
\,_{-1/2}<\tilde{C}|_{-1/2}<D| \Omega'|A>_{-1/2}
|\tilde{B}>_{-1/2}
(C)_{BC} (C)_{AD}\nonumber \\
&=& {\rm Tr}(\Gamma^{11})=0 ~~,\label{zmkb}
\end{eqnarray}
where the following identity has been used:
\begin{eqnarray}
<A|B> = \left( C^{-1} \right)^{AB}~~.
\end{eqnarray}
Thus collecting the contributions of the zero and non-zero modes we
get:
\begin{eqnarray}
Z_{\rm KB}=-2^4\frac{V_{10}}{(8\pi^2\alpha')^{5}} \int_0^\infty
\frac{d\tau}{\tau^6}\, \left[\frac{
f_4(k^2)}{f_1(k^2)}\right]^{8}~~.\label{kbo'1}
\end{eqnarray}

It is useful to write Eq. (\ref{kbo'1}) in the closed string channel
by performing the modular transformation $\tau=1/ 4t$. According to
Eq. (\ref{mtf1}) we have ($q =e^{- \pi t}$):
\begin{eqnarray}
Z_{\rm KB}&=&-2^{10}\frac{V_{10}}{(8\pi^2\alpha')^{5}}
\int_0^\infty d t \, \left[\frac{ f_2(q^2 )}{f_1(q^2)}\right]^{8}
\nonumber\\
&=&-2^{10}\frac{V_{10}}{(8\pi^2\alpha')^{5}} \int_0^\infty d t
\,\frac{1}{2} \left[\frac{ f_2(q
)}{f_1(q)}\right]^{8}~~,\label{kbo'2}
\end{eqnarray}
where the last identity follows after the change of variable
$t\rightarrow t/2$.

 The sum of the contributions from the
Klein bottle in Eq. (\ref{kbo'2}), the M{\"{o}}bius diagram in Eq.
(\ref{freeomf}) for $p=9$ and  the second term in Eq.
(\ref{182clo}) again for $p=9$ corresponding to the R-R spin
structure is equal to:
\begin{eqnarray}
\frac{V_{10}}{(8\pi^2\alpha')^{5}} \int_0^\infty \!\!d t \left[-
2^{10} {\cdot} \frac{1}{2} \left[\frac{ f_2(q
)}{f_1(q)}\right]^{8} + 2^6 N {\cdot}
\frac{1}{2} \left[\frac{ f_2(iq )}{f_1(iq)}\right]^{8} - N^2 {\cdot}
\frac{1}{2} \left[\frac{ f_2(q )}{f_1(q)}\right]^{8}\right].
\label{sum82}
\end{eqnarray}
If we restrict ourselves to the contribution of the massless states
that is obtained by taking the ratio $(f_2 / f_1)^8 = 2^4$, we get a
R-R tadpole given by:
\begin{eqnarray}
- \frac{V_{10}}{(8\pi^2\alpha')^{5}} \left[ N -32  \right]^2  8
\int_0^\infty d t ~~,\label{rrtad}
\end{eqnarray}
that vanishes if we choose $N=32$. This means that type $0'$ is free
from R-R tadpoles if we have a background of $32$ D9
branes~\cite{BFL9904} as in type I theory. Notice that,
as previously discussed, had we chosen $\gamma_{\Omega'_{9}}$ to be
an antisymmetric matrix, we would have obtained Eq. (\ref{tgg}) with
the opposite sign. This would have given a minus sign in front of
the middle term in Eq.(\ref{sum82}) and the cancellation of the R-R
tadpole would not have been possible. The NS-NS tadpole can instead
be extracted from the contribution of the massless states to the
first and last terms of Eq. (\ref{182clo}). It is given by:
\begin{eqnarray}
N^2  \frac{V_{10}}{(8\pi^2\alpha')^{5}} 8 \int_0^\infty d t~~,
\label{nsnstad}
\end{eqnarray}
that cannot be cancelled.\cite{BFL9904}

Let us now compute the interaction between two D$p$ branes in the
closed string channel, by using the boundary state formalism. But,
before we do that, let us give the generalization of Eq.
(\ref{freeom}) to the case in which we have an O$p'$ instead of an
O9-plane orientifold. In this case Eq. (\ref{frer}) is modified as follows:
\begin{eqnarray}
Z_{R;\Omega'I_{9-p'}}^{o}=-\int_0^\infty
\frac{d\tau}{\tau} \rm{Tr}_{ R}\left[ e^{-2\pi\tau L_0}\,
(-1)^{G_{bc}} P_{GSO}\,P_{(-1)^{F_s}}
\frac{\Omega'\,I_{9-p'}}{2}\right]~~,\label{frerI}
\end{eqnarray}
where we have assumed that $p' \geq p$ and
$I_{9-p'}$ is the inversion on $9-p'$ coordinates, i.e:
\begin{eqnarray}
I_{9-p'}\,:\,\,(x^{p'+1},\dots , x^9)\rightarrow (-x^{p'+1},\dots ,-x^9)~~.
\label{i9p}
\end{eqnarray}
The presence in the trace of the reflection operator follows from
being the orientifold  $p'$-plane obtained by performing $9-p'$
T-dualities on $O9$.
In general, $n$ T dualities transform $\Omega'$ into
$\Omega'I_n(-1)^{F_L}$ being $F_L$ the space time fermion number.
The reason why the orientifold projector contains also  the term
$(-1)^{F_L}$ is
because the operator $\Omega'I_{2n}$ squares to unity only for
$n$ even.  In fact $I^2_{2n}$ represents a $2\pi$
rotation in $n$ planes and, for $n$ odd, is equal to
$(-1)^{F_s}=(-1)^{F_L+F_R}$. Therefore:
\begin{eqnarray}
\left[\Omega'I_{2n} (-1)^{F_L}\right]^2=
I_{2n}^2(-1)^{F_L+F_R}=\mathbb{I}\qquad n=1,3~~,
\end{eqnarray}
where we have used the fact that $\Omega' (-1)^{F_L} \Omega'^{-1} =
(-1)^{F_R}$  and $\Omega'^{2}=1$.

$(-1)^{F_L}$  gives an extra minus sign in the R-sector depending
on being the open string considered respectively the right or left
sector of closed string. However, this sign ambiguity is
completely irrelevant in the discussion below and therefore we ignore
it assuming for simplicity in the following that $(-1)^{F_L}$ does not
act on the open string.

$I_{9-p'}$ acts on the NS vacuum as:
\begin{eqnarray}
I_{9-p'}|0\rangle_{-1}=|0\rangle_{-1}~~.
\end{eqnarray}
In the R sector, instead, in order to determine the action of the
reflection operator on the vacuum, we observe that a reflection in
a plane corresponds to a rotation of an angle $\pm \pi$, therefore
we can write for $p'$ odd:
\begin{eqnarray}
I_{9-p'}|s_0\dots s_4\rangle_{-1/2}=e^{\pm i
\pi\sum_{i=(p'+1)/2}^4S^{2i,2i+1}}|s_0\dots s_4\rangle_{-1/2}~~,
\label{Iac}
\end{eqnarray}
being $S^{i,j}$ the zero modes of the Lorentz group generators,
i.e:
\begin{eqnarray}
S^{i,j}=-\frac{i}{2} [\psi^i_0,\,\psi^j_0]~~~,~~~
\sqrt{2}\psi_{0}^{i} \equiv \Gamma^i ~~.
\end{eqnarray}
By introducing the operators $N_{i}$ given in Eq. (\ref{ni56bis})
it is straightforward to verify that $N_i\equiv S^{2i,2i+1}$. In
conclusion we get:
\begin{eqnarray}
I_{9-p'}|s_0\dots s_4\rangle_{-1/2}&=& \prod_{i=(p'+1)/2}^4(\pm 2
iN_i)|s_0\dots
s_4\rangle_{-1/2}\nonumber\\
&=&(\pm)^{\frac{9-p'}{2}}\Gamma^{p'+1}\dots\Gamma^9|s_0\dots
s_4\rangle_{-1/2}~~,\label{acr}
\end{eqnarray}
where we have taken into account that the state $|s_0 \dots
s_4\rangle $  is an eigenstate of the operator $N_i$ with
eigenvalue $s_i =\pm 1/2$. We fix the sign ambiguity in the previous
equation by
observing that, as previously asserted, 9-$p'$ T-dualities have to
be equivalent to the action of $\Omega'I_{9-p'}$.
This means that :
\begin{eqnarray}
(T_{9-p'})^{-1}\Omega'_9T_{9-p'}=\Omega'_{p'} I_{9-p'}
\Longrightarrow -= -(\pm)^{(9-p')/2} (-)^{(9-p')/2}~~,\label{siga}
\end{eqnarray}
where, as explained before, we have neglected the action of
$(-1)^{F_L}$ and denoted
by $\Omega'_A$ the expression of the
world sheet operator given in Eq. (\ref{ramo23}) taken with $A=9$
and $A=p'$.
{From} Eq. (\ref{siga}), we see that the action of $I_{9-p'}$
given in Eq. (\ref{acr}) is compatible with the action of
$\Omega'$ given in Eq. (\ref{ramo23}), only if  Eq. (\ref{acr}) is
taken with the minus sign.

Eq.s (\ref{tracea}), (\ref{tro}), (\ref{trof}),
(\ref{trozm}) and (\ref{trofzm}) are modified as follows ($p$ and $p'$ are odd, with $p' \geq p$):
\begin{eqnarray}
{\rm Tr} \left[ e^{-2\pi\tau (N_{X}+N_{bc})} \Omega' I_{9-p'}\right]&=&
(ik)^{2/3} 2^{\frac{p'-p}{2}}f_2^{p-p'}(ik) f_1^{(p'-p)-8}(ik) {\cdot}
k^{-2} \nonumber \\
 {\rm Tr}_R^{nzm}\left[e^{-2\pi\tau (N_\psi + N_{\beta \gamma})}
\Omega' I_{9-p'}  (-1)^{G_{\beta \gamma}}\right] & = & \frac{(ik)^{-2/3}}{
2^{[8-(p'-p)]/2}} f_2(ik)^{8-(p'-p)}f_1(ik)^{p'-p}{\cdot} k^2
\label{tr2} \nonumber \\
{\rm Tr}_R^{nzm}\left[e^{-2\pi\tau (N_\psi + N_{\beta \gamma})}
\Omega' I_{9-p'}(-1)^F\right]& =& \frac{(ik)^{-2/3}}{ 2^{(p'-p)/2}}
f_1(ik)^{8-(p'-p)}f_2(ik)^{p'-p}{\cdot} k^2 \nonumber  \\
{\rm Tr}_R^{z.m.}\left[\Omega'I_{9-p'}(-1)^{G^{0}_{\beta \gamma}}\right]
&=& -2^4 \delta_{p,p'}\nonumber\\
{\rm Tr}_R^{z.m.}\left[\Omega'I_{9-p'}(-1)^{F_0}\right]& =& -2^4
\delta_{|p-p'|,8}~~.\label{trozmb}
\end{eqnarray}
The last two equations have been derived in Eq.s (\ref{trozm1})
and (\ref{trozm2}). Inserting Eq. (\ref{trozmb}) in Eq.
(\ref{frerI}) one gets:
 \begin{eqnarray}
Z_{R;\Omega'I_{9-p'}}^{o}& = &
\frac{V_{p+1}}{4(8\pi^2\alpha')^{\frac{p+1}{2}}} \mbox{Tr}
\left[\gamma_{\Omega'I_{9-p'}}^T\gamma_{\Omega'I_{9-p'}}^{-1}\right]
\nonumber\\
&\times&\int_{0}^{\infty} \frac{d\tau}{\tau}\tau^{-\frac{p+1}{2}}
e^{-\frac{y^2\tau}{2\pi\alpha'}} \left[
\delta_{p, p'}\left(\frac{
f_2(ik)}{f_1(ik)}\right)^8 + 2^4\delta_{p', p+8} \right]~~.
\label{freeom1}
\end{eqnarray}
where  $\gamma_{\Omega'I_{9-p'}}$ is the matrix representing the
orientifold action on the Chan-Paton factors. The explicit form of
such a matrix can be determined by first observing that, in order to
take into account the brane images under $\Omega'I_{9-p'}$,
the Chan-Paton factors have to be  $2N\times 2N$ matrices. This implies
that also $\gamma_{\Omega'I_{9-p'}}$ has to be a $2N\times 2N$ matrix.
Furthermore, following Ref. \cite{GP9601},  $\gamma_{\Omega'I_{9-p'}}$  has
to be  symmetric or  antisymmetric, i.e:
\begin{eqnarray}
\mbox{Tr}
\left[\gamma_{\Omega'I_{9-p'}}^T\gamma_{\Omega'I_{9-p'}}^{-1}\right]
=\pm 2N ~~.\label{trcpg}
\end{eqnarray}
We can rewrite Eq. (\ref{freeom1}) in
the closed string channel by using Eq.s (\ref{f1f2}) getting:
\begin{eqnarray}
Z_{R;\Omega'I_{9-p'}}^{c}&=& 2^{p' -4}\, \frac{V_{p+1}}
{2(8\pi^2\alpha')^{\frac{p+1}{2}}} \mbox{Tr}
\left[\gamma_{\Omega'I_{9-p'}}^T\gamma_{\Omega'I_{9-p'}}^{-1}\right]\nonumber\\
&\times&\int_{0}^{\infty}  \frac{dt}{t^{(9
-p')/2}}\,e^{-\frac{y^2}{8\pi\alpha't}}
\left[
\delta_{p, p'}\left(\frac{
f_2(iq)}{f_1(iq)}\right)^8 + \delta_{p' , p+8} \right]~~.
\label{freeomf1}
\end{eqnarray}
The boundary state in type $0'$ string, as in type I theory, may
be written as the sum of the usual boundary state describing a
D$p$ brane and of the crosscap state which gives a microscopic
description of the orientifold fixed plane. The M{\"{o}}bius
contribution in the open string vacuum amplitude corresponds, in
fact, in the closed channel to the interaction of the boundary
with the crosscap. But, we have already seen that the contribution
of the M{\"{o}}bius strip in the NS-NS sector is zero. Therefore
we can write:
\begin{eqnarray}
|Dp \rangle_{NS-NS} = |Bp \rangle_{NS-NS} \qquad |Dp\rangle_{R-R}
=|Bp \rangle_{R-R} + | Cp' \rangle_{R-R} \label{b+c}~,
\end{eqnarray}
where the expression of the crosscap describing the orientifold
$p'$-plane is for $p'$ odd:
\begin{eqnarray}
|Cp' \rangle_{R-R}=\frac{1}{2}\left[ |Cp',\,+\rangle_{\rm R-R} +
(-1)^{\frac{9-p'}{2}} |Cp',\,-\rangle_{\rm R-R}\right],
\label{bscrp}
\end{eqnarray}
with \cite{tesia}
\begin{eqnarray}
&&|Cp',\,\eta\rangle=|{Cp'}_X\rangle\,|{Cp'}_{\psi},\,\eta\rangle_{\rm
R-R}\, |{Cp'}_{gh,\,sgh}\rangle \label{cb}\\ &&|Cp_X\rangle= 2^{p'
-4}
\, \frac{{\hat{T}}_{p'}}{2}
\prod_{n=1}^{\infty} e^{-\frac{(-1)^n}{n} \alpha_{-n}
{\cdot}S{\cdot}\tilde\alpha_{-n}} \delta^{9-p'} ( {\hat{q}} - y
)|0\,,0\rangle
\label{cmatt}\\
&&|{Cp'}_{\psi},\,\eta\rangle_{R-R}=\prod_{n=1}^{\infty} e^{i
\eta(-1)^n \psi_{-n}
{\cdot}S{\cdot}\tilde{\psi}_{-n}}|{Cp'}_{\psi}, \eta
\rangle^{(0)}_{R-R}~~,\label{cferm}
\end{eqnarray}
where $ S_{\mu\nu}= (\eta_{\alpha\beta},\,-\delta_{ij})$ with
$\alpha,\beta=0\dots p'$, $i,j=p'+1 \dots 9$ and
\begin{eqnarray}
|{Cp'}_{\psi},0\rangle^{(0)}_{\rm R-R} =
\left[C\Gamma^0\dots\Gamma^p\left(1+i\eta
\Gamma^{11}\right)\right]_{AB} |A\rangle|\tilde{B}\rangle~~,
\label{czm}
\end{eqnarray}
where
the part of the boundary states depending on the ghost and
superghost is the standard one with the modifications  discussed in
Sect. \ref{Boundary
  state}.

The normalization of  $|{Cp'}_X\rangle$ can be fixed
through the open/closed string duality, by requiring the amplitude
${}_{R-R}\langle Bp|\Delta|Cp' \rangle_{R-R} + {}_{R-R}\langle
Cp'| \Delta |Bp \rangle_{R-R}$, where $\Delta$ is the closed
string propagator, to reproduce Eq.(\ref{freeomf1}). In
particular, the normalization in Eq. (\ref{cmatt}) has been
obtained by taking in Eq. (\ref{freeomf1}) the symmetric choice
for the matrices representing the orientifold action on the
Chan-Paton factors. There is, however, an argument that gives the
normalization of the crosscap almost without making any
calculation. In fact, by comparing for instance the second term of
Eq. (\ref{182}), corresponding to the NS $(-1)^F$ spin structure,
with the   second term of Eq. (\ref{182clo}), corresponding to the
RR spin structure, we can get the normalization factor $\frac{
{\hat{T}}_{p'}}{2}$ where ${\hat{T}}_{p'} = \sqrt{\pi} (2 \pi
\sqrt{\alpha'})^{3-p'}$. Then the additional normalization factor
present in boundary state for the crosscap in Eq. (\ref{cmatt})
can be obtained by comparing the first terms in Eq.s
(\ref{freeom1}) and (\ref{freeomf1}). In fact, by comparing these
two terms with those considered above in Eq.s (\ref{182}) and
(\ref{182clo}), we see that we need in the normalization of the
crosscap  an extra factor $[-2^{p' -4}]$~\footnote{The minus sign
can be obtained by comparing Eq. (\ref{cferm}) with the corresponding
one for a D brane given in Eq. (7.232) of Ref.~\cite{DL99121}.}.
This factor precisely agrees with the one derived in Ref.~\cite{MSS}
by observing that the R-R charge of the O9-plane is equal to $-2^5$
times the R-R charge of the D9 brane and that the R-R charge of the O$p$
plane can be obtained  from that of the O9-plane by
performing $(9-p)$ T-dualities. After $(9-p)$ T-dualities the R-R charge
of one of the $2^{9-p}$ orientifold  O$p$-planes will be equal to $-2^5
/2^{9-p} = - 2^{p-4}$ of the R-R charge of a Dp brane that is precisely
the factor that we have obtained by open/closed string
duality~\footnote{We thank J.F. Morales for explaining their
  derivation~\cite{MSS}  to
  us.}. At this point it is clear why we needed to choose the minus
sign in Eq. (\ref{acr}). In fact, had we chosen the opposite sign,
we would not have been able to cancel the phase appearing in the
fourth equation in (\ref{trozmb}) and this would have obliged us to
have this phase in the normalization of the crosscap in
Eq. (\ref{cmatt}). But this is not acceptable because T-duality cannot
change the sign of R-R charge.

{From} the previous equations  we deduce that the boundary state
describing the crosscap in the R-R sector  can be formally
obtained from the boundary of a  Dp$'$ brane  where each
oscillator ($\alpha_n$, $\psi_n$ etc.) is multiplied by a factor
$i^n$ and the overall factor $[-2^{p'-4}]$ must be added by hand.
Keeping in mind these substitutions, we can write the interaction
between the boundary state and the crosscap by slightly  modifying
the expressions that give the interaction between a D$p$ and a
D$p'$ brane. We get:\cite{DL0307,tesia}
\begin{eqnarray}
{}_{R-R}\langle Bp,\,\eta|\Delta|Cp',\,\eta'\rangle_{R-R}=
 2^{p'-4}\, \,V_{{\hat{p}}+1}
\frac{{\hat{T}}_p}{2}\frac{{\hat{T}}_{p'}}{2} \frac{\pi \alpha'}{2}
\int_{0}^{\infty}\!\! dt\,e^{-\frac{y^2}{8\pi\alpha't}}\,
(2 \pi^2 t \alpha')^{-(9- {\tilde{p}})/2} A^{(\eta,\,\eta')},\nonumber
\end{eqnarray}
with
\begin{eqnarray}
A^{(\eta,\,\eta')}& = & 16\, \delta_{\eta\eta',1}\delta_{p,p'}
\prod_{n=1}^{\infty}\!\!\left[\!\frac{ {\rm det} \left( 1+\eta
\eta'S_1 S_2^T (ie^{-\pi t})^{2n} \right)}{{\rm det} \left( 1-S_1
S_2^T (ie^{-\pi t})^{2n} \right)}\frac{  \left( 1- (ie^{-\pi
t})^{2n} \right)^2}{
\left( 1+\eta \eta' (ie^{-\pi t})^{2n} \right)^2}\right]\nonumber\\
&& \hspace*{3cm} +16 \delta_{ |p' -p|,8}\delta_{\eta\eta',-1}~~,\label{ar}
\end{eqnarray}
where  ${\hat{p}} \equiv min (p, p')$ and ${\tilde{p}} \equiv max
(p, p')$.

We have now all the ingredients to compute the interaction between a
D$p$ brane and a crosscap. It is given by:
\begin{eqnarray}
Z^{c} &\equiv& \,_{\rm R-R}\langle Bp|\Delta | Cp' \rangle_{\rm R-R} +
\,_{\rm R-R}\langle Cp'|\Delta |Bp\rangle_{\rm R-R} \nonumber\\
&=& \,_{\rm R-R
}\langle Bp|\Delta| Cp'\rangle_{\rm R-R} + \,_{\rm
R-R}\overline{\langle Bp|\Delta |Cp'\rangle}_{\rm R-R}~~.
\label{zch1}
\end{eqnarray}
We have also taken into account that, due to the structure of the
boundary state, the only non-zero contribution comes from the R-R
sector.

By using Eq. (\ref{ar}) we can separately compute the two
contributions with $p=p'$ and $|p-p'|=8$
 \begin{eqnarray}
\!_{\rm R-R}\langle Bp,\, \eta |\Delta | Cp',\,  \eta'\rangle_{\rm
R-R}&=&  2^{(p-4)}
\frac{V_{p+1}\,N}{(8\pi^2\alpha')^{\frac{p+1}{2} }}
\,\delta_{p,p'}\delta_{\eta\eta',1}\nonumber\\
&\times& \int_0^{\infty} \frac{dt}{t^{\frac{9-
{p}}{2}}}\,e^{-\frac{y^2}{8\pi\alpha't}}
\left(\frac{f_2(iq)}{f_1(iq)}\right)^8 \label{p=p'}
\end{eqnarray}
and
\begin{eqnarray}
\!_{\rm R-R}\langle Bp,\, \eta |\Delta | Cp',\,  \eta'\rangle_{\rm
R-R} & = & - 2^{\frac{p +3p'}{2}  - {\tilde{p}} }
\frac{V_{{\hat{p}}+1}\,N}{(8\pi^2\alpha')^{\frac{p+p'-
{\tilde{p}}+1}{2} }}\, \delta_{|p-p'|,8}\delta_{\eta\eta',-1} \nonumber \\
& \times & \int_0^{\infty}
\frac{dt}{t^{\frac{9-\tilde{p}}{2}}} \,
e^{-\frac{y^2}{8\pi\alpha't}}~~. \label{pp'8}
\end{eqnarray}
In the last two equations we have used the identity:
\begin{eqnarray}
\frac{{\hat{T}}_{p'}}{2} \, \frac{{\hat{T}}_{p}}{2}   \frac{\pi
\alpha'}{2} (2 \pi^2 \alpha')^{-(9- \tilde{p})/2} =
2^{\frac{p+p'}{2}-\tilde{p}}( 8 \pi^2 \alpha'
)^{-(p'+p-\tilde{p}+1)/2}~~.\label{nor89}
\end{eqnarray}
Furthermore, as a consequence of the identities:
\begin{eqnarray}
&&\overline{f}_1(iq)=e^{-i\frac{\pi}{12}} f_1(iq) \qquad
\overline{f}_2(iq)=e^{-i\frac{\pi}{12}} f_2(iq) \nonumber\\
&&\overline{f}_3(iq)=e^{i\frac{\pi}{24}} f_4(iq) \qquad\,\,\,
\overline{f}_4(iq)=e^{i\frac{\pi}{24}} f_3(iq)~~,\label{*f}
\end{eqnarray}
we have:
\begin{eqnarray}
&& \!_{\rm R-R}\langle Bp,\, \eta |\Delta | Cp',\,
\eta'\rangle_{\rm R-R}= \!_{\rm R-R}\langle Cp',\, \eta |\Delta |
Bp,\,  \eta'\rangle_{\rm R-R}~,
\end{eqnarray}
which allows us to write:
\begin{eqnarray}
Z^c & = &\frac{1}{2}\left[  {}_{\rm R-R}\langle Bp,\, + | \Delta |
Cp',\,  +\rangle_{\rm R-R} +(-1)^{9-(p+p')/2} \!_{\rm R-R}\langle
Bp,\, -
| \Delta | Cp',\,  -\rangle_{\rm R-R}\right]\nonumber\\
&& +\frac{1}{2}\left[(-1)^{(9-p')/2}\,_{\rm R-R}\langle Bp,\, +
|\Delta | Cp',\,  -\rangle_{\rm R-R}\right.\nonumber \\
&&+ \left.(-1)^{(9-p)/2}\,_{\rm R-R}\langle Bp,\, -
|\Delta | Cp',\,  +\rangle_{\rm R-R}\right]\nonumber\\
&& = \,\,  2^{\frac{p+3p'}{2} -4 -{\tilde{p}}}  {\times}
\frac{ V_{{\hat{p}}+1}\,N}{ (8\pi^2\alpha')^{\frac{p+ p' -
    {\tilde{p}} + 1}{2} }}\, \nonumber\\
&& \times \int_0^{\infty} \frac{dt}{t^{\frac{9-\tilde{p}}{2}}}
\,e^{-\frac{y^2}{8\pi\alpha't}}\left[ \delta_{p,p'}
\left(\frac{f_2(iq)}{f_1(iq)}\right)^8+ 2^4 \delta_{|p-p'|,8}\right]~~,
\label{dito}
\end{eqnarray}
which reproduces Eq. (\ref{freeomf1}), taken  with $\gamma_{\Omega' I_{9-p'}}$
symmetric, for $p' \geq p$.

In the following we compute the interaction between two crosscaps.
It can be easily extracted from Eq. (\ref{ar}) and one gets:
\begin{eqnarray}
{}_{R-R}\langle Cp', \,\eta|\Delta|Cp',\, \eta'\rangle_{R-R}=
-2^{2(p'-4)} V_{{{p'}}+1} \left(\frac{{\hat{T}}_{p'}}{2} \right)^2
\frac{\pi
\alpha'}{2}\delta_{\eta\eta',1} \nonumber \\
\times \int_{0}^{\infty} \frac{16 \delta_{\eta \eta', 1}
dt}{ (2 \pi^2 t \alpha')^{(9- {{p'}})/2}} \prod_{n=1}^{\infty}
\left(\frac{1+ q^{2n}}{1- q^{2n}} \right)^8~~,\label{crocap}
\end{eqnarray}
that after the transformation $t \rightarrow 2t$ and the use of
the identity given in Eq. (\ref{nor89}) with $p=p'$, it can be
rewritten as follows:
\begin{eqnarray}
{}_{R-R}\langle Cp',\,\eta|\Delta|Cp',\, \eta'\rangle_{R-R} =
-2^{5(p'-5)/2} \, \, \frac{2 V_{{{p'}}+1}\delta_{\eta \eta', 1}}{(
8 \pi^2 \alpha'
)^{(p'+1)/2}} \nonumber \\
\times \int_{0}^{\infty} \frac{dt}{t^{(9-p')/2}} \left[
\left(\frac{f_2 ( q^2)}{f_1 (q^2)} \right)^8 \right]~.
\label{crocap3}
\end{eqnarray}
Taking into account Eq. (\ref{bscrp}) we finally get:
\begin{eqnarray}
{}_{R-R}\langle Cp' |\Delta|Cp'\rangle_{R-R}=- 2^{5(p'-5)/2}
V_{{{p'}}+1}
( 8 \pi^2 \alpha' )^{-(p' +1)/2} \nonumber \\
\times \int_{0}^{\infty} \frac{dt}{t^{(9-p')/2}} \left[
\left(\frac{f_2 ( q^2)}{f_1 (q^2)} \right)^8 \right]~,
\label{crocap4}
\end{eqnarray}
that transformed in the open string channel $(t=1/(4 \tau))$
becomes:
\begin{eqnarray}
Z_{KB} = -2^{(3p' - 19)/2}
\,\frac{V_{p'+1}}{(8\pi^2\alpha')^{(p'+1)/2} } \int_{0}^{\infty}
\frac{d\tau}{\tau^{(p'+3)/2}} \left[  \left(\frac{f_4
(k^2)}{f_1(k^2)}\right)^8 \right] \, ~~.\label{KBre1}
\end{eqnarray}
For $p' =9$ it reproduces Eq. (\ref{kbo'1}). The previous equation
can also be derived in the open string channel by computing the
quantity:
\begin{equation}
Z_{KB}=\int_{0}^{\infty} \frac{d\tau}{\tau} {\rm Tr} \left[
\frac{\Omega'I_{9-p'}}{2} e^{-2\pi\tau(L_0+{\tilde L}_0)}
(-1)^{G_{bc}+{\tilde G}_{bc}+G_{\beta\gamma}+{\tilde
G}_{\beta\gamma}} P_{GSO} \right]~~,\label{KBp}
\end{equation}
where the GSO projection is defined in Eq. (\ref{gso}).
One must use again Eq.s (\ref{traceos}), while Eq. (\ref{tracemo})
becomes
\begin{eqnarray}
\int\frac{d^{10}p}{(2\pi)^{10}}\,Tr[\Omega'I_{9-p'}
e^{-\alpha'\pi\tau{\hat p}^2}]= \int\frac{d^{p' +1}p}{(2\pi)^{p'
+1}}\delta^{p' +1}(0) e^{-\alpha'\pi\tau{p}^2} \nonumber\\
\times  \int d^{9-p'}
p_{\perp}  \delta^{9-p'} ( 2 p_{\perp}) ~~,\label{zebo1}
\end{eqnarray}
where the presence of two different delta functions is due to the
action of the parity operator $I_{9-p'}$ that changes sign to $9-p'$
components of the momentum. By computing the integrals we get:
\begin{eqnarray}
\int\frac{d^{10}p}{(2\pi)^{10}}\,Tr[\Omega'I_{9-p'}
e^{-\alpha'\pi\tau{\hat p}^2}] = \frac{2^{(p'+1)/2}\,\,2^{p'-9}\,
  V_{p'+1}}{(8\pi^2\tau\alpha')^{(p' +1)/2}}; \nonumber \\
V_{p' +1} \equiv (2\pi)^{p' +1} \delta^{p' +1} (0) ~~.\label{zebo2}
\end{eqnarray}
For what concerns the R-R zero modes they are identically zero.
Indeed, by using Eq. (\ref{omegavuo}), together with the action of
$I_{9-p'}$ on the zero modes given in Ref. \cite{D9804} and that
generalize  Eq. (\ref{acr}) to the closed string case, one gets for $p'$ odd:
\begin{eqnarray}
I_{9-p'} \left(|A \rangle_{-\frac{1}{2}} \otimes | \tilde{B}
\rangle_{-\frac{1}{2}}\right)
 & =& \left(
\Gamma^{11}\Gamma^9...\Gamma^{11}\Gamma^{p' +1}
\right)^{A}_{\,\,\,C} \nonumber\\
&\times&\left( \Gamma^{11}\Gamma^9...\Gamma^{11}\Gamma^{p' +1}
\right)^{B}_{\,\,\,D} | C \rangle_{-\frac{1}{2}} \otimes |
\tilde{D} \rangle_{-\frac{1}{2}} ~~,\label{isei}
\end{eqnarray}
and
\begin{equation}
\Omega' I_{9-p'} \left(|A \rangle_{-\frac{1}{2}} \otimes |
\tilde{B} \rangle_{-\frac{1}{2}}\right) = \left(
\Gamma^{9}...\Gamma^{p' +1} \right)^{A}_{\,\,\,C}
 \left(
\Gamma^{0}...\Gamma^{p'} \right)^{B}_{\,\,\,E} | E
\rangle_{-\frac{1}{2}} \otimes | \tilde{C} \rangle_{-\frac{1}{2}} ~~.
\label{omisei}
\end{equation}
Thus evaluating the trace in the partition function, as showed in
Appendix B, yields:
\begin{equation}
{\rm Tr}_{z.m.\, R-R}[\Omega'I_{9-p'}]= {\rm Tr}_{z.m.\,
  R-R}[\Omega'I_{9-p'}  (-1)^{F+ \tilde F}]=0 \label{zmkb1}~~.
\end{equation}
Thus collecting the contributions of the zero and non-zero modes we
get Eq. (\ref{KBre1}).

We conclude this subsection by noticing that the one-loop open
string diagrams involving a D3 brane dressed with an external gauge
field and a stack of $N$ undressed D3 branes get a non-vanishing
contribution only from the annulus diagram (the M{\"{o}}bius strip
is vanishing for $p=3$ as it follows from Eq.s (\ref{185}) and
(\ref{freeom})) and it is exactly equal to the one given in Eq.
(\ref{zetae}).

\subsection{Type $0'$ theory in the orbifold $\mathbb{C}^2/\mathbb{Z}_2$.}
\label{oneloop} We have seen in Eq. (\ref{beta67}) that the one-loop
$\beta$-function of the world-volume gauge theory supported by a
stack of $N$ D3 branes in type $0'$ string theory is zero. For this
reason, such a brane configuration does not represent an interesting
model for exploring the validity of the gauge/gravity
correspondence. In the following, in order to consider
non-supersymmetric gauge theories having a non vanishing one-loop
running coupling constant, we study type $0'$ string theory on the
orbifold ${\rm I\! R}^{1,5}{\times}\mathbb{C}^2/\mathbb{Z}_2$ and in
this background we consider fractional D3 branes sitting at the
orbifold singularity. The $\mathbb{Z}_2$ group has been already
introduced in Sect. \ref{subsez1}.
The gauge theory living on the fractional D3 branes is the
$\mathbb{Z}_2$ invariant subsector of the open string spectrum
introduced in Sect. \ref{openstringspectrum}. In particular, by
using the orbifold actions defined in Eq.s (\ref{z2z3}) and
(\ref{orbf}), it is possible to see that this spectrum, even if
non-supersymmetric, satisfies the Bose-Fermi degeneracy condition.
Indeed it contains an $SU(N)$ gauge field, two scalars in the
adjoint representation of the gauge group, one Dirac fermion in the
two-index symmetric and one Dirac fermion in the two-index
antisymmetric representation of the gauge group. The one-loop
$\beta$-function is therefore given by:
\begin{eqnarray}
\beta(g)=\frac{g^3}{(4\pi)^2} \left[ -\frac{11}{3}N + 2 \frac{N}{6}
+ \frac{4}{3} \left(\frac{N-2}{2}+\frac{N+2}{2}\right)\right]=-2 N
\frac{g^3}{(4\pi)^2}~~,\label{bfo'or}
\end{eqnarray}
that coincides with the $\beta$-function of $ {\cal N}=2$ Super-Yang
Mills.

In order to check the gauge/gravity correspondence in this case, one
can apply the usual strategy, evaluating the one-loop vacuum
amplitude of an open string stretching between a stack of $N$ D3
fractional branes and a further D3 brane dressed with an $SU(N)$
background gauge field. Notice however that the one-loop vacuum
amplitude, given by
\begin{eqnarray}
Z & = &\!\!\!\!\int_{0}^{\infty} \frac{d \tau}{\tau}
\left[
P_{(-1)^{F_s}} \left(\frac{1+\Omega'}{2}\right)  \left(\frac{e+h}{2}
\right)\,{ e}^{- 2 \pi \tau L_0 } (-1)^{G_{bc}} P_{\rm GSO}
  \right] \nonumber \\
& \equiv & Z_{e}^o + Z_{\Omega'}^o~~,
\label{zf}
\end{eqnarray}
simply reduces to the annulus contribution  $Z_{e}^o$ . Indeed the
M{\"{o}}bius strip  $Z_{\Omega'}^o$ is zero because of the trace on
the Chan-Paton factors. In fact, by using the definition of trace in
Eq. (\ref{trcp}), taken in this case with $j,h=1, \dots, N, N+2,
\dots 2N+1$ and $i,k=N+1,2N+2$~\footnote{Notice that we are
interested in computing the one-loop vacuum amplitude of an open
string stretching between the stack of the $N$ undressed branes
(labelled by $1, \dots, N, $ with their ``$\Omega'$-images''
labelled by $N+2, \dots, 2N+1$) and the dressed brane (labelled by
$N+1,2N+2$).} and specifying Eq.s (\ref{gammaomegap}) and
(\ref{gammagamma}) for $p=3$
\begin{eqnarray}
\gamma_{\Omega'_3} =  \left( \begin{array}{cc} 0 &
 \beta \\ i\beta & 0
\end{array} \right)~~~;~~~
\gamma_{\Omega'_3}^{T} \gamma_{\Omega'_3}^{-1} = \left(
\begin{array}{cc} i&
 0  \\ 0 & -i\end{array} \right)~~,
\label{cpfa76}
\end{eqnarray}
where each element of the previous matrix is actually an $(N+1)
{\times} (N+1)$ matrix, we get
\begin{eqnarray}
{\rm Tr} \left[ \gamma_{\Omega'_3}^T \gamma_{\Omega'_3}^{-1} \right]
=\sum_{i=N+1,\,2N+2}\left[\sum_{h=1}^N+\sum_{h=N+2}^{2N+1}\right]
\left(\gamma_{\Omega'_3}^T\right)_{ih} \left(\gamma_{\Omega'_3}^{-1}
\right)_{hi}
=i-i
=0 .
\end{eqnarray}
In conclusion the one-loop vacuum amplitude reduces to the annulus
contribution:
\begin{eqnarray}
Z= Z_{e}^o+ Z_{h}^o~~,\label{Z2}
\end{eqnarray}
with $Z_e^o$ and $Z_h^o$ exactly given respectively in Eq.
(\ref{zetae}) and in  Eq. (\ref{zetatet}). In fact, the expressions
in Eq. (\ref{Z2}) that follow from Eq. (\ref{zf}) without the term
with $\Omega'$ have an additional factor $1/4$ with respect to the
analogous ones in Eq. (\ref{Z1}). But, on the other hand, the
Chan-Paton factor gives a factor $4N$ as we are now going to show.
{From} Eq. (\ref{planar23}) we get:
\begin{eqnarray}
Tr [ \langle ij | hk\rangle ] & =  &\sum_{i, k =N+1, 2N+2} \left[
\sum_{j,h =
  1}^{N} + \sum_{j,h=N+2}^{2N+1} \right] \delta_{jh} \delta_{ik}\nonumber \\
&=& {\rm Tr}\left[ {\mathbb I}_2\right]{\times}{\rm Tr} \left[{\mathbb
I}_{2N}\right]=  4 N ~~.\label{chanpa34}
\end{eqnarray}
Being the expression for the free energy $Z$ in Eq. (\ref{zf})
exactly equal to the one of type IIB theory on the orbifold
$\mathbb{C}^2/ \mathbb{Z}_2$, discussed in Sect.~\ref{subsez1}, it
is clear that the $\beta$-function for the non-supersymmetric theory
living on $N$ D3 branes of type $0'$ theory on the orbifold
$\mathbb{C}^2 / \mathbb{Z}_2$ is equal to the one of ${\cal{N}}=2$
super Yang-Mills  in agreement with Eq. (\ref{bfo'or}).
Moreover, as in the case of type IIB on $\mathbb{C}^2 /
\mathbb{Z}_2$, the only non trivial contribution to the gauge theory
parameters comes from the massless states propagating in the annulus
without threshold corrections. Analogously, in the closed channel
the only non trivial contribution comes from massless closed string
states propagating in the cylinder, without threshold corrections
and it leads exactly to the right values for the gauge theory
parameters at one loop.

\section{Gauge/Gravity Correspondence in Type IIB
on $\Omega' I_6\! (-1)^{F_L}$
} \label{orienti} In this section we study type 0B string theory in
the orientifold $\Omega' I_6 (-1)^{F_L}$, where $\Omega'$ is the
world-sheet parity operator and ${F_L}$ is the space-time
fermion number operator in the left sector.

This theory has a non trivial background made of an orientifold
fixed plane which is, by definition, the set of the points left
invariant by the combined action of $\Omega'$ and $I_6$. In our case
such a plane is located at $x^4=\dots=x^9=0$.

The world-volume gauge theory of $N$ D3 branes in this orientifold
of type $0$B is an example of ${\cal{N}}=4$ orientifold field theory,
\cite{ASV} which is planar equivalent (i.e. equivalent in the limit
$N\rightarrow\infty$ with $\lambda=g^2_{YM}N$ fixed) to ${\cal
{N}}=4$ super Yang-Mills. After discussing the case of type $0$B on
$\Omega' I_6(-1)^{F_L}$ on flat space,  we consider some orbifolds
of this orientifold and, within this framework, we analyze the
world-volume gauge theory living on a stack of $N$ fractional
branes.
We start with discussing the gauge theory living on $N$ fractional
D3 branes in the orbifold $\mathbb{C}^2/ \mathbb{Z}_2$. In the
planar limit this theory, which is non-supersymmetric,  shows some
interesting common features with ${\cal N}=2$ super Yang-Mills. Then
we drive our attention to the more interesting case of the orbifold
$C^3/(\mathbb{Z}_2{\times} \mathbb{Z}_2)$. Here the gauge theory
living on $N$ fractional branes is the one recently discussed by
Armoni-Shifman-Veneziano, that for $N=3$ reduces to  QCD with one
flavour. In Table 2 the spectrum of the world-volume gauge theory of
$N$ D$3$ branes in type 0B/$\Omega' I_6 (-1)^{F_L}$ and its
orbifolds is summarized.

\subsection{Open and closed string spectrum}
\label{opeclo}

Let us determine the spectrum of the massless open string states
attached to $N$ D3 branes at the orientifold plane. The massless
open string states in type 0 are given by Eq.s
(\ref{boson}), (\ref{boson1} and (\ref{ferm}).
On these states we have then to impose the orientifold projection
and select only those states that are invariant under the action of
$\Omega'I_6$. In the NS-sector we have (in the picture -1):
\begin{eqnarray}
\Omega'I_6 \,\,\,  \psi^{\alpha}_{-1/2} |0, k\rangle\rightarrow -
\psi^{\alpha}_{-1/2} |0, k\rangle~~,~~
 \Omega'I_6 \,\,\,\psi^{i}_{-1/2} |0, k\rangle\rightarrow
-\psi^{i}_{-1/2} |0, k\rangle
\end{eqnarray}
and therefore the invariant states satisfy the constraint:
\begin{eqnarray}
\gamma_{\Omega'I_6}\lambda^T_{A, \phi}\gamma_{\Omega'I_6}^{-1} =
-\lambda_{A, \phi} ~~.\label{bos12}
\end{eqnarray}
By using Eq. (\ref{acr}) with $p'=3$, we can write:
\begin{eqnarray}
I_6|s_0\dots s_4\rangle=e^{\pm i\pi( s_2 +s_3+s_4)}|s_0 \dots
s_4\rangle=
 \prod_{i=2}^{4} ( \pm 2i N_i)|s_0 \dots s_4\rangle \nonumber \\
= \pm \Gamma^4\dots \Gamma^9|s_0\dots s_4\rangle~~.\label{act12}
\end{eqnarray}
As previously asserted, we have to take the minus sign in  Eq.
(\ref{act12}) and therefore in the R-sector we have:
\begin{eqnarray}
\Omega'I_6|s_0\dots s_4\rangle= -|s_0\dots s_4\rangle
\Longrightarrow \gamma_{\Omega'I_6}
\lambda^T_{\psi}\gamma_{\Omega'I_6}^{-1} =-\lambda_{\psi}~~.
\label{ferm12}
\end{eqnarray}
In the previous equation we should also consider the action
of the operator $(-1)^{F_L}$. This is irrelevant or gives an extra
minus sign in the R-sector depending on being the open string
considered respectively the right or left sector of the closed
string. However, it is simple to check that this sign ambiguity is
completely irrelevant in determining the spectrum of the massless
states.

In the last part of this section we determine the orientifold action
on the Chan-Paton factors. First we observe that these
 have to be $2N{\times} 2N$ matrices in order to
take into account their images under $\Omega'I_6$. Furthermore,
following Ref. \cite{GP9601}, they have to satisfy the constraint
$\gamma_{\Omega'I_6}=\pm \gamma_{\Omega'I_6}^T$ that implies
\begin{eqnarray}
 \gamma_{\Omega'I_6}=\left( \begin{array}{cc}
                            0            &\mathbb{I}_{N{\times} N}\\
                           \pm\mathbb{I}_{N{\times} N} &  0
                           \end{array}
                      \right) .
\label{g0i6}
\end{eqnarray}
By substituting Eq. (\ref{g0i6}) in  Eq.s (\ref{bos12}) and
(\ref{ferm12}), one gets for the bosonic and fermionic Chan-Paton
factors the following expressions:
\begin{equation}
\lambda_{A, \phi} = \left( \begin{array}{cc}
                      A    & 0 \\
                      0 &  - A^T   \end{array}   \right)
\,\,\,\,\,\,\,\,\,\,\,\,\,\,\, \lambda_{\psi} = \left(
\begin{array}{cc}
                      0    & B \\
                      \pm B^{*} &  0   \end{array}   \right),
\label{equa46}
\end{equation}
where in the last one we have implemented the hermiticity of the
Chan-Paton factors and the matrix $B$ can be chosen to be either
symmetric or antisymmetric depending on how the sign in Eq.
(\ref{g0i6}) is chosen. In particular the symmetric choice for the
matrices given in Eq. (\ref{g0i6}) leads to fermion in the
antisymmetric representation of the gauge group, while the
antisymmetric one to fermions in the symmetric representation. The
number of bosonic degrees of freedom is $8N^2$ which corresponds
to one gauge boson and six real scalars transforming according to
the adjoint representation of $SU(N)$. In the fermionic sector one
has $8N^2\pm 8N$ corresponding to four Dirac fermions in the
two-index symmetric ($+$) or antisymmetric ($-$) representation.
Notice that the spectrum does not satisfy the Bose-Fermi
degeneracy condition that holds in type $0$ and $0'$ theories. In
this case such a degeneracy is present only in the large $N$
limit.

Moreover the spectrum of this theory has the same bosonic content as
${\cal N}=4$ SYM. This is an example of planar
equivalence~\cite{AA9912,ASV} between a supersymmetric model, the
${\cal N}=4$ SYM, which plays the role of the {\it parent} theory
and a non-supersymmetric one, that is the orientifold $\Omega' I_6
(-1)^{F_L}$ of type $0$B, which is the {\it daughter} theory, the
two being equivalent in the large $N$ limit. In Sect. \ref{1lof}, by
using string techniques, we will explicitly see that in this limit
the two theories have the same $\beta$-function.

Let us consider the closed string spectrum. Since $\Omega'$ leaves
invariant the metric and the dilaton, while it changes sign to the
Kalb-Ramond field, it is easy to see that in the  NS-NS sector the
orientifold projection selects the following states:
\begin{equation}
\phi,\,\,\,g_{\alpha\beta},\,\,\,g_{ij},\,\,\,B_{i\alpha}
\,\,\,\,\,\,\,\,\,{\rm with}\,\,\,\,\,\,\alpha, \beta =0,\dots,3
\,\,\,\,\,\,i,j=4,\dots,9~~,\label{spnsns}
\end{equation}
where $\phi$, $g$ and $B$ are respectively the dilaton, graviton and
Kalb-Ramond fields. In the R-R sector the states which are even
under the orientifold projection are
\begin{equation}
{\rm (R+,R+)} \,\,\,\,\,\,\rightarrow\,\,\,\,\,\, C_0,\,\,\,\,
C_{\alpha i},\,\,\,\, C_{0123},\,\,\,\, C_{\alpha \beta ij
},\,\,\,\, C_{ijhk},\,\,\,\,\label{sprr++}
\end{equation}
\begin{equation}
{\rm (R-,R-)} \,\,\,\,\,\,\rightarrow\,\,\,\,\,\, \bar C_{\alpha
\beta},\,\,\,\, \bar C_{ij},\,\,\,\, \bar C_{\alpha \beta \gamma
i},\,\,\,\, \bar C_{\alpha ijk }\,\,\,\,~~.\label{sprr--}
\end{equation}
The previous results follow because, as explained at the end of Sec.
2,
 $\Omega '$ leaves $C_2$, $\bar C_{0}$ and $\bar C_{4}$ invariant and
changes the sign of $\bar C_{2}$, $C_0$ and $C_4$. Notice that the
R-R 5-form field strength surviving the orientifold projection is
the self-dual one ($dC_4=\,^*dC_4$), while the anti-self dual one
($d\bar C_4=-\,^*d\bar C_4$) is projected out.

\subsection{One-loop vacuum amplitude}

In this section we compute the one-loop vacuum amplitude of open
strings stretching between two stacks of $N$ D3 branes in the
orientifold $\Omega' I_6$,
 the action
of the operator $(-1)^{F_L}$ being irrelevant in the open string
calculation, as previously discussed.

The one-loop open string amplitude gets two contributions, the
annulus $Z_{e}$, which encodes the information about the interaction
between the two stacks of branes, and the M\"{o}ebius strip
$Z_{\Omega'I_6}$ which instead describes the interaction of each
stack of $N$ D$3$ branes with the O$3$-plane:
\begin{eqnarray}
Z^{o}&\equiv &  Z_{e}^o + Z_{\Omega'I_6}^o\nonumber\\
&=&\int_{0}^{\infty} \frac{d \tau}{\tau} Tr_{{\rm
    NS-R}} \left[ \frac{e+\Omega'I_6}{2}\,
\, P_{(-1)^{F_s}} \, (-1)^{G_{bc}}\,
P_{GSO} { e}^{- 2 \pi \tau L_0} \right]~~.\label{Z1bis}
\end{eqnarray}
The annulus contribution is equal to the one in Eq. (\ref{zop89})
with $M=N$ and with an extra factor $1/2$ due to the orientifold
projection. For $p=3$ we get:
\begin{eqnarray}
Z^{o}_e = N^2 \frac{V_{4}}{(8\pi^2\alpha')^{2}}\!\int_{0}^{\infty}
\! \frac{d\tau}{2\tau^{3}}
\left[   \left( \frac{f_3(k)}{f_1(k)}\right)^8\!\!-\left(
\frac{f_4(k)}{f_1(k)}\right)^8\!\!  - \left(
\frac{f_2(k)}{f_1(k)}\right)^8 \right] ~~.\label{182bis}
\end{eqnarray}
In particular it vanishes because of the abstruse identity and this
signals the absence of any force among the branes. The contribution
of the M\"{o}bius strip, corresponding to the insertion of $\Omega'
I_6$ in the trace, is instead non-trivial. Let us first compute, for
such a term, the trace over the Chan-Paton factors. By fixing the
standard normalization  $\langle hk|nm\rangle
=\delta_{kn}\delta_{hm}$, one finds:
\begin{eqnarray}
&&{\rm Tr}^{\rm C.P.} \left[ \langle hk|
\Omega'I_6|ij\rangle\right]= {\rm Tr}\left[\gamma_{\Omega'I_6}^{-1}
\gamma_{\Omega'I_6}^{T}\right]\nonumber\\
&&{\rm Tr}^{\rm C.P.} \left[ \langle hk| \Omega'I_6 (-1)^{F_s}|ij
\rangle\right]= {\rm Tr}\left[\gamma_{\Omega'I_6}^{-1}
\gamma^{-1}_{(-1)^{F_s}} \gamma_{\Omega'I_6}^{T}
\gamma^{T}_{(-1)^{F_s}}\right] ~~.\label{cp03}
\end{eqnarray}
Furthermore, from the explicit form of the matrices introduced in
the last expression and given in Eq.s (\ref{gpfs}) and (\ref{g0i6}),
it is straightforward to check that
\begin{eqnarray}
{\rm Tr}\left[\gamma_{\Omega'I_6}^{-1} \gamma^{-1}_{(-1)^{F_s}}
\gamma_{\Omega'I_6}^{T} \gamma^{T}_{(-1)^{F_s}}\right]= - {\rm
Tr}\left[\gamma_{\Omega'I_6}^{-1} \gamma_{\Omega'I_6}^{T}\right] ~~.
\label{cp103}
\end{eqnarray}
This identity implies that the NS contribution to the free energy
vanishes.

A non-vanishing contribution comes from the R sector, where the
trace over the non-zero modes ($n.z.m.$) gives
\begin{eqnarray}
&&{\rm Tr}_R^{n.z.m.}\left[e^{-2\pi\tau (N_\psi + N_{\beta \gamma})
}\Omega'I_6(-1)^{G_{\beta \gamma}}\right] = \frac{(ik)^{-2/3}}{2^4}
f_2^8(ik){\cdot} k^2\label{tro03}\\ &&{\rm
Tr}_R^{n.z.m.}\left[e^{-2\pi\tau (N_\psi + N_{\beta \gamma})}
\Omega'I_6(-1)^F\right] = (ik)^{-2/3} f_1^8(ik) {\cdot} k^2 ,
\label{trof03}
\end{eqnarray}
while the trace over the zero modes ($z.m.$) is given by:
\begin{eqnarray}
&&{\rm Tr}_R^{z.m.}\left[\Omega'I_6(-1)^{G_{\beta
\gamma}^{0}}\right] = {\rm Tr}_R^{z.m.}\left[(-1)^{G_0}\right]
=- 2^4\label{trozm03}\\
&&{\rm Tr}_R^{z.m.}\left[\Omega'I_6(-1)^{F_0}\right] = {\rm
Tr}_R^{z.m.}\left[(-1)^{F_0}\right]=0 ~~.\label{trofzm03}
\end{eqnarray}
By inserting Eqs. (\ref{cp03}), (\ref{cp103})  and (\ref{tro03})
$\div$ (\ref{trofzm03}) in the term with  $\Omega ' I_6$ in Eq.
(\ref{Z1bis}), we get:
\begin{eqnarray}
Z_{\Omega'I_6}^{o}=\frac{V_{4}}{4 (8\pi^2\alpha')^{2}}\, {\rm Tr}
\left[\gamma_{\Omega'I_6}^T \gamma_{\Omega'I_6}^{-1}\right]
\int_{0}^{\infty} \frac{d\tau}{\tau^3}
\left(\frac{ f_2(ik)}{f_1(ik)}\right)^8~~,\label{freeom03}
\end{eqnarray}
where we should use that ${\rm Tr} \left[\gamma_{\Omega'I_6}^T
\gamma_{\Omega'I_6}^{-1}\right]=\pm 2N$. Notice that, because of the
M\"{o}bius strip contribution, the interaction between the $N$ D$3$
branes and the O$3$-plane in this orientifold does not vanish.

Eq. (\ref{freeom03}) is finite in the infrared limit
$\tau\rightarrow \infty$. In order to analyse its behaviour in the
UV regime we perform the modular transformation $\tau=1/4t$ which
leads  into the closed string channel. We get:
\begin{eqnarray}
Z_{\Omega'I_6}^{c}=\frac{V_{4}}{4 (8\pi^2\alpha')^{2}}\, {\rm Tr}
\left[\gamma_{\Omega'I_6}^T \gamma_{\Omega'I_6}^{-1}\right]
\int_{0}^{\infty} \frac{d t}{t^3}
\left(\frac{ f_2(iq)}{f_1(iq)}\right)^8 ~~.\label{freecm03}
\end{eqnarray}
The previous equation reproduces Eq. (\ref{dito}) with $p=p'=3$ for
the symmetric choice of the Chan-Paton factors. The solution
obtained by taking antisymmetric Chan-Paton factors would correspond to
define the crosscap with the opposite sign with respect to the one
in Eq. (\ref{cmatt}).

It is interesting to observe that the latter expression is invariant
under the open/closed string duality and therefore it is also finite
in the limit $t\rightarrow \infty$. In conclusion  Eq.
(\ref{freeom03}) is well-defined both in the IR and UV regimes and
such a property provides a first hint of the absence of R-R tadpoles
in this orientifold. We will come back on this point later.

The M{\"{o}bius amplitude, together with the third term of Eq.
(\ref{182bis}), gives the total fermionic contribution to the
free-energy which at the massless level reduces to:
\begin{eqnarray}
Z^{o}({\rm femionic\,massless})=-(8N^2\pm 8N) \frac{V_{4}}{
(8\pi^2\alpha')^{2}}\,  \int_{0}^{\infty} \frac{d\tau}{\tau^3}~~.
\label{dof4}
\end{eqnarray}
 As usual, the factor
$(8N^2\pm 8N)$ in front of the previous expression counts the number
of the fermionic degrees of freedom of the world-volume gauge
theory, which indeed agrees with the counting of the previous
subsection. As already noticed, we do not have the same number of
bosonic and fermionic degrees of freedom propagating in the loop
and, in particular, the additional $\pm 8N$ fermionic term  comes
from the M\"{o}bius strip, which therefore is responsible of
spoiling the Bose-Fermi degeneracy of the theory.\cite{BFL9906}
This contribution is subleading in the large $N$ limit.

Notice that Eq. (\ref{freeom03}), apart from the Chan-Paton factors
and the substitution $k\rightarrow ik$, is 1/2 of the free energy,
describing in the R-sector the interaction between two D3 branes in
type IIB string theory. It is, by the way, also equal, through the
previous substitutions, to the correspondent expression in type 0
theory, given in Eq. (\ref{free0pp'}).

The existence of the orientifold plane can be a source of
inconsistency of the background if it generates R-R tadpoles which
are not properly cancelled out. In order to check the consistency of the
background in the present case, we have to analyse the field theory
behaviour of the two closed string diagrams involving the
orientifold plane. These are the M\"{o}bius strip, that describes
the interaction between the branes and the orientifold plane, and
the Klein bottle, which gives the self-interaction of the
orientifold plane. We have already evaluated the M\"{o}bius strip
and seen that it does not lead to any tadpole. In order to be sure
about the consistency of the background, we should also consider the
Klein bottle. It can be obtained from Eq. (\ref{KBre1}) for $p' =3$.
One gets:
\begin{eqnarray}
Z_{KB}=-\frac{2^{-5} V_4}{(8 \pi^2\alpha')^2}\int_{0}^{\infty}
\frac{d\tau}{\tau^3} \left(\frac{f_4( k^2)}{f_1( k^2)}\right)^8.
\label{KBre}
\end{eqnarray}
In order to write the previous expression in the closed string
channel, one has to perform the modular transformation
$\tau\rightarrow \frac{1}{4t}$ and to use the modular transformation
properties of the functions $f_{i}$ obtaining:
\begin{equation}
Z_{KB}=-\frac{ 2^{-5} V_4}{(8\pi^2\alpha')^2}\int_{0}^{\infty}
\frac{dt}{t^3} \left(\frac{f_2( q^2)}{f_1(q^2)}\right)^8,
\label{KBretree}
\end{equation}
which is finite in the limit $t\rightarrow \infty$ leading to no R-R
tadpoles! One can conclude that the O$3$-plane does not generate any
R-R tadpole and therefore the background is perfectly consistent, as
expected, being the space transverse to the orientifold plane non
compact.

\subsection{One-loop vacuum amplitude with
an external field} \label{1lof}

Let us consider, in the open channel, the one-loop vacuum amplitude
of an open string stretching between a D3 brane dressed with a
constant $SU(N)$ gauge field and a stack of $N$ undressed D3 branes.
The gauge field is chosen as in Eq. (\ref{effe}).
The presence of the external field modifies Eq. (\ref{freeom03}) as
follows:
\begin{eqnarray}
Z^{o}(F)_{\Omega'I_6}&=&
-\frac{2\,{\mbox Tr}\left[\gamma_{\Omega'I_6}^T
\gamma_{\Omega'I_6}^{-1}\right]}{(8\pi^2\alpha')^2 } \int d^4x
\sqrt{-{\rm det} (\eta+\hat{F}) } \int_{0}^{\infty}
\frac{d\tau}{\tau} {\rm e}^{-\frac{y^2\tau}{2\pi\alpha'}}
\sin\pi\nu_f \sin\pi\nu_g\nonumber\\
&& \times \frac{ f_2^4(ik) \Theta_2\left(i\nu_f\tau|i\tau +1/2 \right)
\Theta_2 \left(i\nu_g\tau|i\tau+1/2 \right) }{f_1^4(ik)
\Theta_1\left(i\nu_f\tau|i\tau+1/2 \right) \Theta_1
\left(i\nu_g\tau|i\tau+1/2 \right)} ~~.\label{ZF03}
\end{eqnarray}
Notice that in the previous expression we should have put $y=0$
because all branes are located at the orientifold point. However we
keep $y \ne 0$ because it provides a natural infrared cutoff. Eq.
(\ref{ZF03}) describes the M\"{o}bius strip with the boundary on the
dressed brane. The trace over the Chan-Paton factors gives $\pm 2$
counting the dressed brane and its image under $\Omega'I_6$. The
overall factor 2, instead, is a consequence of the fact that in the
trace we have to sum over two different but equivalent open string
configurations: the first one in which only the end-point of the
string parametrized by the world-sheet coordinate $\sigma=0$ is
charged under the gauge group, and the other one in which the gauge
charge is turned on instead at the other end-point at $\sigma=\pi$.

In the last part of this subsection, in order to explore the
gauge/gravity correspondence in this non-supersymmetric model, we
evaluate the threshold corrections to the running coupling constant.

The starting point is again Eq. (\ref{ZF03}) that we now expand up
to the quadratic order in the gauge fields without performing any
field theory limit (more details on the calculation are contained
 \ref{appo1} obtaining $( k = {e}^{- \pi \tau} ))$:
\begin{eqnarray}
\frac{1}{g^2_{\rm YM}} =\pm \frac{1}{(4\pi)^2} \int_{0}^{\infty}
\frac{d\tau}{\tau} e^{-\frac{y^2\tau}{2\pi\alpha'} }
 \left(  \frac{f_2(ik)}{f_1(ik)}\right)^8 \left[\frac{1}{3\tau^2} +k
\frac{\partial}{\partial k} \log f_2^4(ik)\right] ~~,\label{gym03}
\end{eqnarray}
where the upper sign refers to the antisymmetric choice of the
Chan-Paton factors while the lower sign to the symmetric one.

We can now perform the field theory limit corresponding to $\tau
\rightarrow \infty$, $\alpha' \rightarrow 0$ keeping the quantity
$ \sigma = 2 \pi \alpha' \tau$ fixed. In this way from Eq.
(\ref{gym03}) we get:
\begin{eqnarray}
\frac{1}{g^2_{\rm YM}}= \mp \frac{16}{ 3(4\,\pi)^2 }
\int_{1/\Lambda^2}^{\infty} \frac{d\sigma}{\sigma} {\rm
e}^{-{\mu}^2\,\sigma}\,~~,\label{run52}
\end{eqnarray}
where $\Lambda$ is an UV cut-off  and $\mu = \frac{y}{2 \pi
\alpha'}$ is an IR one. By using Eq. (\ref{comple67}) and adding the
contribution of the tree diagrams we get the following expression
for the running coupling constant:
\begin{equation}
\frac{1}{g^2_{\rm YM} (\mu)}= \frac{1}{g^2_{\rm YM} (\Lambda )} \mp
\frac{1}{3\pi^2} \log
 \frac{\mu^2}{\Lambda^2} ~~.
\label{run78}
\end{equation}
Finally from Eq. (\ref{run78}) one reads the expected
$\beta$-function~\cite{BFL9904}
\begin{eqnarray}
\beta(g_{YM})={\pm}\frac{g_{YM}^{3}}{(4\pi)^2} \frac{16}{3} ~~.
\label{beta}
\end{eqnarray}
As already observed, in the planar limit $N\rightarrow\infty$ with
$\lambda=g_{\rm YM}^2 N$ fixed, the ratio $\beta(g_{\rm YM})/g_{\rm
YM}$ reduces to zero and coincides with the one of its parent theory
${\cal N}=4$ SYM.

It is also interesting to write down the one-loop amplitude given by
Eq. (\ref{ZF03}) in the closed string channel by performing the
modular transformation $\tau=1/4t$, as shown in \ref{app0}:
\begin{eqnarray}
Z^{c}(F)_{\Omega'I_6}&=& \pm\frac{1}{(8\pi^2\alpha')^2 } \int d^4x
\sqrt{-{\rm det} (\eta+\hat{F}) } \int_{0}^{\infty} \frac{dt}{t^3}
{\rm e}^{-\frac{y^2}{8\pi\alpha't}}
\sin\pi\nu_f \sin\pi\nu_g\nonumber\\
&&\frac{ f_2^4(iq) \Theta_2\left(\frac{\nu_f}{2}|it +\frac{1}{2}
\right) \Theta_2 \left(\frac{\nu_g}{2}|it +\frac{1}{2} \right)
}{f_1^4(iq) \Theta_1\left(\frac{\nu_f}{2}|it +\frac{1}{2} \right)
\Theta_1 \left(\frac{\nu_g}{2}|it +\frac{1}{2} \right)} ~~.
\label{ZF05}
\end{eqnarray}
Expanding this equation up to the second order in the external field
gives ($q = {e}^{-\pi t}$):
\begin{eqnarray}
\frac{1}{g^2_{\rm YM}} =\pm \frac{1}{(4\pi)^2} \int_{0}^{\infty}
\frac{dt}{4t^3} e^{-\frac{y^2}{8\pi\alpha't} } \left(
\frac{f_2(iq)}{f_1(iq)}\right)^8 \left[\frac{4}{3} + \frac{1}{
\pi}\partial_t \log f_2^4(iq)\right] ~~.\label{gymc03}
\end{eqnarray}
Under the inverse modular transformation $t=1/4\tau$ this equation
perfectly reproduces the expression obtained in the open channel
(Eq. (\ref{gym03})), as one can easily check by using Eq.
(\ref{f1f2}). The field theory limit of the previous expression,
realized as $t\rightarrow \infty$ and $\alpha'\rightarrow 0$ with
$s=2\pi \alpha't$ fixed, gives a vanishing contribution .
One could be led to conclude that the gauge/gravity correspondence
does not work in this non-supersymmetric model. However, in the
planar limit ($N\rightarrow \infty$ and $g_{\rm YM}^2N$ fixed) the
theory has a vanishing $\beta$-function, as already noticed after
Eq. (\ref{beta}). Therefore one can conclude that in the large $N$
limit, where the M\"{o}bius strip contribution is suppressed and
the gauge theory recovers the Bose-Fermi degeneracy in its spectrum,
the gauge/gravity correspondence holds and admits a consistent
interpretation in terms of open/closed string duality.

\subsection{Orbifold $\mathbb{C}^2/\mathbb{Z}_2$ }

The gauge theory living on the fractional D3 branes is the
$\mathbb{Z}_2$ invariant subsector of the open string spectrum
introduced in Sect. \ref{opeclo}. By using Eq.s (\ref{z2z3}) and
(\ref{orbf}) it is easy to see that the spectrum contains one
$SU(N)$ gauge field, two real scalars in the adjoint representation
of the gauge group and two Dirac fermions in the two-index symmetric
(or antisymmetric) representation. Notice that the spectrum has a
{\it common sector} \cite{ASV} with ${\cal N}=2$ SYM, namely the
bosonic one. However, because of the fermionic contributions which
are different, the one-loop $\beta$-function of our theory contains
a subleading correction in $1/N$ with respect to ${\cal N}=2$
$\beta$-function:
\begin{eqnarray}
\beta(g_{YM})= \frac{g_{YM}^{3}}{(4 \pi)^2} \left[- \frac{11}{3} N +
2 \frac{N}{6} + 2 \frac{4}{3} \frac{N\pm2}{2}  \right] =
\frac{g^3_{YM}N}{(4\pi)^2} \left[ -2 \pm\frac{8}{3N}\right].
\label{betaz2}
\end{eqnarray}
In the  large $N$ limit the subleading term in $1/N$  is suppressed
and the two $\beta$-functions coincide. This circumstance signals
the existence of a planar equivalence between the two theories at
one-loop and suggests the possibility of an extension of such
equivalence at all perturbative orders.
\begin{table}[h]
\label{Tabella2}
\caption{Spectrum of the $SU(N)$ world-volume gauge theory
of $N$ D$3$ branes on the top of the O$3$-plane. }
\vskip 0.5cm
{\begin{tabular}
{@{}c|ccc@{}}
{\,\,\,\,\,\,\,\,\,\,\,\,\,\,\,\,\,\,\,\,\,}         &
{\it $0B/\Omega'I_6(-1)^{F_L}$ } &{\it $0B/\Omega'I_6(-1)^{F_L}$}
  & {\it $0B/\Omega'I_6(-1)^{F_L}$ }
\\
\,\,\,\,\,\,\,\,\,\,\,\,\,\,\,\,\,\,\,\,\,&{\it on flat space} &{\it
on} $\mathbb{C}^2/\mathbb{Z}_2$&{\it on}
$\mathbb{C}^3/(\mathbb{Z}_2{\times}\mathbb{Z}_2)$\\
\hline
     {\it Gauge vectors}   & \,\,{Adj ($2N^2$ d.o.f.) }  &  \,\,{Adj ($2N^2$ d.o.f.)} &
      \,\,{Adj ($2N^2$ d.o.f.)}
\\
\hline
{ \it Scalars}     &{ $6 {\times}$ Adj ($6N^2$ d.o.f.)}     & { $2
{\times}$ Adj ($2N^2$ d.o.f.)} &{--}
\\
\hline {\it Dirac fermions} & {$4 {\times} \symm$  or $4{\times}
\asymm$ }& {$2{\times} \symm$ or $2 {\times} \asymm$} &{$\symm$ or
$\asymm$ }
\\\\
 {}&($8N^2{\pm}8N$ d.o.f.)&($4N^2{\pm}4N$ d.o.f.)&($2N^2{\pm}2N$
d.o.f.)\\
\hline {\it 1-loop $\beta$-function}  &
$\pm\frac{g^3_{YM}}{16\pi^2}\frac{16}{3}$& $\frac{N
g^3_{YM}}{16\pi^2}\left(-2{\pm}\frac{8}{3N}\right)$ & $\frac{N
g^3_{YM}}{16\pi^2}\left(-3{\pm}\frac{4}{3N}\right)$
\\\hline
{\it Parent theory} &{${\cal N}=4$ SYM }&{ ${\cal N}=2$
SYM}&{${\cal N}=1$ SYM }\\
\hline
\end{tabular}}
\end{table}
The one-loop vacuum amplitude of an open string stretching between a
stack of $N$ undressed D$3$ branes and a dressed one is given by:
\[
Z =  \int_{0}^{\infty} \frac{d \tau}{\tau} Tr_{{\rm
    NS-R}} \left[ \left(\frac{1 +h}{2}\right) \left(\frac{e+\Omega'I_6}{2}
\right) P_{(-1)^{F_s}}
\right.
\]
\begin{equation}
{\times} \left.  (-1)^{G_{bc}} P_{GSO}
{ e}^{- 2 \pi \tau L_0} \right]
 \equiv  Z_{e}^o + Z_{\Omega'I_6}^o + Z_{he}^o + Z_{h\Omega'I_6}^o~~,
\label{Z1orb}
\end{equation}
where the trace over the Chan-Paton factors has been understood and
we have used the following notation:
\begin{eqnarray}
Z_{e}^o&\equiv&\frac{1}{2}\left(Z_{e}^o+Z_{e(-1)^{F_s}}^o\right)\,\,\,\,\,\,\,\,\,\,\,\,\,\,\,\,\,\,\,
Z_{he}^o\equiv\frac{1}{2}\left(Z_{he}^o+Z_{he(-1)^{F_s}}^o\right)
\nonumber\\
Z_{\Omega'I_6}^o&\equiv&\frac{1}{2}\left(Z_{\Omega'I_6}^o+Z_{\Omega'I_6(-1)^{F_s}}^o\right)
\,\,\,\,\,\,\,\,\,\,\,\,\,\, Z_{h\Omega'I_6}^o\equiv
\frac{1}{2}\left(Z_{h\Omega'I_6}^o+Z_{h\Omega'I_6(-1)^{F_s}}^o\right)\,.
\label{defs}
\end{eqnarray}
However notice that  the term $(-1)^{F_s}$ gives a non vanishing
contribution to the trace on the Chan-Paton factors, only if it
appears together with the projector $\Omega'I_6$, namely in the
terms in the second line  of Eq. (\ref{defs}).

The first two terms of Eq. (\ref{Z1orb}) are the ones we have
already computed in the previous section (apart from an additional
factor $\frac{1}{2}$ coming from the orbifold projection). Here we
need just to evaluate the last two terms. The third term turns out
to be
equal to the one appearing in the pure orbifold calculation given in
Eq. (\ref{zetatet}), as follows from the fact that
\begin{equation}
\label{traccia} {\rm Tr}\langle
ij|e|nm\rangle=\delta_{jj}\delta_{mm}=4N\,\,\,\,\,\,\,\,\,\,\,\,
{\rm Tr}\langle ij|(-1)^{F_S}|nm\rangle=0~~,
\end{equation}
where the indices $i,m = 1,...,N, N+3,...,2N+2 $ enumerate
respectively the stack of $N$ undressed branes and their images,
while the indices $j,n = N+1, N+2 $ indicate the dressed brane and
its image.

Analogously, the last term in Eq. (\ref{Z1orb}) can be obtained from
Eq. (\ref{zetatet}) with the substitution $k \rightarrow ik$.
As noticed after Eq. (2.5) of Ref.~\cite{LMP}, in the twisted
sector, only the NS and NS$(-1)^F$ and R$(-1)^F$ sectors contribute
to the amplitude. However, the presence of the type $0$B projector
$\frac{1+(-1)^{F_s}}{2}$ makes the  NS and NS$(-1)^F$ contributions
to vanish because of Eq.s (\ref{cp03}) and (\ref{cp103}). Thus the
only twisted sector which gives a non vanishing contribution to the
interaction is the $R(-1)^F$ which is equal to
\begin{eqnarray}
Z_{h\Omega'I_6}^o= \mp \frac{2i }{32\pi^2}\int d^4x\,
F_{\alpha\beta}^a {\tilde F}^{a\,\alpha\beta} \int_{0}^\infty
\frac{d\tau}{\tau} e^{-\frac{y^2 \tau}{2\pi\alpha'}} ~~.\label{teta2}
\end{eqnarray}
The overall factor 2 again takes into account the two inequivalent
configurations that we have to consider in evaluating the trace, as
discussed after Eq. (\ref{ZF03}).

Then the coefficient of the gauge kinetic term in this theory comes
only from the second and third term of Eq. (\ref{Z1orb})
corresponding to the untwisted M{\"{o}}bius strip and twisted
annulus and turns out to be:
\begin{eqnarray}
\frac{1}{g_{YM}^{2}}& = & - \frac{1}{16 \pi^2}
\int_0^\infty\frac{d\tau}{\tau}e^{-\frac{y^2\tau}{2\pi\alpha'}}
\left\{ \left[ 2N \mp \frac{8}{3} \right] +{\bf \Delta} \right\}~~,
\label{runthreop}
\end{eqnarray}
where the first two terms give  the massless states contributions to
the running coupling constant coming from the twisted annulus
(leading term in $N$) and the untwisted Moebius strip (subleading
term in $N$). The last term, given by:
\begin{eqnarray}
{\bf \Delta}= \pm \frac{1}{2}\left[ \frac{16}{3} -\left(
\frac{f_2(ik)}{f_1(ik)}\right)^8 \left(\frac{1}{3\tau^2} +
k\partial_k \log f_2^4(ik)\right) \right]~~,\label{runthreop1}
\end{eqnarray}
contains the threshold correction due to the massive string states,
which are subleading in $N$ because they come entirely from the
untwisted Moebius strip.

In the field theory limit only the massless states contributions
survive and one gets:
\begin{equation}
\frac{1}{g_{YM}^{2}} = \frac{1}{16 \pi^2} \left[ 2N \mp \frac{8}{3}
\right] \log
  \frac{\mu^2}{\Lambda^2}~~,
\label{runorior}
\end{equation}
consistently with our previous calculation in Eq. (\ref{betaz2}).
Moreover from Eq.s (\ref{zetatet}) and (\ref{teta2}), following the
same procedure as in Ref. \cite{LMP}, one can read also the $\theta$
angle which receive contributions only from the massless states
propagating in the twisted annulus and Moebius strip and turns out
to be
\begin{eqnarray}
\theta_{YM} = -2\theta(N\pm 2)~~,\label{theta45bis}
\end{eqnarray}
where  $\theta$ is the phase of the complex cut-off $\Lambda
e^{-i\theta}$. Notice that, differently from what happens for the
running coupling constant, neither the leading nor the subleading
term in $N$ are affected by threshold corrections.

The gauge theory so obtained shares some common features with ${\cal
N}=2$ SYM. As previously noticed, the running coupling constant and
the $\beta$-function of this theory, in the large $N$ limit,
reproduce those of ${\cal N}=2$ SYM. Moreover also the $\theta_{YM}$
angle in Eq. (\ref{theta45}) in the large $N$ limit reduces to the
one of ${\cal N}=2$ SYM, implying that, in the planar limit, the two
theories are very close to each other. This connection appears as
the natural extension to the case ${\cal N}=2$ of the Armoni,
Shifman and Veneziano planar equivalence for ${\cal N}=1$, in which
the {\em parent} theory is the ${\cal N}=2$ SYM   and the {\em
daughter} theory is the world-volume theory of $N$ fractional branes
of our orbifold.

Finally, it is useful to rewrite the previous expressions in the
closed string channel.   By using Eq.s (\ref{mod1}) and (\ref{the2})
and the well-known modular transformation properties  of the
$\Theta$-functions, we can rewrite Eq.s (\ref{zetatet}),
(\ref{teta2}) and $Z^o_{\Omega'I_6}$ in the closed string channel.
The other terms in the free energy are irrelevant in the forthcoming
discussion because they are vanishing in the field theory limit.
{From} the annulus we obtain:
\begin{eqnarray}
Z_{he}^{c} & = & \frac{N}{(8 \pi^2 \alpha')^2} \int d^4x \sqrt{
-\mbox{det}(\eta+\hat{F})} \int_{0}^\infty \frac{dt}{t} {\rm
e}^{-\frac{y^2}{2\pi\alpha't}} \left[ \frac{ 4\, \sin \pi \nu_f
\sin \pi \nu_g} {\Theta_{4}^2(0|it) \Theta_{1}(\nu_f|it)\Theta_{1}
(\nu_g|it)} \right]
\nonumber \\
{} && \left[ \Theta_{2}^2(0|it)\Theta_{3}(\nu_f |i t)
\Theta_{3}(\nu_g |i t) - \Theta_{3}^2(0|i t) \Theta_{2}(\nu_f |i t)
\Theta_{2}(\nu_g |i t) \right]
\nonumber \\
&&- \frac{iN}{32\pi^2}\int d^4x\, F_{\alpha\beta}^a {\tilde
F}^{a\,\alpha\beta} \int_{0}^\infty \frac{d t}{t} e^{-\frac{y^2
}{2\pi\alpha't}}~~,\label{zetatetc}
\end{eqnarray}
while from the M\"{o}bius strip:
\begin{eqnarray}
Z^{c}_{\Omega'I_6}(F)&=& \mp\frac{1}{(8\pi^2\alpha')^2 } \int d^4x
\sqrt{-{\rm det} (\eta+\hat{F}) } \int_{0}^{\infty} \frac{dt}{4t^3}
{\rm e}^{-\frac{y^2}{8\pi\alpha't}}
\sin\pi\nu_f \sin\pi\nu_g\nonumber\\
&&\frac{ f_2^4(iq) \Theta_2\left(\frac{\nu_f}{2}|it +\frac{1}{2}
\right) \Theta_2 \left(\frac{\nu_g}{2}|it +\frac{1}{2} \right)
}{f_1^4(iq) \Theta_1\left(\frac{\nu_f}{2}|it +\frac{1}{2} \right)
\Theta_1 \left(\frac{\nu_g}{2}|it +\frac{1}{2} \right)}\label{zoi6}\\
Z_{h\Omega'I_6}^c(F)&= &\mp  \frac{2i }{32\pi^2}\int d^4x\,
F_{\alpha\beta}^a {\tilde F}^{a\,\alpha\beta} \int_{0}^\infty
\frac{d t}{t} e^{-\frac{y^2}{8\pi\alpha't}} \,\, ~~.\label{teta2cl}
\end{eqnarray}
By expanding the previous expressions up to quadratic terms in the
external field and isolating only the terms depending on the gauge
field, we have from the annulus:
\begin{eqnarray}
Z_h^c(F)\!\!&\rightarrow&\!\!\left[- \frac{1}{4} \int d^4 x
F_{\alpha \beta}^{a} F^{a\,\alpha \beta }\right]
 \left\{
- \frac{N}{8 \pi^2} \int_{0}^{\infty} \frac{dt}{t} {\rm
e}^{-\frac{y^2}{2\pi\alpha't}}
 \right\}  \nonumber\\
&-&  iN \left[ \frac{1}{32\pi^2} \int d^4x F^a_{\alpha\beta}\tilde
F^{a\,\alpha\beta} \right] \int_{0}^{\infty} \frac{dt}{t}{\rm
e}^{-\frac{y^2}{2\pi\alpha't}} ~~.\label{F285bis}
\end{eqnarray}
The last two equations are exact at string level even if they
receive a contribution only from the massless closed string states.
For $y=0$ both of them are left invariant  under open/closed string
duality  and for this reason  one expects to obtain, from the closed
channel, the planar contribution to the $\beta$-function and the
complete expression of the $\theta$-angle.

By expanding  the Moebius strip amplitude to the second order in the
background field, one gets:
\begin{eqnarray}
& Z_{\Omega'I_6}^c(F)  \rightarrow  \left[-
\frac{1}{4}\! \int\! d^4 x F_{\alpha \beta}^{a} F^{a\,\alpha \beta
}\right] & \nonumber \\
&\times  \left\{\! \pm \frac{1}{(4\pi)^2}\! \int_{0}^{\infty}
\frac{dt}{8t^3} e^{-\frac{y^2}{8\pi\alpha't} } \left(
\frac{f_2(iq)}{f_1(iq)}\right)^8 \left[\frac{4}{3} + \frac{1}{
\pi} \partial_t \log f_2^4(iq)\right]\!\! \right\}~~.&
\label{teta2cl1}
\end{eqnarray}
In this case the massless pole in the open channel is not left
invariant under open/closed string duality and  by performing the
field theory limit on such expression, we obtain a vanishing result.

Summarizing, from Eq.s (\ref{F285bis}) and (\ref{teta2cl1}) we get
the following expression for the running coupling constant at the
full closed string level:
\begin{eqnarray}
\frac{1}{g_{YM}^{2}} & = &
- \frac{N}{8 \pi^2} \int_0^\infty\frac{dt}{t}e^{-\frac{y^2}{2\pi\alpha't}}\nonumber\\
&& \pm\frac{1}{16 \pi^2}
\int_0^\infty\frac{dt}{8^3}e^{-\frac{y^2}{8\pi\alpha't}} \left(
\frac{f_2(iq)}{f_1(iq)}\right)^8 \left[\frac{4}{3} + \frac{1}{ \pi}
\partial_t \log f_2^4(iq)\right]~~,\label{runthrecl}
\end{eqnarray}
where the first line corresponds to the massless states contribution
coming from the twisted cylinder $Z_{h}^c$, while the second line
corresponds to the threshold corrections coming from
$Z_{\Omega'I_6}^c$ which are subleading in $N$. By performing the
field theory limit, only the first line survives and therefore one
can conclude that the closed channel is able to reproduce only the
contribution of the planar diagrams to the $\beta$ function in Eq.
(\ref{betaz2}), but not also the contribution of the non planar
ones.

In conclusion, we can assert that the gauge/gravity correspondence
certainly holds in the planar limit. However, some non planar
information can  be still obtained from the closed channel as the
example of the $\theta$-angle has showed.

\subsection{Orbifold ${\mathbb C^3}/(\mathbb{Z}_2{\times}
  \mathbb{Z}_2$)}

Let us consider now the orbifold $\mathbb{C}^3/(\mathbb{Z}_2{\times}
\mathbb{Z}_2)$ of our orientifold theory. The action of this
orbifold has been discussed in Section 3.

The states left invariant are one gauge vector and one Dirac fermion
in the two-index symmetric (or anti-symmetric) representation of the
gauge group. Also in this case the theory has a common bosonic
sector with a supersymmetric model, that is ${\cal N}=1$ SYM. It is
simple to check that the $\beta$-function for this theory is, at
one-loop:
\begin{eqnarray}
\beta(g_{YM})= \frac{g_{YM}^{3}}{(4 \pi)^2} \left[- \frac{11}{3}N +
\frac{4}{3} \frac{N \pm 2}{2} \right]= \frac{g^3_{YM}N}{(4\pi)^2}
\left[ -3\pm \frac{4}{3N}\right] ~~,\label{betaz2xz2}
\end{eqnarray}
which differs from the one of ${\cal N}=1$ SYM because of the
subleading term in $1/N$.

Notice that for $N=3$ the two-index antisymmetric representation is
equal to the fundamental one.
Therefore, with the antisymmetric choice, {\it the world-volume
gauge theory living on a stack of $N$ fractional branes in the
orbifold $\mathbb{C}^3/(\mathbb{Z}_2{\times} \mathbb{Z}_2)$ of the
orientifold type $0$B/ $\Omega' I_6 (-1)^{F_L}$, for $N=3$ is
nothing but one flavour QCD}. This is an alternative and simpler
stringy realization  of the Armoni-Shifman-Veneziano model. In Ref.
\cite{ASV} and references therein, the same gauge theory is
realized, in the framework of type 0A theory, by considering a stack
of $N$ D$4$ branes on top of an orientifold O4-plane, suspended
between orthogonal NS $5$ branes. It would be interesting to exploit
the relation between the two models which should be connected by a
simple T-duality.

Besides the stringy realization, the gauge theory we end up with is
related by planar equivalence to ${\cal N}=1$ SYM. In the language
of Armoni, Shifmann and Veneziano, the symmetric $(+)$ and
antisymmetric $(-)$ choices correspond to the $S$ and $A$
orientifold theories of ${\cal N}=1$ SYM. This opens the way to a
very interesting extension of many predictions of supersymmetric
parent theory to the non-supersymmetric daughter theory, which holds
in the large $N$  limit.\cite{ASV}

As discussed in Sect. 3.3, the orbifold
$\mathbb{C}^3/(\mathbb{Z}_2{\times} \mathbb{Z}_2)$ can be seen as
obtained by three copies of the orbifold $\mathbb{ C}^2/\mathbb{
Z}_2$ where the $i$-th $\mathbb{Z}_2$ contains the elements $(1,
h_i)$ $(i=1, \dots 3)$.

In particular we consider the one-loop vacuum amplitude of an open
string stretching between a stack of $N_I$ ($I=1, \dots, 4$) branes
of type $I$ and a D3-fractional brane of type $I=1$ dressed with an
$SU(N)$ gauge field. In this case the amplitude turns out to be
given by the sum of eight terms:
\[
Z =  \int_{0}^{\infty} \frac{d \tau}{\tau} Tr_{{\rm
 NS-R}} \left[ \left(\frac{1 +h_1+h_2+h_3}{4}\right)
\left(\frac{e+\Omega'I_6}{2} \right)P_{(-1)^{F_s}}\right.
\]
\begin{equation}
{\times} \left.  (-1)^{G_{bc}}P_{GSO} { e}^{- 2 \pi \tau L_0}
\right]
 \equiv  Z_{e}^o + Z_{\Omega'I_6}^o + \sum_{i=1}^{3}
\left[  Z_{h_ie}^o + Z_{h_i\Omega'I_6}^o\right]~~.\label{Z2orb}
\end{equation}
Here the first two terms are the same as the ones of the previous
orbifolds except for a further factor 1/2 due to the orbifold
projection. The terms $ Z_{h_ie}^o$ turn out to be
\begin{eqnarray}
Z_{h_i}^o & = & \frac{f_i(N)}{2\,(8\pi^2\alpha')^2} \int d^4x \sqrt{
-\mbox{det}(\eta+\hat{F})} \int_{0}^\infty \frac{d \tau}{\tau}{\rm
e}^{-\frac{y^2\tau}{2\pi\alpha'}} \frac{4\,\sin\pi\nu_f
\sin\pi\nu_g}{ \Theta_{2}^2(0|i \tau) \Theta_{1}(i \nu_f \tau| i
\tau) \Theta_{1}(i \nu_g \tau|i \tau)}
\nonumber\\
&& \times \left\{\Theta_{3}^2(0|i \tau) \Theta_{4}(i \nu_f \tau|i
\tau)\Theta_{4}(i \nu_g \tau |i \tau) -\Theta_{4}^2(0|i \tau)
\Theta_{3}(i \nu_f \tau|i \tau)
\Theta_{3}(i \nu_g \tau |i \tau)\right\}\nonumber\\
&&-\frac{i\, f_i(N)}{64\pi^2} \int d^4x F^a_{\alpha\beta}\tilde F^{a
\, \alpha\beta}\int_{0}^{\infty} \frac{d\tau}{\tau} {\rm
e}^{-\frac{y^2\tau}{2\pi\alpha'}}~~,
\label{ztet1bis}
\end{eqnarray}
where the  functions $f_i(N)$ are given in Eq. (\ref{coupling}). As
in the previous orbifold case, all the bosonic terms  of
$Z_{h_i\Omega'I_6}$ vanish because of the contribution to the trace
of the  projector $P_{(-1)^{F_s}}$, while the $R(-1)^F$ sector gives
\begin{eqnarray}
Z_{h_i\Omega'I_6}^o= \mp \frac{2i}{64\pi^2} \int d^4x
F^a_{\alpha\beta}\tilde F^{a \, \alpha\beta}\int_{0}^{\infty}
\frac{d\tau}{\tau}{\rm e}^{-\frac{y^2\tau}{2\pi\alpha'}}~~.
\label{teta3}
\end{eqnarray}
By extracting the coefficient of the gauge kinetic term from the
field theory limit of the amplitude in Eq. (\ref{Z2orb}) and
specializing to the case $N_1=N,\,N_2=N_3=N_4=0$ we get:
\begin{equation}
\frac{1}{g_{YM}^{2}} = \frac{1}{16 \pi^2} \left[3 N \mp \frac{4}{3}
\right] \log  \frac{\mu^2}{\Lambda^2} ~~,\label{runorior1}
\end{equation}
while the theta angle $\theta_{\rm YM}$ turns out to be
\begin{eqnarray}
\theta_{\rm YM}=-(N\pm 2)\theta \label{theta451bis} ~~.
\end{eqnarray}
We can repeat the same analysis in the closed string channel by
transforming under open/closed string duality Eq. (\ref{Z2orb}) and
performing all the steps explained in the last subsection. However,
being Eq.s (\ref{ztet1}), (\ref{teta3}) and $Z_{\Omega'I_6}^o$
coincident, apart from an overall factor, with Eq.s (\ref{zetatet})
and (\ref{teta2}) we get the same conclusions as we did in that
subsection. In the closed string channel one is able to capture, in
the field theory limit, only the planar contribution to the
$\beta$-function and the complete expression for the $\theta$-angle.
Gauge/gravity correspondence in these non-supersymmetric models has
a full consistent interpretation in terms of open/closed string
duality only in the large $N$ limit even if some non planar results
are still present in the closed channel.

\section{Conclusion}

In this article we have investigated the conditions that have to be
satisfied for making the gauge/gravity correspondence to be at
work, both in supersymmetric and non-supersymmetric string
theories. The supersymmetric theories, that we have examined, are
${\cal N}$=1,2 SQCD, showing that the gauge/gravity correspondence
is a consequence of the open/closed string duality {\em if} the
threshold corrections, i.e. the contribution of the massive string
states to the  gauge coupling constant and $\theta_{YM}$-angle,
vanish. Indeed when this happens the contribution of the massless
open string states is necessarily mapped, under open/closed string
duality, into the one of the massless closed string states and
this provides the reason of why it is possible to get
gauge-theories quantities from supergravity. This equivalence between
massless states in the two channels has an interesting consequence. In
fact, on the one hand, it is clear from  Eq. (\ref{gau57}) that the
massless open string states generate an UV divergence for small
values of the modular parameter $\tau$. On the other hand,
open/closed string duality together with  the absence
of threshold corrections, transforms such a behaviour in an IR
logarithmic divergence in the closed string channel.
Such a  divergence in the closed string channel
has a precise physical meaning: it corresponds to the propagation in
the two transverse dimensions not included among those on which the
orbifold acts, of some bulk field. We can therefore conclude that
the vanishing of the threshold corrections
determines the propagation in two dimensions of some bulk fields
which contribute, through the holographic identities, to the
gauge-theory parameters. This is exactly what happens in the
supersymmetric models taken in consideration where those bulk fields
are in fact the twisted fields.

The same analysis has then been performed for non-supersymmetric models.
In particular, we have considered D3 branes stuck at the
fixed point of orientifolds of type 0 string theory. The gauge theory
living on the  world-volume
of such branes provides an example of the so-called {\em orientifold field
theories}. The most interesting model  has been obtained by taking fractional
D3 branes in type 0/$\Omega' I_{6} (-1)^{F_{L}}$ in the orbifold
$\mathbb{C}^{3}/(\mathbb{Z}_{2} \times \mathbb{Z}_{2})$.
In fact in this case we get  QCD with one flavour.
In these models we have shown that, for the $\theta_{YM}$-angle, the
threshold corrections are absent, while, for the gauge coupling
constant, they are vanishing  only in the large $N$ limit,
As a consequence, the $\theta$-angle can be exactly obtained from the
closed string channel, while the gauge coupling constant can be
obtained from the closed string channel only for large $N$. Again
their values are given by massless closed string fields propagating in
two dimensions.

\label{conclu}
\begin{table}[h]
\caption{Summary}
\vskip 0.5cm
{\small
{\begin{tabular}
{c|cccc}
THEORY          & GAUGE/GRAVITY               & BOSE/FERMI & NO
FORCE & THRESHOLD
\\
         &                & degeneracy &  & CORRECTIONS
\\ \hline
Bosonic     & No       &  No    & No & $\neq 0$       \\ \hline type
0B & No & No    & No & $\neq 0$
\\ \hline

Dyonic branes  & Yes & Yes    & Yes & $= 0$
\\
in  type 0B &  & & & \\
\hline 0'B & Yes & Yes    & Yes & $= 0$
\\ \hline
0B/$\Omega' I_6(-1)^{F_L}$
 & Yes  & Yes    & Yes
 & $= 0$
\\
($g_{YM}$)& for $N\rightarrow\infty$&for $N\rightarrow\infty$  &for
  $N\rightarrow\infty$ & for $N\rightarrow\infty$\\
\hline 0B/$\Omega' I_6(-1)^{F_L}$
 & Yes & No & Yes
 & $= 0$ \\
 ($\theta_{YM}$)& also for  finite $N$ & for finite $N$ & for finite $N$ &also for finite $N$
\\
\hline
\end{tabular}}}
\end{table}

{\bf{Acknowledgments}} We thank  M. Bill\'{o}, M. Frau, A. Lerda and
I. Pesando for useful discussions and, in particular, A. Lerda for
his help in deriving a boundary state invariant under $\Omega'$. R.M. thanks
also W. M\"uck for discussions.
R.M. and F.P. thank Nordita for their kind hospitality in different
stages of this work. One of us (P.DV.) thanks C{\'{e}}sar G{\'{o}}mez
and the participants at the workshop ``Gravitational Aspects of Branes
and Strings'', held in Feb. 2005 in Miraflores (Spain), for many stimulating
questions during the discussion session that helped to arrive at a more
clear version of this paper.  We thank Adi Armoni for many useful comments on the first version of our paper.

\appendix
\section{$\Theta$-functions}
\label{app0}

The $\Theta$-functions which are the solutions of  the heat
equation:
\begin{eqnarray}
\frac{\partial}{\partial\tau}
\Theta\left(\nu|i\tau\right)=\frac{1}{4\pi} \partial_{\nu}^2
\Theta(\nu|i\tau) \label{iden}
\end{eqnarray}
are given by
\begin{eqnarray} &&\Theta_1(\nu
|it)\equiv\Theta_{11}(\nu,|it) =-2 q^{\frac{1}{4}}\sin\pi\nu
\prod_{n=1}^\infty \left[(1-q^{2n}) (1-e^{2i\pi\nu}q^{2n})
(1-e^{-2i\pi\nu}q^{2n}) \right]
\nonumber\\
&&\Theta_2(\nu |it)\equiv\Theta_{10}(\nu,|it) =2
q^{\frac{1}{4}}\cos\pi\nu \prod_{n=1}^\infty \left[(1-q^{2n})
(1+e^{2i\pi\nu}q^{2n})(1+e^{-2i\pi\nu}q^{2n}) \right]
\nonumber\\
&&\Theta_3(\nu,|it)\equiv\Theta_{00}(\nu,|it) =\prod_{n=1}^\infty
\left[(1-q^{2n})(1+e^{2i\pi\nu}q^{2n-1}) (1+e^{-2i\pi\nu}q^{2n-1})
\right]
\nonumber\\
&&\Theta_4(\nu,|it)\equiv\Theta_{01}(\nu,|it) =\prod_{n=1}^\infty
\left[(1-q^{2n})(1-e^{2i\pi\nu}q^{2n-1}) (1-e^{-2i\pi\nu}q^{2n-1})
\right],
\end{eqnarray}
with $q=e^{-\pi\tau}$. The modular transformation properties of the
$\Theta$ functions are
\begin{eqnarray}
&&\Theta_1(\nu |it) =i\,\Theta_{1}(-i\frac{\nu}{t}
|\frac{i}{t})e^{-\pi\nu^2 /t} t^{-\frac{1}{2}}
\nonumber\\
&&\Theta_{2,\,3,\,4}(\nu|it)
=\Theta_{4,\,3,\,2}(-i\frac{\nu}{t}|\frac{i}{t})e^{-\pi\nu^2/t}
t^{-\frac{1}{2}}~~.\label{modtras}
\end{eqnarray}

It is also useful to define the $f$-functions and their
transformation properties:
\begin{eqnarray}
  \label{f1}
  f_1\equiv q^{{1\over 12}} \prod_{n=1}^\infty (1 - q^{2n})~~;
\end{eqnarray}
\begin{eqnarray}
  \label{f2}
  f_2\equiv \sqrt{2}q^{{1\over 12}} \prod_{n=1}^\infty (1 + q^{2n})~~;
\end{eqnarray}
\begin{eqnarray}
  \label{f3}
  f_3\equiv q^{-{1\over 24}} \prod_{n=1}^\infty (1 + q^{2n -1})  ~~;
\end{eqnarray}
\begin{eqnarray}
  \label{f4}
  f_4\equiv q^{-{1\over 24}} \prod_{n=1}^\infty (1 - q^{2n -1}) ~~.
\end{eqnarray}
In the case of a real argument $q=e^{- \pi t}$ they transform as
follows under the modular transformation $t\rightarrow 1/t$:
\begin{eqnarray}
\label{mtf1} f_1(e^{-\frac{\pi}{t}})=\sqrt{t}f_1(e^{-\pi
t})~~~~;~~~~ f_2(e^{-\frac{\pi} {t}})=f_4(e^{-\pi t})~~~~;
f_3(e^{-\pi t})=f_3(e^{-\frac{\pi}{ t}}),
\end{eqnarray}
while for complex argument one gets:\cite{FGLS}
\begin{eqnarray}
f_1(ie^{-\pi t})=(2t)^{-1/2}f_1(ie^{-\frac{\pi}{4 t}})&&
f_2(ie^{-\pi
  t})=f_2(ie^{-\frac{\pi}{4 t}})
\label{f1f2}
\\
f_3(ie^{-\pi t})=\,\,\,\,\,\,\,e^{i\pi/8}f_4(ie^{-\frac{\pi}{4
t}})&& f_4(ie^{-\pi t})=e^{-i\pi/8} f_3(ie^{-\frac{\pi}{4 t}})~.
\label{f3f4}
\end{eqnarray}
The following relations are also useful:
\begin{eqnarray}
\Theta_{2,3,4} ( 0| it ) = f_1(e^{-\pi t}) \,f_{2,3,4}^{2} (e^{-\pi
t})~~;~~\lim_{\nu \rightarrow 0} \frac{\Theta_{1} (
  \nu | it )}{2 \sin \pi \nu} = - f_{1}^{3}(e^{-\pi t})~.
\label{usere34}
\end{eqnarray}
It is also useful to give an alternative representation of the
$\Theta$-functions:
\begin{eqnarray}
\Theta \left[
\begin{array}{ll}
a \\
b
\end{array}\right]
\left(\nu |t\right) =\sum_{n= -\infty}^{\infty} {e}^{ 2 \pi i \left[
\frac{1}{2} (n +
    \frac{a}{2})^2 t + ( n + \frac{a}{2} )( \nu + \frac{b}{2} \right)]},
\label{theab98}
\end{eqnarray}
where $a,b$ are rational numbers. It is easy to show that
\begin{equation}
\Theta \left[
\begin{array}{ll}
1 \\
1
\end{array}\right]
\left(\nu |t\right) = - i \sum_{n= - \infty}^{\infty} (-1)^n {e}^{i
\pi
  t (n-\frac{1}{2})^2} {e}^{i \pi \nu (2n-1)} \equiv
\Theta_1 \left(\nu |t\right) \label{the198}
\end{equation}
and
\begin{equation}
\Theta \left[
\begin{array}{ll}
1 \\
0
\end{array}\right]
\left(\nu |t\right) =  \sum_{n= - \infty}^{\infty}  {e}^{i \pi
  t (n-\frac{1}{2})^2} {e}^{i \pi \nu (2n-1)} = \Theta_2 \left(\nu |t\right)
\equiv - \Theta_1 \left( \nu + \frac{1}{2} |t\right) .
\label{the298}
\end{equation}
{From} the definition in Eq. (\ref{theab98}) it is easy to derive
the following identities:
\begin{eqnarray}
\Theta \left[
\begin{array}{ll}
a \\
b
\end{array}\right]
\left(\nu+\frac{\epsilon_1}{2}t+\frac{\epsilon_2}{2}|t\right)
=e^{-\frac{i\pi t\epsilon_1^2}{4}} e^{-\frac{i\pi
\epsilon_1}{2}(2\nu+b)} e^{-\frac{i\pi
\epsilon_1\epsilon_2}{2}}\Theta \left[
\begin{array}{ll}
a +\epsilon_1 \\
b +\epsilon_2
\end{array}\right](\nu|t) .
\label{theide98}
\end{eqnarray}
and
\begin{equation} \frac{1}{2}\sum_{a,b=0}^1
(-1)^{a+b+ab}\prod_{i=1}^4\Theta \left[
\begin{array}{ll}
a+h_i \\
b+g_i
\end{array}\right](\nu_i)=-\prod_{i=1}^4\Theta
\left[
\begin{array}{ll}
1-h_i \\
1-g_i
\end{array}\right](\nu'_i)~~,\label{kir}
\end{equation}
with $\sum_i h_i=\sum_i g_i=0$ and  $\nu'_i\equiv
\frac{1}{2}\left(-\nu_i+\sum_{j\neq i}\nu_j\right)$. Eq.
(\ref{theide98}) can be used to write a different expression for the
$\Theta$-function. In fact, by applying such equation with $a=b=0$,
we get:
\begin{eqnarray}
\Theta \left[
\begin{array}{ll}
0 \\
0
\end{array}\right]
\left(\nu+\frac{\epsilon_1}{2}t+\frac{\epsilon_2}{2}|t\right)
=e^{-\frac{i\pi t\epsilon_1^2}{4}} e^{-i\pi \epsilon_1 \nu }
e^{-\frac{i\pi \epsilon_1\epsilon_2}{2}}\Theta \left[
\begin{array}{ll}
\epsilon_1 \\
\epsilon_2
\end{array}\right](\nu|t) ~.
\label{ttf}
\end{eqnarray}
Redefining $\epsilon_1\equiv a$ and $\epsilon_2\equiv b$, we have:
\begin{eqnarray}
\Theta \left[
\begin{array}{ll}
a \\
b
\end{array}\right]\!\!
\left(\nu |i\,t \right)=e^{-\frac{ \pi t a^2}{4}}e^{i\pi
b(\nu+\frac{a}{2})} \prod_{n=1}^{+\infty}\!\!
\left(1-q^{2n}\right)\!\! \left( 1+e^{2i\pi(\nu +\frac{b}{2})}
q^{2n-1+a}\right) \nonumber \\
\times \left( 1+e^{ -2i\pi(\nu +\frac{b}{2})}
q^{2n-1-a}\right)~~.\label{theta3}
\end{eqnarray}
Under an arbitrary modular transformation $t\rightarrow
\frac{at+b}{ct+d}$,  $\Theta_1$ transforms as follows
\begin{equation}
\Theta_1\left(\frac{\nu}{ct+d}|\frac{at+b}{ct+d}\right)
=\eta'\,\Theta_{1}({\nu} |t)e^{i\pi c\nu^2 /(ct+d)}
(ct+d)^{\frac{1}{2}}~~,\label{genmod}
\end{equation}
where $\eta'$ is an eighth-root of unity. It implies the following
transformation of $\Theta_1$:
\begin{eqnarray}
\Theta_1\left(-\frac{\nu}{2} |\frac{t}{4}-\frac{1}{2}\right)
=\frac{1}{\eta'}\,\Theta_{1}\left(-\frac{\nu}{t}
|\frac{1}{2}-\frac{1}{t}\right)e^{-i\pi \nu^2 /t}
\left(\frac{2}{t}\right)^{\frac{1}{2}}~~,\label{mod1a}
\end{eqnarray}
that is obtained from Eq. (\ref{mod1a}) by first making the
 substitutions $ \nu \rightarrow - \frac{\nu}{2}$ and $ t \rightarrow
 \frac{t}{4} - \frac{1}{2}$ and then choosing $ a=d=1, c=2$ and
 $b=0$. By performing in the previous equation the substitution $ t
 \rightarrow 4i t$ we get the following equation:
\begin{equation}
\Theta_1\left(-\frac{\nu}{2} |it -\frac{1}{2}\right)
=\frac{1}{\eta'}\,\Theta_{1}\left( \frac{i\nu}{4t}
|\frac{1}{2}+\frac{i}{4t}\right)e^{-\pi \nu^2 /(4t)}
\left(\frac{1}{2it}\right)^{\frac{1}{2}} ~~.\label{the1a}
\end{equation}
Finally, by using Eq. (\ref{genmod}) with $a=b=d=1$ and $c=0$ we can
write:
\begin{eqnarray}
\Theta_1\left(\nu| t +1\right)=\eta' \Theta_1\left(\nu| t \right) ~.
\label{mod11}
\end{eqnarray}
This latter equation allows us to write
\begin{eqnarray}
\Theta_1\left(-\frac{\nu}{2} |it -\frac{1}{2}\right)=\frac{1}{\eta'}
\Theta_1\left(-\frac{\nu}{2} |it +\frac{1}{2}\right) \label{mod33}
\end{eqnarray}
that, inserted in Eq. (\ref{the1a}), leads to:
\begin{eqnarray}
\Theta_1\left(-\frac{\nu}{2} |it +\frac{1}{2}\right)
=\Theta_{1}\left( \frac{i\nu}{4t}
|\frac{1}{2}+\frac{i}{4t}\right)e^{-\pi \nu^2 /(4t)}
\left(\frac{1}{2it}\right)^{\frac{1}{2}} ~~.\label{mod1}
\end{eqnarray}

In order to get the analogous transformation property of $\Theta_2$,
we use the following relation:
\begin{eqnarray}
\Theta_2\left(-\frac{\nu}{2} |\frac{t}{4}-\frac{1}{2}\right)
=-\Theta_{1}\left(\frac{1-\nu}{2}|\frac{t}{4}-\frac{1}{2}\right) ~.
\label{mod2}
\end{eqnarray}
Then, by applying Eq. (\ref{genmod})  with
$\nu\rightarrow\frac{1-\nu}{2}$,
$t\rightarrow\frac{t}{4}-\frac{1}{2}$ and  $a=d=1,\,c=2,\,b=0$, to
the second term of the previous equation, it can be rewritten as
\begin{eqnarray}
\Theta_1\left(\frac{1-\nu}{2} |\frac{t}{4}-\frac{1}{2}\right)
=\frac{1}{\eta'}\,\left(\frac{2}{t}\right)^{1/2}\,e^{-\frac{i\pi(1-\nu)^2}{
t}}\,
\Theta_{1}\left(\frac{1-\nu}{t}\,|\frac{1}{2}-\frac{1}{t}\right)
\end{eqnarray}
and then, by substituting it in  Eq.(\ref{mod2}) and sending
$t\rightarrow 1 /t$, one gets:
\begin{eqnarray}
\Theta_2\left(-\frac{\nu}{2} |\frac{1}{4t}-\frac{1}{2}\right)
=-\frac{1}{\eta'}\,(2t)^{1/2}\,e^{-i\pi(1-\nu)^2 t}\,
\Theta_{1}\left((1-\nu) t\,|\frac{1}{2}-t\right) ~~.\label{mod2b}
\end{eqnarray}
Let us consider the $\Theta_1$ in the second term of the previous
equation. By defining in it $t'\equiv\frac{1}{2}-t$ and then
$\nu'\equiv -\nu\left(\frac{1}{2}-t'\right)$ we can rewrite it as
\begin{eqnarray}
\Theta_{1}\left((1-\nu)t\,|\frac{1}{2}-t\right)=
\Theta_{1}\left(\nu'-t'+\frac{1}{2}\,|t'\right)~~.\label{mod2cb}
\end{eqnarray}
Therefore, by using Eq. (\ref{theide98}) with $a=b=1$,
$\epsilon_2=1, \epsilon_1=-2$, we can write Eq.(\ref{mod2cb}) as
follows:
\begin{eqnarray}
\Theta_{1}\left((1-\nu) t\,|\frac{1}{2}-t\right)=i e^{i\pi(1-2\nu)
t} \Theta_{2}\left(-\nu t\,|\frac{1}{2}-t\right)~~,\label{mod2bb}
\end{eqnarray}
where we have restored the variables $\nu$ and $t$ and used the
following identity:
\begin{eqnarray}
\Theta \left[
\begin{array}{ll}
-1 \\
\,\,\,\,2
\end{array}\right](\nu|t)=-\Theta \left[
\begin{array}{ll}
1 \\
0
\end{array}\right](\nu|t)=-\Theta_2(\nu|t)~~,
\end{eqnarray}
that can be easily derived starting from the general expression of
the
 $\Theta$-function given in Eq.
(\ref{theab98}). Then by inserting  Eq. (\ref{mod2bb}) in Eq.
(\ref{mod2b}) we get
\begin{eqnarray}
\Theta_2\left(-\frac{\nu}{2} |\frac{1}{4t}-\frac{1}{2}\right)
=-\frac{i}{\eta'}\,\Theta_{2}\left(-\nu t
|\frac{1}{2}-t\right)e^{-i\pi \nu^2 t} (2t)^{\frac{1}{2}} ~~.
\label{mod4}
\end{eqnarray}
Furthermore, by performing the substitution $ t \rightarrow -
\frac{i}{4t}$, Eq. (\ref{mod4})  becomes
\begin{eqnarray}
\Theta_2\left(-\frac{\nu}{2} | {it} -\frac{1}{2}\right)
=-\frac{i}{\eta'}\,\Theta_{2}\left(\frac{i \nu}{4 t} |\frac{1}{2} +
\frac{i}{4t} \right)e^{- \pi \nu^2 /(4t)} (2it)^{-\frac{1}{2}} ~~.
\label{the2a}
\end{eqnarray}
Finally, by rewriting  $\Theta_2$-function in terms of $\Theta_1$ by
means of Eq. (\ref{the298}) and using Eq. (\ref{mod11}), we can
write:
\begin{eqnarray}
\Theta_2\left(-\frac{\nu}{2} | {it} -\frac{1}{2}\right)=
\frac{1}{\eta'} \Theta_2\left(-\frac{\nu}{2} | {it}
+\frac{1}{2}\right) ~~.\label{pippo}
\end{eqnarray}
The last identity allows us to write:
\begin{eqnarray}
\Theta_2\left(-\frac{\nu}{2} | {it} +\frac{1}{2}\right)
=-i\Theta_{2}\left(\frac{i \nu}{4 t} |\frac{1}{2} + \frac{i}{4t}
\right)e^{- \pi \nu^2 /(4t)} (2it)^{-\frac{1}{2}} ~~.\label{the2}
\end{eqnarray}

\section{Derivation of some results}
\label{appo1}

In this Appendix we explicitly derive many equations of the previous
sections.

In order to derive the coefficient of the gauge kinetic term in the
open string channel, we need the following expansions of the
$\Theta$-functions up to the quadratic order in the gauge fields:
\begin{eqnarray}
\Theta_n\left[i\nu_f\tau|i\tau\right]& \simeq&
\Theta_n\left[0|i\tau\right] +2 \frac{\tau^2}{\pi}\partial_\tau
\Theta_n\left[0|i\tau\right]f^2\nonumber\\
&=& f_1(k)f_n^2(k) \left[1- 2\tau^2 k\frac{\partial}{\partial k}
\log [f_1(k)f_n^2(k)]\, f^2\right] \label{thex}
\end{eqnarray}
for  $n=2,3,4$ and
\begin{eqnarray}
&&\frac{\sin\pi\nu_f}{\Theta_1\left[i\nu_f\tau|i\tau\right]}\simeq
\frac{i}{2\tau f_1^3(k)}\left[ 1+\left(\frac{1}{6}+ \tau^2
k\frac{\partial}{\partial k} \log f_1^2(k)\right)f^2\dots\right]
\label{th1ex}\\
&&\Theta_1^2(i\frac{\tau}{2}(\nu_f-\nu_g) |i\tau)\simeq
-\tau^2(if-g)^2 f_1^6(e^{-\pi\tau})\label{exp3}
\end{eqnarray}
for $\Theta_1$, together with
\begin{eqnarray}
\sqrt{ -\mbox{det}(\eta+2\pi\alpha'{F})}=  1 - \frac{1}{2} ( f^2 -
g^2) + \dots  ~~~~;~~~~ f^2 - g^2 = - \frac{(2 \pi \alpha' )^2}{4}
F^2 \label{exp77}
\end{eqnarray}
and
\begin{equation}
\sum_{n=1}^{\infty} \frac{ n q^{2n}}{1 - q^{2n}} =
\sum_{n=1}^{\infty} \frac{ q^{2n}}{(1 - q^{2n})^2} ~~.\label{rel82}
\end{equation}
Eqs. (\ref{thex}) and (\ref{th1ex}) may be proved by using the
following relations:
\begin{eqnarray}
&&\sum_{n=1}^\infty \frac{k^{2n-1}}{(1-k^{2n-1})^2} = -k
\frac{d}{dk}\left[ \frac{1}{2} \log \prod_{n=1}^\infty (1-k^{2n})+
\log \prod_{n=1}^\infty (1-k^{2n-1})\right]~,
\nonumber\\
&&\sum_{n=1}^\infty \frac{k^{2n-1}}{(1+k^{2n-1})^2} = k
\frac{d}{dk}\left[\frac{1}{2} \log \prod_{n=1}^\infty
(1-k^{2n})+\log \prod_{n=1}^\infty
(1+k^{2n-1})\right]~,\nonumber\\
&&\sum_{n=1}^\infty \frac{k^{2n}}{(1+k^{2n})^2}=  k
\frac{d}{dk}\left[\frac{1}{2} \log \prod_{n=1}^\infty
(1-k^{2n})+\log \prod_{n=1}^\infty
(1+k^{2n})\right]~,\nonumber\\
&&\sum_{n=1}^{\infty} \frac{k^{2n}}{(1-k^{2n})^2}= - k
\frac{d}{dk}\left[\frac{1}{2}
\log \prod_{n=1}^\infty (1-k^{2n}) \right]~.\nonumber\\
&& \label{rel1}
\end{eqnarray}

For selecting the coefficient of the gauge-kinetic term in the
closed channel one needs the following expansions of the
$\Theta$-functions at the presence of an external field
\begin{eqnarray}
\Theta_n\left(\nu |it\right)&=
&\Theta_n\left(0 |it\right)
-\frac{2}{\pi}\partial_\tau\Theta_n\left(0 |it\right)\,f^2 \qquad n=2,3,4\nonumber\\
&=&f_1(e^{-\pi t})f_n^2(e^{-\pi
t})\left[1-\frac{2}{\pi}\,f^2\partial_\tau \log\left(f_1(e^{-\pi
t})f_n^2(e^{-\pi t})\right) \right] \label{exf1}
\end{eqnarray}
and
\begin{eqnarray}
\frac{\sin\pi\nu_f}{\Theta_1\left(\nu_f|it\right)}\simeq
-\frac{1}{2f_1^3(q)}\left\{ 1-2f^2 q\partial_q
\log\prod_n\left(1-q^{2n}\right)\right\} \label{exf2}
\end{eqnarray}
for $\Theta_1$, which has been used to obtain Eq. (\ref{cl34}).

The Euler-Heisenberg action in Eq. (\ref{263bis}) is obtained
through the use of the following expressions which hold for
$\tau\rightarrow\infty$ and $\alpha'\rightarrow 0$ :
\begin{equation}
\Theta_1 ( i \nu \tau| i \tau + \frac{1}{2} ) \rightarrow - 2 i
(ik)^{1/4}  \sinh \pi \nu \tau~~,~~ \Theta_2 ( i \nu \tau| i \tau +
\frac{1}{2} ) \rightarrow  2 (ik)^{1/4}  \cosh \pi \nu \tau
\label{the12}
\end{equation}
\begin{equation}
\nu_f \rightarrow - 2 \alpha' i {\hat{f}}~~,~~ \nu_g \rightarrow - 2
\alpha'  {\hat{g}} \label{nuf}
\end{equation}
and
\begin{equation}
f_1 (ik) \rightarrow (ik)^{1/12}~~~,~~~f_2 (ik) \rightarrow \sqrt{2}
(ik)^{1/12} \label{flim89}
\end{equation}
which, together with
\begin{equation}
\sqrt{-{\rm det} (\eta+\hat{F}) } \sin \pi \nu_f \sin \pi \nu_g = i
(2 \pi \alpha')^2 {\hat{f}}{\hat{g}}~~~, \label{equa23}
\end{equation}
lead to Eq. (\ref{263bis}).

In order to derive Eq. (\ref{gym03}) we need to use  the expansions
of the $\Theta$-functions given in Eqs. (\ref{thex}) and
(\ref{th1ex}) in which the second argument is $i\tau+\frac{1}{2}$
instead of $i\tau$. The effect of the previous shift  is simply to
change the argument of the $f_i$-functions in Eqs. (\ref{thex}) and
(\ref{th1ex}) from $k$ to $ik$.
By inserting these equations with an imaginary argument in Eq.
(\ref{ZF03}), we get $(k= {e}^{-\pi \tau})$:
\begin{eqnarray}
Z^F&\simeq& \pm 4 \frac{1}{(8\pi^2\alpha')^2} \int d^4 x
\int_{0}^{\infty} \frac{d\tau}{\tau} e^{-\frac{y^2\tau}{2\pi\alpha'}
} \left[1- \frac{1}{2} (f^2-g^2) \right] \left[\frac{i}{ 2 \tau
f_1^3 (ik)} \right]^2
\left[  \frac{f_2^8(ik)}{f_1^2(ik)}\right]\nonumber\\
&{\times}&\!\!\left[1 + \left(\frac{1}{6} +\tau^2k
    \frac{\partial}{\partial k}
\log  f_1^2(ik) \right) (f^2-g^2)\right] \left[ 1+2(f^2-g^2)\frac{
\tau^2}{\pi} \partial_\tau
\log  \left( f_1(ik) f_2^2(ik) \right) \right]\nonumber\\
&=&\mp \frac{1}{(8\pi^2\alpha')^2} \int d^4 x \int_{0}^{\infty}
\frac{d\tau}{\tau} e^{-\frac{y^2\tau}{2\pi\alpha'} }
\left(  \frac{f_2(ik)}{f_1(ik)}\right)^8\nonumber\\
&{\times}& \left[\frac{1}{\tau^2} + \left(- \frac{1}{3\tau^2} +
\frac{2}{\pi} \partial_\tau\log f_2^2(ik)\right) (f^2-g^2)\right]~.
\end{eqnarray}
{From} it we can easily obtain Eq. (\ref{gym03}) if $f (g) = 2 \pi
\alpha' {\hat{f}} ({\hat{g}})$ is taken into account..

In order to write the amplitude in Eq. (\ref{ZF03}) in the closed
channel we need to perform the modular transformation $\tau=1/4t$
that gives
\begin{eqnarray}
Z^{c}(F)_{\Omega'I_6}&=&\mp \frac{1}{(8\pi^2\alpha')^2 } \int d^4x
\sqrt{-{\rm det} (\eta+\hat{F}) } \int_{0}^{\infty} \frac{dt}{t^3}
\sin\pi\nu_f \sin\pi\nu_g\nonumber\\
&&\frac{ f_2^4(ie^{-\pi t})
\Theta_2\left(i\frac{\nu_f}{4t}|\frac{i}{4t} +\frac{1}{2} \right)
\Theta_2 \left(i\frac{\nu_g}{4t}|\frac{i}{4t}+\frac{1}{2} \right)
}{f_1^4(ie^{-\pi t})
\Theta_1\left(i\frac{\nu_f}{4t}|\frac{i}{4t}+\frac{1}{2} \right)
\Theta_1 \left(i\frac{\nu_g}{4t}|\frac{i}{4t}+\frac{1}{2}\right)} ~~.
\label{ZF04}
\end{eqnarray}

Eq. (\ref{ZF04}) is obtained from Eq. (\ref{ZF03}) by using Eq.s
(\ref{f1f2}) and by changing variable from $\tau$ to $t = \frac{1}{4
\tau}$. Finally, by using Eq.s (\ref{mod1}) and (\ref{the2}) one
gets Eq. (\ref{ZF05}) from Eq. (\ref{ZF04}).

Let us write now the formulas for the $\Theta$-functions that are
needed to get the previous equation. We show that this is equal to
Eq. (\ref{ZF05}).

In order to obtain Eq. (\ref{gymc03}) we have used the following
expansions in the external field
\begin{eqnarray}
\Theta_n\left(\frac{\nu_f}{2} |it+\frac{1}{2}\right)& \simeq&
\Theta_n\left[0|it+\frac{1}{2}\right] -\frac{2}{\pi}\partial_t
\Theta_n\left(0|it+\frac{1}{2}\right)\frac{f^2}{4}\nonumber\\
&=& f_1(iq)f_n^2(iq) \left[1+ 2q\partial_q \log [f_1(iq)f_n^2(iq)]\,
\frac{f^2}{4}\right] \label{thexcl}
\end{eqnarray}
for $n=2,3,4$, and
\begin{eqnarray}
\frac{\sin\pi\nu_f}{\Theta_1\left(\frac{\nu_f}{2}|it+\frac{1}{2}\right)}\simeq
-\frac{1}{f_1^3(iq)}\left\{
1+f^2\left[\frac{1}{8}-q\partial_q\frac{1}{2}
\log\prod_n\left(1-(iq)^{2n}\right)\right]\right\} \label{th1excl}
\end{eqnarray}
for $\Theta_1$.

In the following we give some equations useful in computing the traces over zero
modes in the  R sector ($p'$ odd):
\begin{eqnarray}
{\rm Tr}_R^{z.m.}\left[\Omega' I_{9-p'} (-1)^{G^{0}_{\beta \gamma}} \right]
& = & -{\rm Tr}\left[ \Gamma^{p'+1}\dots \Gamma^9 \Gamma^{p+1}\dots \Gamma^9 \right]
{\rm Tr}\left[ (-1)^{G^{0}_{\beta \gamma}}\right] \nonumber \\
& = &  -2^4 (-1)^{(9-p)/2} \delta_{p, p'}~~,
\label{trozm1}
\end{eqnarray}
while
\begin{eqnarray}
&&{\rm Tr}_R^{z.m.}\left[\Omega' I_{9-p'} (-1)^{F^0}\right]
=-{\rm Tr}\left[ \Gamma^{11}\Gamma^{p'+1}\dots \Gamma^9 \Gamma^{p+1}\dots
\Gamma^9\right]
{\rm Tr}\left[\mathbb{I}_{gh.}\right]\nonumber\\
&&=-\lim_{x \rightarrow 1} \frac{(2i)^4(-2)}{1-x^2}
{\rm Tr}\left[ \prod_{k=0}^4 x^{2N_k}\prod_{k=0}^4 N_k
\prod_{i=(p'+1)/2}^4 (2i N_i)\prod_{i=(p+1)/2}^4 (2i N_i)
\right] \nonumber \\
&& = -2^4  \delta_{|p- p'|,8}~~,
\label{trozm2}
\end{eqnarray}
where we have used Eq.s 9\ref{ni56bis}). The previous equations
have been used for  deriving Eq.s (\ref{trozmb}).

In order to define properly the trace over zero modes in the R-R sector in Eq.
(\ref{zmkb}) and (\ref{zmkb1}), let us start from considering the trace of the
identity matrix $\mathbb I$ in the $2^{d/2}$-dimensional spinor
representation:
\begin{equation}
{\rm Tr}[{\mathbb I}] = 2^{d/2}
\end{equation}
and observe that  $\mathbb I$ can be considered of course as the product
$C^{-1}C$, being $ C$ the charge conjugation operator. Hence we have:
\begin{equation}
2^{d/2}= {\rm Tr}[{\mathbb I}] =
\sum_{A,B} \left( C^{-1} \right)^{AB}
\left( C \right)_{BA}=
\sum_{A,B} <A|B>  \left( C \right)_{BA}~~,
\end{equation}
 where we have used
\begin{equation}
<A|B> = \left( C^{-1} \right)^{AB} ~~.
\end{equation}
This shows that:
\begin{equation}
{\rm Tr}[{\mathbb I}] = \sum_{A,B} <A| {\mathbb I} |B> C_{BA}
\end{equation}
 and therefore for any operator ${\cal O}$ one has:
\begin{equation}
{\rm Tr}[{\cal O}] = \sum_{A,B} <A| {\cal O} |B> C_{BA}~~.
\end{equation}
Let us apply this definition to ${\rm Tr} \left[ \Omega^{'} I_{9-p'} \right]$:
\begin{equation}
{\rm Tr} \left[ \Omega' I_{9-p'} \right] = \sum_{A,B}
<\tilde{C}| <D|  \Omega^{'} I_{9-p'}|A> |\tilde{B}> (C)_{BC}
(C)_{AD} ~~.\label{tra}
\end{equation}
We know that, for $p$ odd, the following equation holds:
\begin{eqnarray}
&&\hspace*{3cm}\Omega' I_{9-p'} \left[ |A>_{-\frac{1}{2}}|\tilde{B}>_{-\frac{1}{2}} \right]  \nonumber \\
&&=\left( \Gamma^{9} \dots \Gamma^{p'+1} \right)^{A}_{F} \left(
\Gamma^{9} \dots \Gamma^{p'+1} \Gamma^{11} \right)^{B}_{E}
|E>_{-\frac{1}{2}} |\tilde{F}>_{-\frac{1}{2}} (C)_{BC}
 (C)_{AD}.
\end{eqnarray}
Hence the trace (\ref{tra}) becomes:
\begin{eqnarray}
{\rm Tr} \left[ \Omega' I_{9-p'} \right] & = & {}_{-\frac{1}{2}}<
\tilde{C}| {}_{-\frac{1}{2}}<D| \left( \Gamma^{9} \dots
\Gamma^{p'+1} \right)^{A}_{F} \left( \Gamma^{9} \dots
\Gamma^{p'+1} \Gamma^{11} \right)^{B}_{E}
|E>_{-\frac{1}{2}} |\tilde{F} >_{-\frac{1}{2}}(C)_{BC} (C)_{AD} \nonumber\\
 & = &   (C^{-1})^{CF}  (C^{-1})^{DE}
 \left( \Gamma^{9} \dots \Gamma^{p'+1} \right)^{A}_{F}
(\left( \Gamma^{9} \dots \Gamma^{p'+1} \Gamma^{11} \right)^{B}_{E}
(C_{BC})(C_{AD}) \nonumber \\
 & = & \left[ \Gamma^{9} \cdots \Gamma^{p'+1} \right]^{A}_{F}
\left[\left(C^{-1}\right)^T\right]^{FC} \left[C^{T}\right]_{CB}
[\left( \Gamma^{9} \dots \Gamma^{p'+1} \Gamma^{11} \right]^{B}_{E}
\left[\left(C^{-1}\right)^T\right]^{ED} \left[ C^{T} \right]_{DA} \nonumber \\
 & = &  {\rm Tr} \left[ \Gamma^{9} \dots \Gamma^{p'+1} (C^{-1})^T C \Gamma^{9} \dots
\Gamma^{p'+1} \Gamma^{11} (C^{-1})^T C \right] \nonumber \\
 & = & {\rm Tr}
\left[ \Gamma^{9} \dots \Gamma^{p'+1} \Gamma^{9}
 \dots \Gamma^{p'+1} \Gamma^{11} \right] = 0~,
\end{eqnarray}
where we have used $ (C^{-1})^T = -C^{-1}$ and
${\rm Tr} \left[ \Gamma^{11} \right] = 0$.

In the following we compute the superghost zero modes contribution to the
interaction between two branes and  we  show that it
coincides with the results known in literature.

The zero-mode part  of the supeghost  boundary state is given by:
\begin{eqnarray}
|B_{\rm sgh},\eta\rangle_{\rm R-R}^{(0)}\!\!&=&\!\!\frac{1}{\sqrt{2}}\left[
 \frac{ e^{i\eta \gamma_0
\tilde\beta_0}}{1+i\eta}|0\rangle_{-\frac{1}{2}}\otimes |\tilde{0}
\rangle_{-\frac{3}{2}} +
\frac{ e^{-i\eta \tilde\gamma_0
\beta_0}}{1-i\eta}|\tilde{0}\rangle_{-\frac{1}{2}}\otimes |0\rangle_{-\frac{3}{2}}
\right]\nonumber\\
&=&\frac{1}{\sqrt{2}}\left[|B_{\rm sgh}^{(1)},\eta\rangle_{\rm R-R}^{(0)}
+|B_{\rm sgh}^{(2)},\eta\rangle_{\rm R-R}^{(0)}\right].
 \label{nbsgzma}
\end{eqnarray}

Consistently, we also have:
\begin{eqnarray}
\,^{(0)}_{\rm R-R}\langle\eta, B_{\rm sgh}|&=&\frac{1}{\sqrt{2}} \left[
{}_{-\frac{1}{2}} \langle \tilde{0}|\otimes {}_{-\frac{3}{2}} \langle{0}|
\frac{e^{-i\eta\beta_0 \tilde{\gamma}_0}}{1-i\eta}+
{}_{-\frac{3}{2}}\langle \tilde{0}| \otimes {}_{-\frac{1}{2}} \langle 0|
\frac{e^{i\eta\tilde{\beta_0} \gamma_0}}{1+i\eta}
\right] \nonumber\\
&=&\frac{1}{\sqrt{2}}\left[\,^{(0)}_{\rm R-R}\langle\eta, B_{\rm sgh}^{(1)}|+
+\,^{(0)}_{\rm R-R}\langle\eta,  B_{\rm sgh}^{(2)}|\right].
\label{nbsgzmb}
\end{eqnarray}
Let us now compute
\begin{eqnarray}
\,^{(0)}_{\rm R-R}\langle\eta_2, B_{\rm sgh}|B_{\rm sgh},
\eta_1\rangle_{\rm R-R}^{(0)}&\equiv
&\frac{1}{2}\lim_{x\rightarrow 1} \left[\,^{(0)}_{\rm
R-R}\langle\eta_2, B_{\rm sgh}^{(1)}| x^{-2\gamma_0\beta_0}
|B_{\rm sgh}^{(1)},\eta_1\rangle_{\rm R-R}^{(0)}\right.\nonumber\\
&+&\left.
\,^{(0)}_{\rm R-R}\langle\eta_2, B_{\rm sgh}^{(2)}| x^{-2\gamma_0\beta_0}
|B_{\rm sgh}^{(2)},\eta_1\rangle_{\rm R-R}^{(0)}\right].
\label{isgh}
\end{eqnarray}
The first contribution to the previous expression is the one given in
Ref.s \cite{DLII} and \cite{BDFLPRS}, i.e:
\begin{eqnarray}
\,^{(0)}_{\rm R-R}\langle\eta_2, B_{\rm sgh}^{(1)}| x^{-2\gamma_0\beta_0}
|B_{\rm sgh}^{(1)},\eta_1\rangle_{\rm R-R}^{(0)}&=&
\frac{1}{1+i(\eta_1-\eta_2)+\eta_1\eta_2}\left(\frac{1}{1+\eta_1\eta_2x^2}\right)\nonumber\\
&=&\frac{\delta_{\eta_1\eta_2;1}}{2(1+x^2)}+\frac{\delta_{\eta_1\eta_2;-1}}
{i(\eta_1-\eta_2)(1-x^2)}~.
\label{isgh1}
\end{eqnarray}
Let us now compute the second term in Eq. (\ref{isgh}):
\begin{eqnarray}
&&\,^{(0)}_{\rm R-R}\langle\eta_2, B_{\rm sgh}^{(2)}| x^{-2\gamma_0\beta_0}
|B_{\rm sgh}^{(2)},\eta_1\rangle_{\rm R-R}^{(0)}\nonumber
\\
&&=
{}_{-\frac{3}{2}}\langle \tilde{0}| \otimes {}_{-\frac{1}{2}}\langle 0|
\frac{e^{i\eta_2\tilde{\beta_0} \gamma_0}}{1+i\eta_2} x^{-2\gamma_0\beta_0}
\frac{ e^{-i\eta_1 \tilde\gamma_0
\beta_0}}{1-i\eta_1}|\tilde{0}\rangle_{-\frac{1}{2}} \otimes |0\rangle_{-\frac{3}{2}}~.
\label{isgh2}
\end{eqnarray}
By observing that:
\begin{eqnarray}
[\gamma_0\beta_0,\,\beta_0^n]=n\beta_0^n
\label{com}
\end{eqnarray}
and using the identity:
\begin{eqnarray}
e^A B e^{-A}=B+[A,\,B]+\frac{1}{2}[A,\,[A,\,B]]+\dots
\label{id0}
\end{eqnarray}
we can compute:
\begin{eqnarray}
x^{-2\gamma_0\beta_0}\beta_0^n x^{2\gamma_0\beta_0}=x^{-2n}\beta_0^n
\label{id1}
\end{eqnarray}
from which it follows:
\begin{eqnarray}
x^{-2\gamma_0\beta_0} e^{-i\eta\tilde{\gamma}_0 \beta_0} x^{2\gamma_0\beta_0}
=e^{-i\eta x^{-2} \tilde{\gamma}_0 \beta_0}~~.
\label{id2}
\end{eqnarray}
Furthermore, since
\begin{eqnarray}
\gamma_0|0\rangle_{-\frac{3}{2}} =0,
\end{eqnarray}
we can write:
\begin{eqnarray}
&&\hspace*{-0.5cm}\,^{(0)}_{\rm R-R}\langle\eta_2, B_{\rm sgh}^{(2)}|
x^{-2\gamma_0\beta_0}
|B_{\rm sgh}^{(2)},\eta_1\rangle_{\rm R-R}^{(0)}\nonumber
\\
&&\hspace*{-0.5cm}=\left( \frac{1}{x^2}\right){}_{-\frac{3}{2}}\langle \tilde{0}|
\otimes {}_{-\frac{1}{2}}\langle 0|
\frac{e^{i\eta_2\tilde{\beta_0} \gamma_0}}{1+i\eta_2} x^{-2\gamma_0\beta_0}
\frac{ e^{-i\eta_1 \tilde\gamma_0
\beta_0}}{1-i\eta_1 }x^{2\gamma_0\beta_0}|\tilde{0}\rangle_{-\frac{1}{2}} \otimes
|0\rangle_{-\frac{3}{2}}
\nonumber\\
&&\hspace*{-0.5cm}=\left( \frac{1}{x^2}\right){}_{-\frac{3}{2}}\langle \tilde{0}|
\otimes {}_{-\frac{1}{2}}\langle 0|
\frac{e^{i\eta_2\tilde{\beta_0} \gamma_0}}{1+i\eta_2}
\frac{ e^{-i\eta_1 x^{-2} \tilde\gamma_0
\beta_0}}{1-i\eta_1 }|\tilde{0}\rangle_{-\frac{1}{2}} \otimes |0\rangle_{-\frac{3}{2}}
\nonumber\\
&&\hspace*{-0.5cm}=\frac{1}{1-i(\eta_1-\eta_2)+\eta_1\eta_2}
\left(\frac{1}{x^2+\eta_1\eta_2}\right)=
\frac{\delta_{\eta_1\eta_2;1}}{2(1+x^2)}+\frac{\delta_{\eta_1\eta_2;-1}}{i(\eta_2-\eta_1)(x^2-1)},
\end{eqnarray}
which concides with Eq. (\ref{isgh1}).

In conclusion we have:
\begin{eqnarray}
\,^{(0)}_{\rm R-R}\langle\eta_2, B_{\rm sgh}| x^{-2\gamma_0\beta_0}
|B_{\rm sgh},\eta_1\rangle_{\rm R-R}^{(0)}=
\frac{\delta_{\eta_1\eta_2;1}}{2(1+x^2)}+\frac{\delta_{\eta_1\eta_2;-1}}{i(\eta_1-\eta_2)(1-x^2)},
\end{eqnarray}
which coincides with the results given in Ref.s \cite{DLII} and
\cite{BDFLPRS}.

Finally, we give the action of the operator $I_{9-p'}$ with $p'$ odd on the boundary
state:
\begin{eqnarray}
&&I_{9-p'}\left[C\Gamma^0\dots \Gamma^{p'}(1+\eta\Gamma^{11})\right]_{AB}
|A\rangle|\tilde{B}\rangle\nonumber\\
&&=
\left[C\Gamma^0\dots \Gamma^{p'}(1+\eta\Gamma^{11})\right]_{AB}
\left( \Gamma^9\Gamma^8\cdots\Gamma^{p'+1}\right)_{\,\,\,D}^A
\left( \Gamma^9\Gamma^8\cdots\Gamma^{p'+1}\right)_{\,\,\,C}^B
|D\rangle|\tilde{C}\rangle\nonumber\\
&&=\left[\left( \Gamma^9\Gamma^8\cdots\Gamma^{p'+1}\right)^T
C\Gamma^0\dots \Gamma^{p'}(1+\eta\Gamma^{11})
\Gamma^9\Gamma^8\cdots\Gamma^{p'+1}\right]_{AB}|A\rangle|\tilde{B}\rangle\nonumber\\
&&=\left[C\Gamma^0\dots \Gamma^{p'}(1+\eta\Gamma^{11})\right]_{AB}
|A\rangle|\tilde{B}\rangle ~~.
\label{tpeo}
\end{eqnarray}
In the latter two equations we have used the following
identities:
\begin{eqnarray}
\left(\Gamma^{11}\right)^T=-C\Gamma^{11}C^{-1} \qquad \left(\Gamma^{M}\right)^T
=-C\Gamma^{M}C^{-1}\,\,\,\,\,M=0\dots9~~.
\end{eqnarray}

\section{Euler-Heisenberg actions}
\label{euheise}

We start by summarizing the calculation of the Euler-Heisenberg
action for an arbitrary gauge theory containing $N_s$ real scalars
and $N_f$ Dirac fermions and described by the following Lagrangian
in $d$ space-time dimensions:
\begin{eqnarray}
L = - \frac{1}{4g^2} F_{\mu \nu}^a F_{\mu \nu}^a + \frac{1}{2}
(D_{\mu}\Phi)^i  (D_{\mu} \Phi )^i + i {\bar{\Psi}}^i \gamma^{\mu}
D^{ab}_{\mu}  \Psi^i ~~.\label{lagra29}
\end{eqnarray}
If we expand it around a  background ${\bar{A}}_{\mu}^{a} $ solution
of the classical equations of motion, assuming that the fluctuation
${\cal{A}}$ satisfies the background gauge condition,
\begin{eqnarray}
A_{\mu}^{a} = {\bar{A}}_{\mu}^{a} + {\cal{A}}_{\mu}^{a}~~,~~ (
{\bar{D}}_{\mu} {\cal{A}}_{\mu} )^a =0 \label{expa391}
\end{eqnarray}
and keeping only up to the quadratic terms in the fluctuations, we
get the Euler-Heisenberg effective action:
\begin{eqnarray}
S_{EH} =\frac{1}{2} Tr \log \Delta_1 + ( \frac{N_s}{2} -1 )  Tr \log
\Delta_0 - N_f Tr \log \Delta_{1/2}~~,\label{euhei}
\end{eqnarray}
where we have neglected the contribution of the classical action,  and
\begin{eqnarray}
(\Delta_1 )^{ab}_{\mu \nu} = - ( {\bar{D}}^2 )^{ab} \delta_{\mu \nu}
+ 2 f^{acb} {\bar{F}}_{\mu \nu}^{c} \hspace{.5cm} (\Delta_0 )^{ij} =
- ( {\bar{D}}^2 )^{ij} \hspace{.5cm} (\Delta_{\frac{1}{2}})^{ij} = i
\gamma^{\mu} ({\bar{{D}}_{\mu} })^{ij}.\label{opera32}
\end{eqnarray}
${\bar{{D}}_{\mu} }$ is the covariant derivative computed in the
classical background. The determinant in Eq. (\ref{euhei}) can be
explicitly computed if we assume that the field strength
corresponding to the background ${\bar{A}}$ is constant. In this
case we get:
\begin{eqnarray}
S_{EH} = N I_1 + c_s I_0 + c_f I_{1/2}~~,\label{euhei2}
\end{eqnarray}
where
\begin{eqnarray}
I_0
= \frac{1}{(4\pi)^{d/2}} \int d^d x \int_{0}^{\infty} \frac{d
  \sigma}{\sigma^{1+d/2}} {e}^{- \sigma m^2}
{\cdot} \frac{{\hat{f}} \sigma}{\sin ({\hat{f}}
  \sigma)} {\cdot} \frac{{\hat{g}} \sigma}{\sinh {(\hat{g}}
  \sigma)}
\label{scala56}
\end{eqnarray}
for the scalar,
\[
I_1
= \frac{N}{(4\pi)^{d/2}} \int d^d x \int_{0}^{\infty} \frac{d
  \sigma}{\sigma^{1+d/2}} {e}^{- \sigma m^2} {\times}
\]
\begin{eqnarray}
{\times} \frac{{\hat{f}} \sigma}{\sin ( {\hat{f}}
  \sigma)} {\cdot} \frac{{\hat{g}} \sigma}{\sinh ({\hat{g}}
  \sigma)} \left[ d-2 + 4\sinh^2 ({\hat{g}} \sigma) - 4\sin^2 ({\hat{f}}
\sigma) \right] \label{glu56}
\end{eqnarray}
for the gluon that includes also the contribution of the
Faddev-Popov ghost and
\begin{eqnarray}
I_{1/2}
= - \frac{ 2^{[d/2]}}{(4\pi)^{d/2}} \int d^d x \int_{0}^{\infty}
\frac{d
  \sigma}{\sigma^{1+d/2}} {e}^{- \sigma m^2}
\frac{{\hat{f}} \sigma}{\sin ({\hat{f}}
  \sigma)} \frac{{\hat{g}} \sigma}{\sinh ({\hat{g}}
  \sigma)} cos ({\hat{f}} \sigma)  cosh ({\hat{g}} \sigma)
\label{ferm56}
\end{eqnarray}
for a complex fermion. We have introduced an infrared cut-off $m$
and we have taken the constant field strength with the following
form:
\begin{eqnarray}
\bar{F}_{\alpha \beta } =  \left(
\begin{array}{ccccc}
0  & {\hat{f}} & 0 & 0 & {\cdot}\\
-{\hat{f}} & 0 & 0 & 0 & {\cdot}\\
0  & 0 & 0 & {\hat{g}} & {\cdot}   \\
0  & 0 &- {\hat{g}}& 0 & {\cdot} \\
{\cdot}  & {\cdot} & {\cdot} & {\cdot} & {\cdot}
\end{array}
\right) ~~.\label{effebar}
\end{eqnarray}
All the other matrix elements are zero. The constants $c_s$ and
$c_f$ set the normalization of the generators of the $SU(N)$ gauge
group, namely $Tr ( T^A  T^B ) = c \delta^{AB}$ and $[\frac{d}{2}] =
\frac{d}{2}$ if $d$ is even and $[\frac{d}{2}] = \frac{d-1}{2}$ if
$d$ is odd. Remember that $c=N$ in the adjoint representation of
$SU(N)$ as we have used in Eq. (\ref{euhei2}) for the gluon
contribution. Eq.s (\ref{scala56}) and (\ref{ferm56}) are derived in
Appendix B of Ref.~\cite{MNP} and Eq. (\ref{glu56}) can be derived
in a similar way.

In the following we perform the field theory limit of the various
one-loop open string contributions with a constant gauge field that
we constructed for different models. By this limit we mean:
\begin{eqnarray}
\tau \rightarrow \infty~~,~~\alpha' \rightarrow 0~~;~~ \sigma \equiv
2 \pi \alpha' \tau\,\,,\,\, {\hat{f}}\,\,,\,\, {\hat{g}} \,\, fixed.
\label{ftlimit}
\end{eqnarray}
We start by considering the bosonic string. In this case the annulus
diagram with a constant gauge field  is given in Eq.s
(\ref{zetaebos}) and (\ref{op56}).   In the case of the bosonic
string the field theory limit is not well-defined because we have an
open string  tachyon and we have to eliminate its contribution by
hand keeping only the contribution of the massless open string
states. In the field theory limit we can use the expressions:
\begin{eqnarray}
\nu_f  \sim - 2i \alpha ' {\hat{f}}~~~,~~~\nu_g  \sim - 2  \alpha '
{\hat{g}}~~, \label{nfng}
\end{eqnarray}
\begin{eqnarray}
\Theta_1 ( i \nu_f \tau| i \tau) \sim -2 k^{1/6} f_1 (k) \sin (
{\hat{f}} \sigma ) \left[ 1 - 2 \cos (2 {\hat{f}} \sigma) k^2 +
\dots \right] \label{the178}
\end{eqnarray}
and
\begin{eqnarray}
\Theta_1 ( i \nu_g \tau| i \tau) \sim 2 i  k^{1/6} f_1 (k) \sinh (
{\hat{g}} \sigma ) \left[ 1 - 2 \cosh (2 {\hat{g}} \sigma) k^2 +
\dots \right]~~, \label{the179}
\end{eqnarray}
where the dots denote the contribution of the massive states. By
using the previous equations together with Eq. (\ref{equa23}), we
get for the untwisted contribution in Eq. (\ref{zetaebos}) the
following expression:
\begin{eqnarray}
(Z^{o}_{e})^{ftl}_{bos} = \frac{N}{ 2( 4 \pi)^2} \int d^4 x
\int_{0}^{\infty} \frac{d \sigma}{\sigma^3} {e}^{- m^2 \sigma}
\frac{{\hat{f}}\sigma}{\sin {\hat{f}} \sigma}
\frac{{\hat{g}}\sigma}{\sinh {\hat{g}} \sigma}\nonumber
\\ \times  \left[ 24 - 4 \sin^2
({\hat{f}} \sigma) +  4  \sinh^2 ({\hat{g}}
  \sigma) \right]~,
\label{zoe}
\end{eqnarray}
where $ m \equiv \frac{y}{2 \pi \alpha'}$. If we remember that in
our case we are considering a D3 brane in a $26$-dimensional space,
we see that the previous equation, multiplied by a factor 2 due to
the missing orbifold projector, corresponds to the four-dimensional
Euler-Heisenberg action for a gluon and $22$ adjoint scalars:
\begin{eqnarray}
(Z^{o}_{e})^{ftl}_{bos} =  N( I_1 + 22 I_0 )~. \label{ftl92}
\end{eqnarray}
The field theory limit of the twisted contribution can be performed
in the same way. When we sum it to the untwisted one we get:
\begin{eqnarray}
(Z^{ftl})_{bos} = N(  I_1 + N_s I_0 ) ~~,\label{zftl}
\end{eqnarray}
that is equal to the Euler-Heisemberg action for a gluon and $N_s$
scalar in the adjoint representation of $SU(N)$.

Let us now perform the field theory limit on the sum of the
untwisted sector in Eq. (\ref{zetae}) and of the twisted sector one
in Eq. (\ref{zetatet}).  We have to use the following equations
valid in the field theory limit:
\begin{eqnarray}
\Theta_3 (i \nu \tau | i \tau ) \sim 1 + 2 \cosh (2 \pi \nu \tau)
k~~~,~~~\Theta_4 (i \nu \tau | i \tau ) \sim 1 - 2 \cosh (2 \pi \nu
\tau) k \label{the45}
\end{eqnarray}
\begin{eqnarray}
\Theta_1 (i \nu \tau | i \tau ) \sim-2 k^{1/4} \sin (i \pi \nu
\tau)~~,~~ \Theta_2 (i \nu \tau | i \tau ) \sim 2 k^{1/4} \cos ( i
\pi \nu \tau )~~.\label{the46}
\end{eqnarray}
In particular the field theory limit is given by:
\begin{eqnarray}
Z^o\equiv -\frac{N}{(4\pi)^2} \int d^4 x  \int_{0}^{\infty}
\frac{d\sigma}{\sigma} e^{- m^2 \sigma } \frac{{\hat{f}} \sigma }{
\sin
  ({\hat{f}} \sigma)} {\cdot} \frac{ {\hat{g}} \sigma}{\sinh ({\hat{g}}
  \sigma )}
\left( {{\hat{Z}} }^o_e + {{\hat{Z}}}^o_h \right)~,
\end{eqnarray}
where:
\begin{eqnarray}
{\hat{Z}}^o_e&\simeq&  - \frac{1}{4k} \left[(1+4k)(1+2
\cos(2{\hat{f}} \sigma)k +2
\cosh ( 2{\hat{g}} \sigma)k)\right.\nonumber\\
&-&\left.(1-4k)(1-2 \cos({2\hat{f}} \sigma )k -2 \cosh
(2{\hat{g}\sigma}) k ) -16 k \cos({\hat{f}} \sigma)
\cosh ({\hat{g}} \sigma) \right]\nonumber\\
&=&- \frac{1}{2} \left[ \left( 8 + 4 \sin^2 ({\hat{f}} \sigma) - 4
\sinh
  ({\hat{g}} \sigma )\right)   - 8 \cos ({\hat{f}} \sigma )
\cosh ({\hat{g}} \sigma )\right] \\
{\hat{Z}}^o_h&\simeq& 2  \left[  {\rm sin}^2 ({\hat{f}} \sigma) -
{\rm sinh}^2 ({\hat{g}} \sigma) \right]~.
\end{eqnarray}
This means that the contribution of the untwisted sector multiplied
by a factor $2$, in order to get rid of the factor $1/2$ of the
orbifold projection, is equal to:
\[
(Z^{o}_{e})^{ftl} = \frac{N}{(4 \pi)^2} \int d^4 x \int_{0}^{\infty}
\frac{d \sigma}{\sigma^3} {e}^{- m^2 \sigma}
\frac{{\hat{f}}\sigma}{\sin
    ({\hat{f}} \sigma) } {\cdot} \frac{{\hat{g}}\sigma}{\sinh
    ({\hat{g}} \sigma)} {\times}
\]
\begin{eqnarray}
{\times} \left[ 8 + 4 \sinh^2 ({\hat{g}} \sigma ) - 4 \sin^2
({\hat{f}} \sigma )  -  8 \cos ({\hat{f}} \sigma ) \cosh ({\hat{g}}
\sigma ) \right]~~,\label{zetaeeh}
\end{eqnarray}
where the first three terms in the last line come from the NS sector
and correspond to one gluon and six scalars, while the last term
comes from the R sector and corresponds to $4$ Majorana fermions.
Both the scalars and the fermions are in the adjoint representation
of $SU(N)$. In conclusion the previous equation can be written as:
\begin{eqnarray}
(Z^{o}_{e})^{ftl} = N( I_1 + 6 I_0 + 2 I_{1/2})~~,\label{n=4}
\end{eqnarray}
that is the Euler-Heisenberg action of ${\cal{N}}=4$ super
Yang-Mills. It is also equal to what one gets for a D3 brane in type
0' theory because the contribution of the M{\"{o}}bius diagram is
vanishing for a D3 brane (see Eq.s (\ref{185}) and (\ref{freeomf})).

The field theory limit of the twisted contribution in Eq.
(\ref{zetatet}) is given by:
\begin{eqnarray}
(Z^{o}_{h})^{ftl} = -\frac{N}{(4 \pi)^2} \int d^4 x
\int_{0}^{\infty} \frac{d \sigma}{\sigma^3} {e}^{- m^2 \sigma}
\frac{{\hat{f}}\sigma}{\sin
    ({\hat{f}} \sigma)} \frac{{\hat{g}}\sigma}{\sinh
    ({\hat{g}} \sigma)} \nonumber \\
\times \left[2 \sin^2 ({\hat{f}} \sigma ) -
2 \sinh^2 ({\hat{g}} \sigma ) \right]~~,\label{zetateteh}
\end{eqnarray}
with in addition the $\theta$ term that we omit to write here.

Multiplying the contribution in Eq. (\ref{zetaeeh}) with a factor
$1/2$ due to the orbifold projection and summing it to the twisted
one in Eq. (\ref{zetateteh}), we get the Euler-Heisenberg action of
${\cal{N}}=2$ super Yang-Mills:
\begin{eqnarray}
Z^{ftl}_{{\cal{N}}=2}  = N( I_1 + 2 I_0 + I_{1/2})~~.\label{n=2}
\end{eqnarray}
In the last part of this Appendix we perform the field theory limit
 of the orientifold $\Omega' I_6$
discussed in Sect. (\ref{1lof}). The annulus diagram for a D3 brane
of this orientifold is equal to the untwisted part of the annulus
diagram for a D3 brane of type IIB theory on the orbifold
$\mathbb{C}^2/ \mathbb{Z}_2$. In the field theory limit one gets the
expression
 in Eq. (\ref{zetaeeh}).

The contribution of the M{\"{o}}bius diagram to the Euler-Heisenberg
action can be obtained from Eq. (\ref{ZF03}) using in the field
theory limit (see Eq.(\ref{ftlimit})) Eq.s
(\ref{the12})$\div$(\ref{equa23}). We get:
\begin{eqnarray}
Z^F\simeq\mp \frac{16}{(4\,\pi)^2 } \int d^4 x  \int_{0}^{\infty}
\frac{d\sigma}{\sigma} {\rm e}^{-\frac{y^2}{(2 \pi
\alpha')^2}\,\sigma}\,\, \hat{f}\,\hat{g}\,\,\frac{ \cos(\sigma
\hat{f}) \cosh(\sigma \hat{g})}{ \sin(\sigma \hat{f}) \sinh(\sigma
\hat{g})} ~~.\label{263bis}
\end{eqnarray}
Adding Eq.s (\ref{zetaeeh}) and (\ref{263bis}) we get the total
contribution for the theory described by the orientifold $\Omega '
I_6$ that is equal to:
\begin{eqnarray}
S_{EU}^{or} = N ( I_{1} + 6 I_0 ) + 4 \frac{N \pm 2 }{2} I_{1/2}~~,
\label{finori}
\end{eqnarray}
that is the correct Euler-Heisenberg action for a system of one
gluon, six adjoint scalars, but not the correct one for four Dirac
fermions transforming according to the two-index (anti)symmetric
representation of $SU(N)$.

\end{document}